\documentclass[12pt,german,english,twoside,a4paper%
,openright 										
]{report}

\usepackage[ngerman, english]{babel}	
\usepackage[latin1]{inputenc}		
\usepackage[T1]{fontenc}				
\usepackage{graphicx}						

\usepackage{amsmath,amssymb}
\usepackage[footnotesize,sc]{caption}
\usepackage[numbers,sort&compress]{natbib}

\usepackage{fancyhdr}

\newcommand{\dcauthorpre}{Herr Dipl.-Phys.} 
\newcommand{\dcauthorsurname}{Leder} 
\newcommand{\dcauthorname}{Björn} 
\newcommand{\dcauthoradd}{geboren am 22.10.1978 in Altenburg}

\newcommand{\dctitle}{The Schr\"odinger functional for Gross-Neveu models} 
\newcommand{\dcsubtitle}{~}  

\newcommand{\dcapprovala}{Dr. Rainer Sommer} 
\newcommand{\dcapprovalb}{Prof. Dr. Ulrich Wolff} 
\newcommand{\dcapprovalc}{Dr. Peter Weisz} 

\newcommand{\dcdegree}{doctor rerum naturalium\\(Dr. rer. nat.)} 
\newcommand{\dcsubject}{Physik} 
\newcommand{\dcfaculty}{Mathematisch-Naturwissenschaftlichen Fakult\"at I}
\newcommand{\dcuniversity}{Humboldt-Universit\"at zu Berlin}
\newcommand{\dcdean}{Prof. Dr. Christian Limberg}
\newcommand{\dcpresident}{Prof. Dr. Christoph Markschies}

\newcommand{\dcdatesubmitted}{26. Januar 2007} 
\newcommand{\dcdateexam}{18. April 2007} 

\newcommand{\dckeydea}{Chirales Gross-Neveu Modell}
\newcommand{\dckeydeb}{Ginsparg-Wilson Fermionen}
\newcommand{\dckeydec}{Schr\"odinger Funktional}
\newcommand{\dckeyded}{Gitterst\"orungstheorie}

\newcommand{\dckeywordsde}{\vfill \raggedright {\textbf{Schlagw\"orter:}}\\ \dckeydea, \dckeydeb, \dckeydec, \dckeyded \\}

\newcommand{\dckeyena}{Chiral Gross-Neveu model}
\newcommand{\dckeyenb}{Ginsparg-Wilson fermions}
\newcommand{\dckeyenc}{Schrödinger functional}
\newcommand{\dckeyend}{Lattice perturbation theory}

\newcommand{\dckeywordsen}{\vfill \raggedright {\textbf{Keywords:}}\\ \dckeyena, \dckeyenb, \dckeyenc, \dckeyend \\}

\newcommand{\dcpdfsubject}{Dissertation}							
\usepackage{ifpdf}

\ifpdf
\usepackage[%
	pdftitle={\dctitle},
	pdfauthor={\dcauthorsurname\ \dcauthorname},
	pdfsubject={\dcpdfsubject}, 
	pdfkeywords={\dckeydea, \dckeydeb, \dckeydec, \dckeyded},
	pdfpagemode=UseOutlines,
	linktocpage=true,
  colorlinks=true,					
	linkcolor=black,					
	filecolor=black,					
	urlcolor=black,						
	citecolor=black,					
	pdftex=true,              
	plainpages=false,         
	hypertexnames=false,      
	pdfpagelabels=true,       
	hyperindex=true]{hyperref}
\else

	\usepackage[dvips,linktocpage]{hyperref}
\fi

\newcommand{\psibar}{\overline{\psi}}

\newcommand{\rhobar}{\overline{\rho}}
\newcommand{\etabar}{\overline{\eta}}
\newcommand{\zetabar}{\overline{\zeta}}

\newcommand{\gmu}{\gamma_\mu}
\newcommand{\gnu}{\gamma_\nu}
\newcommand{\go}{\gamma_0}
\newcommand{\gl}{\gamma_1}
\newcommand{\gi}{\gamma_i}
\newcommand{\gfive}{\gamma_5}
\newcommand{\dmu}{\partial_\mu}
\newcommand{\dmub}{{\partial^\ast}_{\mspace{-11.0mu} \mu}}
\newcommand{\db}[1]{{\partial^\ast}_{\mspace{-11.0mu} #1}}
\newcommand{\dk}{\partial_k}
\newcommand{\dkb}{{\partial^\ast}_{\mspace{-11.0mu} k}}
\newcommand{\md}{\widetilde{\partial}}
\newcommand{\mdmu}{\md_\mu}

\newcommand{\e}{\mathrm{e}}
\newcommand{\rd}{\mathrm{d}}
\newcommand{\rD}{\mathrm{D}}
\newcommand{\rO}{\mathrm{O}}
\newcommand{\<}{\langle}
\newcommand{\z}{\rangle}
\newcommand{\GammaI}{\Gamma_{\mathrm{I}}}
\newcommand{\GammaR}{\Gamma_{\mathrm{R}}}

\newcommand{\p}{{p_1}}
\newcommand{\tr}{\,\mathrm{tr}}
\newcommand{\trf}{\,\mathrm{tr}_f}
\newcommand{\trd}{\,\mathrm{tr}_d}
\newcommand{\trdf}{\,\mathrm{tr}_{d,f}}
\newcommand{\TRdf}[1][NO]{\,\trdf\left\{#1\right\}}
\newcommand{\TRd}[1][NO]{\,\trd\left\{#1\right\}}
\newcommand{\TRf}[1][NO]{\,\trf\left\{#1\right\}}

\newcommand{\Tr}[1][NO]{\,\mathrm{Tr}\left\{#1\right\}}

\newcommand{\moverset}[2]{\overset{\scriptstyle #1}{\phantom{\scriptstyle #2}}\mspace{-14.0mu} #2}
\newcommand{\oover}[1]{\,\overset{\scriptscriptstyle\circ}{\phantom{\scriptstyle #1}}\mspace{-9.0mu} #1}
\newcommand{\po}{\oover{p}}
\newcommand{\+}{{\scriptscriptstyle +}}

\newcommand{\pp}{p^\+}
\newcommand{\pmu}{p_\mu}
\newcommand{\Sp}{\widetilde{S}}
\newcommand{\li}{\left}
\newcommand{\re}{\right}
\newcommand{\ren}[1]{\li(#1\re)_\mathrm{R}}
\newcommand{\ev}[1]{\bigl\<#1\bigr\z}
\newcommand{\threevector}[1]{\mathbf{{#1}}}
\newcommand{\vx}{\threevector{x}}
\newcommand{\vy}{\threevector{y}}
\newcommand{\vp}{\threevector{p}}
\newcommand{\phat}{\hat{p}}
\newcommand{\vphat}{\hat{\vp}}
\newcommand{\eqsref}[2]{(\ref{#1}--\ref{#2})}
\newcommand{\Dw}{D_\text{W}}
\newcommand{\Dgw}{D_\text{N}}
\newcommand{\weq}[1]{\stackrel{\text{\eqref{#1}}}{=}}
\newcommand{\eqw}[1]{\stackrel{#1}{=}}
\newcommand{\Op}{\mathcal{O}}
\newcommand{\tD}{\widetilde{D}}
\newcommand{\Oss}{O_{SS}}
\newcommand{\Opp}{O_{PP}}
\newcommand{\Ovv}{O_{VV}}
\newcommand{\Opss}{O'_{SS}}
\newcommand{\Oppp}{O'_{PP}}
\newcommand{\Opvv}{O'_{VV}}
\newcommand{\Ohss}{\hat{O}_{SS}}
\newcommand{\Ohpp}{\hat{O}_{PP}}
\newcommand{\Ohvv}{\hat{O}_{VV}}

\newcommand{\Ohpvv}{\hat{O}'_{VV}}
\newcommand{\gp}{{g'}}
\newcommand{\deltap}{{\delta'}}
\newcommand{\gfiveh}{\hat{\gamma}_5}
\newcommand{\eq}[1]{Eq.\ \eqref{#1}}
\newcommand{\tab}[1]{Table \ref{#1}}
\newcommand{\unity}{1\mspace{-4.5mu} \mathrm{l}}
\newcommand{\tw}{{\text{w}}}
\newcommand{\tgw}{{\text{gw}}}

\newtheorem{lemma}{Lemma}
\newtheorem{definition}{Definition}
\newtheorem{theorem}{Theorem}

\usepackage{calc}

\fancyheadoffset[LE,RO]{\marginparsep+\marginparwidth}

\fancypagestyle{plain}{%
  \fancyhf{}
	\fancyhead{} 
}

\begin{document}

\pagenumbering{roman}


%
%
%
%

\author{von \\ \dcauthorpre\ \dcauthorname\ \dcauthorsurname\ \\ \dcauthoradd}

\title{ \vspace{-5cm}\dctitle \\ 
\vspace{0.5cm}
\large{\dcsubtitle} \\ 
\vspace{0.5cm} {\Large{DISSERTATION}}\\ 
\vspace{0.5cm} \large{zur Erlangung des akademischen Grades \\ 
\dcdegree\\ im Fach \dcsubject \\ 
\vspace{0.5cm} eingereicht an der \\ 
\dcfaculty \\ 
\dcuniversity \\}}
\date{\vspace{0.5cm}
\raggedright{
Pr\"asident der Humboldt-Universit\"at zu Berlin:\\
\dcpresident \vspace{-0.3cm}
}\vspace{0.5cm}\\
\raggedright{
Dekan der \dcfaculty:\\
\dcdean \vspace{-0.3cm}
}\vspace{0.5cm}\\
\raggedright{
Gutachter:
\begin{enumerate} 
\item{\dcapprovala} \vspace{-0.3cm}
\item{\dcapprovalb} \vspace{-0.3cm}
\item{\dcapprovalc} \vspace{-0.3cm}
\end{enumerate}} \vspace{0.5cm}
\raggedright{
\begin{tabular}{lll}
eingereicht am: &  &\dcdatesubmitted\\ 
Tag der m\"undlichen Pr\"ufung: & & \dcdateexam
\end{tabular}}\\ 
}
\maketitle



\selectlanguage{english}

\begin{abstract}
\setcounter{page}{2} 
Gross-Neveu type models with a finite number of fermion flavours are studied on a two-dimensional Euclidean space-time lattice. 
The models are asymptotically free and are invariant under a chiral symmetry. These similarities to QCD make them
perfect benchmark systems for fermion actions used in large scale lattice QCD computations.
The Schr\"odinger functional for the Gross-Neveu models is defined for both, Wilson and Ginsparg-Wilson fermions, and shown to 
be renormalisable in 1-loop lattice perturbation theory.

In two dimensions four fermion interactions of the Gross-Neveu models have dimensionless coupling constants. The symmetry properties
of the four fermion interaction terms and the relations among them are discussed.
For Wilson fermions chiral symmetry is explicitly broken and additional terms must be included in the action. Chiral symmetry is restored
up to cut-off effects by tuning the bare mass and one of the couplings. The critical mass and the symmetry restoring coupling
are computed to second order in lattice perturbation theory.

This result is used in the 1-loop computation of the renormalised couplings and the associated  beta-functions. 
The renormalised couplings are defined in terms of suitable boundary-to-boundary correlation functions.
 In the computation the known first order coefficients of the beta-functions
are reproduced. One of the couplings is found to have a vanishing beta-function.
The calculation is repeated for the recently proposed  Schr\"odinger functional with exact chiral symmetry, i.e. Ginsparg-Wilson fermions.
The renormalisation pattern is found to be the same
as in the Wilson case. Using the regularisation dependent finite part of the renormalised couplings, the ratio of the Lambda-parameters is computed.
\\
\dckeywordsen				
\end{abstract}


\selectlanguage{german}

\begin{abstract}
\setcounter{page}{3} 
In dieser Arbeit werden Gross-Neveu Modelle mit einer endlichen Anzahl von Fermiontypen auf einem zweidimensionalen 
Euklidischen Raumzeitgitter betrachtet.
Modelle dieses Typs sind asymptotisch frei und invariant unter einer chiralen Symmetrie. Aufgrund dieser Gemeinsamkeiten mit QCD
sind sie sehr gut geeignet als Testumgebungen f\"ur Fermionwirkungen die in gro{\ss}angelegten Gitter-QCD-Rechnungen benutzt werden.
Das Schr\"odinger Funktional f\"ur die Gross-Neveu Modelle wird definiert f\"ur Wilson und Ginsparg-Wilson Fermionen.  In 1-Schleifenst\"orungstheorie
wird seine Renormierbarkeit gezeigt. 

Die Vier-Fermionwechselwirkungen der Gross-Neveu Modelle habe dimensionslose Kopplungskonstanten in zwei Dimensionen. Die Symmetrieeigenschaften
der Vier-Fermionwechselwirkungen und deren Beziehungen untereinander werden diskutiert. Im Fall von Wilson Fermionen ist die chirale Symmetrie explizit 
gebrochen und zus\"atzliche Terme m\"ussen in die Wirkung aufgenommen werden. Die chirale Symmetrie wird durch das Einstellen der nackten Masse und einer der 
Kopplungen bis auf Cut-off-Effekte wiederhergestellt. Die kritische Masse und die symmetriewiederherstellende Kopplung
werden bis zur zweiten Ordnung in Gitterst\"orungstheorie berechnet.

Dieses Resultat wird in der 1-Schleifenberechnung der renormierten Kopplungen und der zugeh\"origen Betafunktionen benutzt. Die renormierten Kopplungen werden definiert mit Hilfe von geeignete Rand-Rand-Korrelatoren. Die Rechnung reproduziert die bekannten f\"uhrenden Koeffizienten der  Betafunktionen. Eine der Kopplungen hat eine verschwindende Betafunktion. Die Rechnung wird mit dem vor kurzem vorgeschlagenen Schr\"odinger Funktional mit exakter chiraler
Symmetrie, also Ginsparg Wilson Fermionen, wiederholt. Es werden die gleichen Divergenzen gefunden, wie im Fall von Wilson Fermionen. Unter Benutzung des 
regularisierungsabh\"angigen, endlichen Teils der renormierten Kopplungen werden die Verh\"altnisse der Lambda-Parameter bestimmt.\\
\dckeywordsde
\end{abstract}

\selectlanguage{english}				
\setcounter{page}{4}  					



\selectlanguage{english}

\tableofcontents

\listoffigures

\listoftables
\cleardoublepage


\pagenumbering{arabic}



\chapter{Introduction}

The theory of strong interactions, quantum chromodynamics (QCD), is a gauge theory with few parameters and is though assumed to describe many phenomenons,
such as the mass spectrum of the hadrons and scattering processes involving quarks and gluons. Experimental data is available from the low energy
regime of the light meson masses to the high energy regime of hadron-hadron scattering  \cite{Yao:2006px}.

At high energies the fundamental degrees of freedom, the quarks and the gluons, are only weakly coupled.
In such a situation a perturbative treatment,
where the interactions of the quarks and gluons are small corrections, is justified. Indeed, perturbative calculations successfully describe the data, for
example, of deep inelastic hadron-lepton scattering. At low energies the coupling becomes large and the interactions are not small but dominant. Therefore the perturbative approach is not applicable in this regime.  
In addition, the relevant degrees of freedom are no longer the fundamental quarks and gluons, but the lightest bound states (the light mesons, i.e. pions).
Chiral  perturbation theory has been developed to accommodate this. But as an effective theory it has to be matched to the experiments,  thus losing 
 the appeal of a first principle  computation.
 
 The lattice discretisation of quantum field theories is a powerful method that was first applied to QCD long ago by Wilson
  \cite{Wilson:1974sk}. Since then lattice QCD has proved to allow for non-perturbative calculations from first principle and to connect the low and high
	energy regime (see	\cite{DellaMorte:2004bc} for an example).
	First of all the lattice regulates the theory, in which the inverse lattice spacing $1/a$ serves as a sharp momentum cut-off
	and thus renders the theory ultraviolet finite. In the infrared a non-vanishing mass or a finite volume scheme with specific boundary
	conditions like the Schrödinger functional (SF) provides a lower bound on the modes in the theory.
	(The merits of the SF are described in Chapter \ref{schroendinger_functional}.)
	 In this way a quantum field
	theory on the lattice is mathematically well defined without reference to perturbation theory. Nevertheless perturbation theory can be used at this
	point. Being in general more complicated than similar studies in the continuum, lattice perturbation theory is needed, for example,
  to translate lattice results into the language of continuum renormalisation schemes like $\overline{\text{MS}}$ (minimal subtraction),
	 that are mostly used by experimentalists.
	Furthermore,	in the weak coupling regime, it serves as guidance and cross check for non-perturbative methods.
	 
	If the metric of the lattice theory is the Euclidean one,
	the path integral representation of quantum field theory is accessible to numerical evaluation via Monte Carlo simulations.
	Today this is the prominent approach to extract physics from lattice QCD. Since the days of Wilsons proposal there has been substantial progress 
	in the understanding of lattice QCD as well as in the algorithms that are used to perform the simulations. Still, the by definition limited computational
	resources are the main obstacle to accomplish the goal of producing predictions without any compromise. 
	
	Along the way so called ``toy models'' were studied. The term refers to quantum field theories that are simpler but in some respects similar to QCD.
	They were used, for example, to test new methods \cite{Luscher:1991wu} or to conjecture the phase structure of lattice QCD \cite{Aoki:1983qi}.
	One reason to consider these simpler models is that there might be analytical tools at hand that allow one to solve the model exactly. Often another 
	advantage is that Monte Carlo simulations of the toy model are much more cheaper. The	knowledge and results gained in the simpler models 
	can then be used to argue in the more complicated theory. Or assumptions and approximations used in lattice QCD without a chance to prove their
	validity there, can be applied and confronted with the exact result and/or high precision numerical data in the simpler theory.
	 
	For example, in order to	make predictions about the real world, the lattice discretisation, as any
	regulator, has to be removed in the end. Numerical simulations of lattice QCD are only possible at finite lattice spacing $a$. In practice one computes the 
	observable of interest for a number of lattice spacings and extrapolates to the continuum limit. If the measured points show a significant dependence on 
	the lattice spacing one has to assume a functional form to perform the extrapolation. The only known prescription of the lattice artefacts goes back
	to the work of Symanzik \cite{Symanzik:1979ph,Symanzik:1981hc,Symanzik:1983dc,Symanzik:1983gh}.
	His conclusions for the functional form of lattice artefacts in lattice field theories are based on an
	effective theory and perturbation theory. Using this form in the extrapolation step of lattice QCD computations introduces a possible source of systematic
	errors in a presumably first principle computation. In the two dimensional and asymptotically free non-linear sigma models these aspects can 
	be studied with high precision Monte Carlo simulations and analytic tools \cite{Knechtli:2005jh}.  The message for lattice QCD is clear: try to avoid the
	extrapolation step. This can be achieved partly by eliminating the leading lattice artefacts
	by implementing Symanzik's improvement program \cite{Symanzik:1983dc,Symanzik:1983gh}.
	
	At the time of writing this thesis a number of collaborations of lattice physicists are simulating full
	lattice QCD with two or three light quarks and have presented first results
	\cite{DellaMorte:2005kg,Schaefer:2005qg,Bernard:2006wx,Hasenfratz:2006xi,
		DelDebbio:2006cn,DelDebbio:2007pz,Boucaud:2007uk}. The main
	difference between the approaches followed by these collaborations is in the used fermion action.
	Beside the already mentioned Wilson fermions, there are Wilson twisted mass, overlap, staggered
	fermions and the fixed point action. There are also differences in the gauge action and the
	various kinds of improvement applied, but here we concentrate on the fermionic part of the action
	(see \cite{Jansen:2003nt} for an overview).
	
	Some of these fermion actions are more theoretically sound than others. The hope is, based on
	universality arguments, that in the continuum limit, where the correlation length diverges,
	differences on the scale of the lattice spacing are unimportant. Clearly, a numerical test 
	of this presumed agreement in the continuum limit is desirable. Due to the restricted numerical
	resources of today this is impossible in lattice QCD in the near future. In such a situation a
	two dimensional non-trivial fermionic quantum field theory could serve as a benchmark system.
	If the actions agree in the toy model, it would not be a proof for QCD, but some confidence would
	be gained.
	If they do not agree, it would be clear that there are serious problems and that most probably also
	lattice QCD simulations are affected.
		
	In this work we study models of self-coupled fermions in two dimensional space-time. 
	There are several theories of this kind in two dimensions referred to as Gross-Neveu \cite{Gross:1974jv}
	and Thirring models \cite{Thirring:1958in}. We consider here the first type.
	Among them the most similar to QCD is the	chiral Gross-Neveu model (CGN)  with $N$ types (in
	the following referred to as flavours) of fermions. They are coupled through quartic
	interaction terms. Since in two dimensions the fermion fields have mass dimension $1/2$, the corresponding
	couplings are dimensionless. The CGN shares with QCD 
	the features asymptotic freedom and a continuous chiral symmetry (in the massless theory).
	The continuum CGN has been studied in perturbation theory up to three loops 
	\cite{Mitter:1974cy,Destri:1988vb,Bondi:1989nq}. Beyond perturbation theory
	the $S$-matrix and the particle spectrum are known to some extent
	\cite{Dashen:1975xh,Kurak:1978su,Andrei:1979sq,Andrei:1979un,Forgacs:1991nk}. 
	Many properties	of the model can be studied in the limit of infinite many flavours (large-$N$ limit),
	which is also sensitive to the non-perturbative nature of the model. In this limit the
	model is asymptotically free and a fermion mass is dynamically generated
	\cite{Gross:1974jv,Witten:1978qu}.
	
	On the lattice the model has been studied so far almost exclusively in the large-$N$ limit
	\cite{Aoki:1985jj,Izubuchi:1998hy}. In this thesis we define the CGN with a finite number of fermion
	flavours 
	on the lattice. For the fermion action we use Wilson's version \cite{Wilson:1974sk} since it is the
	most rigorous and theoretically sound one. After the theory has been established in this way,
	it can be used
	to check other actions. As a first application we analyse a recently proposed Dirac operator 
	\cite{Luscher:2006df} that
	is expected to be compatible with the Schrödinger functional boundary conditions and, at the same time,
	is a solution to the Ginsparg-Wilson relation
	\cite{Ginsparg:1981bj}
	in the bulk of the lattice (up to exponentially decreasing	tails). A Dirac operator that satisfies
	this relation has better chiral properties than standard Wilson fermions and is thus
	better suited in cases were chiral symmetry plays an important role. Since Ginsparg-Wilson fermions
	are very expensive in terms of computational costs, precise studies in two dimensions are very welcome.
	We define observables suitable for Monte Carlo simulations \cite{Korzec:2005ed} and compute them in
	first order lattice perturbation theory. This is the first computation with Ginsparg-Wilson fermions
	in the Schrödinger functional beyond the free theory.
	 
	This thesis is organised as follows. 
	In Chapter \ref{lattice_pt} we give a short review of the perturbative renormalisation and discretisation
	of quantum field theories. In Chapter \ref{chiral_symmetry} aspects of chiral symmetry on the lattice
	are addressed
	and	the chiral properties of the Dirac operators used in this thesis are outlined. As indicated above
	we define the theory on a lattice with boundaries. The specific form of the boundary conditions 
	and the implications of the presence of the boundaries on the renormalisability are discussed
	thoroughly in Chapter \ref{schroendinger_functional}. The two dimensional fermionic model we utilise
	in this work is carefully defined in Chapter \ref{selfcoupled}. In particular, we are concerned with the
	symmetries of the model, its lattice formulation and renormalisability. Since Wilson fermions explicitly
	break chiral symmetry, it has to be assured that the symmetry is recovered in the continuum limit. 
	The employed strategy and the result are presented in Chapter \ref{chiral_sym_restoration}. 
	We use this result in Chapter \ref{renormalised_coupling} to define renormalised couplings.
	A next-to-leading order computation is then carried out for Wilson and Ginsparg-Wilson fermions.
	We draw conclusions and give an outlook in Chapter \ref{conclusions}.
\chapter{Lattice perturbation theory}
\label{lattice_pt}

 In this Chapter we introduce the basic concepts used in this thesis. Since we want to use lattice perturbation
 theory, we have to discuss perturbative renormalisation (Section \ref{renormalisation}). After these general
 remarks we list our notation and conventions for the lattice computation (Section \ref{discretisation}). 
 We close the Chapter with some remarks on the continuum limit and the analysis of lattice diagrams (Section \ref{continuum_limit}).

%
%
%
%

\section{Renormalisation}
\label{renormalisation}
	
	The bare couplings and masses that appear as parameters in the classical action of a quantum field theory
	are not the couplings and masses which are measured in experiments. Experimentalists rather gather data of
	cross sections and transition 
	amplitudes. These quantities have to be computed in the theory that is supposed to describe the
	phenomenons. For each parameter in the
	action one input measurement is needed. Once all the parameters are fixed the theory can be used for
	predictions.
	
	The crucial point of course is how many parameters are there in the action. The more parameter the less predicting power
	does the theory have. The number of terms in an action and thus the number of bare parameters is mainly restricted by symmetries and
	dimensional analysis.
	
	Computing a cross section or transition amplitude yields a relation between an observable and the bare
	parameters of the theory.
	The observable itself may now be called coupling. In order to avoid confusion one calls it renormalised coupling since it is 
	a redefinition of the bare coupling. In the same way all other couplings and masses may be redefined.
	The renormalised quantities
	may be regarded as the physical parameters of the theory because all observables can be expressed in
	terms of them.
	
	In perturbation theory the necessity for renormalisation is encountered in the form of ultraviolet infinities when calculating loop
	corrections.
	To keep physical amplitudes finite these infinities have to be absorbed in a redefinition of the parameters and fields order by order
	in the perturbative expansion.
	
	In order to handle the terms producing the infinities they first have to be rendered finite. In lattice perturbation theory the 
	inverse lattice	spacing $1/a$ provides an ultraviolet cut-off to the theory. This regularisation has to be removed before comparing
	with experiment. On the lattice this amounts to taking the continuum limit $a\to 0$. In this process the
	ultraviolet divergences show up and the renormalisation has to be implemented. 
	
\subsection{Mass independent renormalisation scheme}
\label{MIndependentRenScheme}

	Let us consider a quantum field theory with one mass and one coupling constant that has been regularised on an infinite lattice,
	say QCD with $N$ mass degenerated quarks.
	All information of the theory is contained in the $n$-point Green's functions. Any unrenormalised Green's function 
	$\Gamma(p;g_0,m_0,1/a)$
	will then depend on the momenta of the external lines collectively labelled with $p$, on the bare mass $m_0$ and coupling constant $g_0$
	and on the ultraviolet cut-off $1/a$.
	
	QCD is a renormalisable quantum field theory.
	A renormalisation scheme is given through conditions that define the renormalised mass $m_R$,
	renormalised coupling	$g_R$. Often also a wave function renormalisation factor $Z_i$ for each type $i$
	of fields in the theory (i.e. in QCD one for the quark fields
	and one for the gluon fields) is introduced. This is not a necessity but convenient in 
	the course of computations.
	Since we will consider massless field theory, with a renormalisation scale $\mu$,
	a mass independent renormalisation scheme is needed to 
	avoid infrared divergences \cite{Weinberg:1951ss}. 
	The renormalisation conditions are then posed at the scale $\mu$ and
	vanishing renormalised mass. For the renormalised parameters one expects
	\begin{align}
  		Z_i &= Z_i(g_0,a\mu)\,,\\
  		g_R &= g_0\,Z_g(g_0,a\mu)\,,\\
  		m_R &= m_q\,Z_m(g_0,a\mu)\,, \quad m_q=m_0-m_c \label{renMass}\,,
  \end{align}
  where $m_c$ accounts for the additive mass renormalisation needed if chiral symmetry is broken by the regularisation
   (cf. Section \ref{wilson_fermions}). If the regularisation does not violate chiral invariance the
   renormalised mass vanishes at zero bare mass.
 
 	The renormalized 
	coupling in lattice QCD, for example, may be defined through demanding the triple gluon vertex function to take its tree level
	value at momenta of order $\mu$. Then the renormalised Green's functions have finite continuum limits. They are functions of the
	 renormalised coupling and mass and are related to the bare ones as
	\begin{equation}\label{renormGammalimit}
  		\Gamma_R(p;g_R,m_R,\mu) = Z_\Gamma(g_0,a\mu) \Gamma(p;g_0,m_0,1/a)\,,
  \end{equation}
  where $Z_\Gamma$ depends on the number and types of the external lines.
  (Note that there will also be some 
  dependence on a gauge fixing parameter as in the continuum. However, this dependence disappears
  when physical quantities are computed and is not important for the aspects considered here. See
  \cite{Capitani:2002mp} for a complete review of lattice perturbation theory.) 
  
  Eq. \eqref{renormGammalimit} really only holds in the continuum limit. 
  At finite cut-off, that is finite lattice spacing $a$, perturbation theory states that the renormalised Green's functions
  are cut-off independent only up to terms of order $a$ 
	\begin{equation}\label{renormGamma}
  		\Gamma_R(p;g_R,m_R,\mu,a\mu) = \Gamma_R(p;g_R,m_R,\mu) + \rO\li(a(\ln a)^k\re)\,,
  \end{equation}
  at $k$-loop order \cite{Symanzik:1979ph}. These terms are called {\it scaling violations}. Since they are small near the continuum limit we suspend their discussion
  until Section \ref{continuum_limit} and neglect them in the following.

\subsection{Renormalisation group equations}\label{RGequations}

	Since $g_R$ and $m_R$ depend on the renormalisation scale $\mu$ while $\Gamma$ does not, differentiation on both sides of
	 \eqref{renormGammalimit}
	yields the so called renormalisation group equations
	\begin{equation}\label{RGE}
  		\li\{ \mu\frac{\partial}{\partial \mu} + \beta(g_R) \frac{\partial}{\partial g_R}
  			 + \tau(g_R) m_R \frac{\partial}{\partial m_R}  - \gamma_\Gamma(g_R) \re\}\Gamma_R = 0\,,
  \end{equation}
  where
	\begin{align}
  		\beta(g_R) &\equiv \mu\frac{\partial}{\partial \mu} g_R(g_0,a\mu)\,,\label{RGfunctions1}\\
  		\tau(g_R) &\equiv \mu\frac{\partial}{\partial \mu} \ln Z_m(g_0,a\mu)\,,\\
  		\gamma_\Gamma(g_R) &\equiv \mu\frac{\partial}{\partial \mu} \ln Z_\Gamma(g_0,a\mu)\label{RGfunctions2}\,.
  \end{align}
  The renormalisation group functions $\beta$, $\tau$ and $\gamma_\Gamma$ are the so called beta-function
  for the coupling and the
  anomalous dimensions of the mass and the Green's function. (If two types of fields appear in $\Gamma$, say $n_1$ of type one and $n_2$ of type
  two, we have to take $Z_\Gamma=Z_1^{n_1} Z_2^{n_2}$. Then $\gamma_\Gamma$ is the sum
  of the anomalous dimension of the two types $\gamma_\Gamma = n_1\gamma_1 + n_2 \gamma_2$.)  
  Note that the coefficient functions \eqsref{RGfunctions1}{RGfunctions2} must be independent of $a$ because they appear
  in a differential equation of an cut-off independent quantity. Since they are dimensionless they must also be independent of $\mu$.
  Thus they only depend on the renormalised coupling $g_R$.
    
  The functions $\beta$, $\tau$ and $\gamma_i$ can be calculated in perturbation theory as a power series
  in the renormalised coupling. In the case of QCD the beta-function
	\begin{equation}
  		\beta(\alpha_s) = b_0 \alpha_s^2 + b_1 \alpha_s^3 + b_2 \alpha_s^4 + \rO(\alpha_s^5)\,,
  \end{equation}
  is known to tree loops. The first two coefficients, for example, are
  \cite{Gross:1973id,Politzer:1973fx,Jones:1974mm,Caswell:1974gg}
	\begin{equation}\label{QCDbeta}
  		b_0 = -\li(11-\frac{2N}{3}\re)\, \quad b_1 = -\li(102-\frac{38N}{3}\re)\,.
  \end{equation}
  (Note that in many textbooks and publications another definition of the renormalised coupling of QCD is
  used. The relation to the one used here is $\alpha_s = g_R^2/4\pi$.)

\subsection{Asymptotic freedom}
\label{asymptotic_freedom}
	
	From the shape of the beta-function the behaviour of the renormalised coupling at high energies may be deduced. The example above is
	characterised by a negative $\beta(g)$ for small $g\ge 0$ and leads to a vanishing renormalised coupling as $\mu\to\infty$. This 
	behaviour is called asymptotic freedom. There are three other possible scenarios, we do not list them here but refer to the diverse
	textbooks on the topic
	\cite{Weinberg:1996kr,Montvay:1994cy,Peskin:1995ev}.
	To make the above statement more explicit and general, assume a beta-function that is negative for
	small positive $g$
	\begin{equation}\label{leadingbeta}
  		\beta(g) \stackrel{g\to 0}{\to} - b g^n\,,\quad b>0\,,
  \end{equation}
	where $g^n$ is the power of the coupling in front of the lowest-order divergent diagram
	contributing to $\beta(g)$ and therefore is always greater one. The
	renormalisation group equation is then
	\begin{equation}
  		\mu\frac{\rd}{\rd\mu}g(\mu) = - b g^n(\mu)\,.
  \end{equation}
  Given the renormalised coupling at some scale $\mu$ the coupling at the energy $E$ can be calculated by
  integrating this equation. The solution is
	\begin{equation}
  		g(E) = g(\mu)\li[1+(n-1)\,b\,\ln(E/\mu)\,(g(\mu))^{(n-1)}\re]^{-1/(n-1)}\,.
  \end{equation}
	For $E\to\infty$ this solution becomes independent of $g(\mu)$
	\begin{equation}
  		g(E) \stackrel{E\to\infty}{\to} \li[(n-1)\,b\,\ln(E/\mu)\re]^{-1/(n-1)}\,.
  \end{equation}
	Thus starting from a value justifying the approximation \eqref{leadingbeta} $g(E)$ always tends to zero for $E\to\infty$. On the other hand
	at small $E$ the coupling may become large $g(E)>1$. Thus perturbation theory becomes unreliably at small energies and non-perturbative 
	methods are needed.
	
	Along similar steps it can be shown that the effective dimensionality of operators and fields is given by dimensional analysis up to 
	logarithmic corrections \cite{Weinberg:1996kr}.
	
	In this context it is worth mentioning that the first two coefficients of the beta-function are independent of how exactly the
	renormalised coupling is defined as long as for small bare coupling
	 $g_R=g_0+\rO(g_0^2)$.\footnote{Note that notation might be misleading here. With $g_R$ a renomalised coupling like $\alpha_s$ of QCD is meant. The more familiar 
	renormlised coupling	of QCD, that is called $g_R$, the beta-function woud start with a third power. See also the note under \eqref{QCDbeta}}
	To see this assume two renormalised couplings $g_A$ and $g_B$. Since there are no other dimensionless parameters $g_A$ is a function only of
	$g_B$. We can expand the one in powers of the other
	\begin{equation}\label{gAexpandgB}
  		g_A(g_B)=g_B + c_1 g_B^2 + \rO(g_B^3)\,,
  \end{equation}
  or
	\begin{equation}\label{gBexpandgA}
  		g_B(g_A)=g_A - c_1 g_A^2 + \rO(g_A^3)\,,
  \end{equation}
  where the leading order coefficient is fixed by the condition that $g_A$ and $g_B$ at leading order are equal to the bare coupling. The two 
  beta-functions can be related through
	\begin{equation}\label{defbetafunc}
  		\beta_A(g_A) \weq{RGfunctions1} \mu\frac{\rd}{\rd\mu}g_A 
  			\weq{gAexpandgB} \mu\frac{\partial g_B}{\partial \mu} \frac{\partial g_A}{\partial g_B}
  				\weq{RGfunctions1} \beta_B(g_B) \frac{\partial g_A}{\partial g_B}\,.
  \end{equation}
  The beta-function has an expansion in the renormalised coupling
	\begin{equation}\label{beta_expand}
  		\beta_B(g_B)=b_0^B g_B^2 + b_1^B g_B^{3} + \rO(g_B^{4})\,,
  \end{equation}
  where the leading power is two but the argument holds for arbitrary leading power greater unity.
	In terms of $g_A$ this becomes 
	\begin{equation}
  		\beta_B(g_A)=b_0^B g_A^2 + (b_1^B - 2 c_1 b_0^B) g_A^{3} + \rO(g_A^{4})\,,
  \end{equation}
  and the derivative is
	\begin{equation}
  		\frac{\partial g_A}{\partial g_B} = 1 + 2c_1 g_B + \rO(g_B^2) = 1 + 2c_1 g_A + \rO(g_A^2) \,.
  \end{equation}
  Now the right hand side of \eqref{defbetafunc} can be evaluated in terms of $g_A$
	\begin{align}
  		\beta_A(g_A) & = \li[b_0^B g_A^2 + (b_1^B - 2 c_1 b_0^B) g_A^{3} + \rO(g_A^{4})\re]
  												 \cdot \li[1 + 2c_1 g_A + \rO(g_A^2)\re]\,,\\
  								 & = b_0^B g_A^2 + b_1^B g_A^{3} + \rO(g_A^{4})\,,
  \end{align}
	proving that the first two coefficients are universal in the sense that they neither depend on the regularisation nor on the the
	renormalisation scheme. For the first coefficient this is a direct consequence of demanding the 
	renormalised couplings to coincide at leading order (that is, at leading order they coincide with the unrenormalised coupling). The second
	order coefficients coincide because the expansion of beta-function starts at the next-to-leading order.
	
	Finally we introduce the $\Lambda$-parameter 
	\begin{equation}\label{Lambda}
  		\Lambda  = \mu (b_0 g_R^2)^{-b_1/(2b_0^2)}\e^{-1/(2b_0 g_R^2)}
														\cdot \exp\li\{ -\int_0^{g_R}\rd g\; \li[\frac{1}{\beta(g)} + \frac{1}{b_0 g^3} + \frac{b_1}{b_0^2 g} \re] \re\}\,.
  \end{equation}
	In the massless theory the $\Lambda$-parameter is the only dimensionfull parameter. 
	It is the standard solution of the renormalisation group equation for physical quantities
	\begin{equation}\label{RGE2}
  		\li\{ \mu\frac{\partial}{\partial \mu} + \beta(g_R) \frac{\partial}{\partial g_R} \re\}\,P(\mu,g_R) = 0\,,
  \end{equation}
	which expresses the abitrariness of the refernce scale $\mu$.
	
	For each the two renormalised couplings $g_A$ and $g_B$ a $\Lambda$-parameter can be defined. Given the relation \eqref{gAexpandgB} between the couplings
	the ratio of the $\Lambda$-parameters is then a pure number \cite{Celmaster:1979km}
	\begin{equation}\label{LambdaRatio}
  		\Lambda_A/\Lambda_B  = \exp\li\{ \frac{1}{2 b_0} \li(\frac{1}{g_B^2} - \frac{1}{g_A^2} \re)  + \rO(g_A^2) \re\} = \exp\li\{ \frac{c_1}{b_0}\re\}\,.
  \end{equation}

\subsection{Multiple couplings}
\label{multiple_coupling}

	So far we considered theories with a single dimensionless coupling. It is not difficult to generalize the concepts to multiple
	such couplings \cite{Weinberg:1996kr}. There will be as many renormalised couplings $g_l$ as bare couplings. Green's functions depend on all
	these couplings and in \eqref{renormGammalimit} $g_0$ and $g_R$ may collectively refer to them. Then for each $g_l$ there is a
	renormalisation group	equation
	\begin{equation}\label{RGEmultiple}
  		\mu\frac{\rd}{\rd \mu} g_l(\mu) = \beta_l(g(\mu))\,,
  \end{equation}
	with $\beta_l$ depending in general on all the renormalised couplings $g_l$. In the case of one
	coupling the beta-function determines the asymptotic behaviour of this coupling. In the case of
	multiple couplings the beta-functions $\beta_l$ determine the
	asymptotic trajectories in the space spanned by the couplings $g_l$.
	Clearly, there are many possibilities now. Let us concentrate
	on the prominent case of trajectories approaching a fixed point in $g$-space. A fixed point $g_(\mu)=g^*_l$ is defined through a mutual
	zero of the beta-functions
	\begin{equation}
  		\beta_l(g*) = 0\,.
  \end{equation}
	Shifting $g_l\to g_l-g^*_l$ by the fixed point, Taylor-expanding $\beta_l(g-g^*)$ and ignoring terms $\rO((g-g^*)^2)$
	 eq. \eqref{RGEmultiple} becomes
	\begin{equation}\label{RGEmultiple2}
  		\mu\frac{\rd}{\rd \mu} [g_l(\mu) - g^*_l] = \sum_k M_{lk} [g_k(\mu) - g^*_k]\,,
  \end{equation}
	with the matrix $M$ given by
	\begin{equation}
  		M_{lk} = \li(\frac{\partial \beta_l(g)}{\partial g_k}\re)_{g=g^*}\,.
  \end{equation}
  Suppose that the eigenvalues of this matrix are non-degenerate. Surely, that is not always true but it is the generic case. Then the
  eigenvectors $v^m$
 	\begin{equation}
  		\sum_k M_{lk} v^m_k = \lambda^m v^m_l\,,
  \end{equation}
  form a complete set and can be used to express the solution to \eqref{RGEmultiple2} 
 	\begin{equation}
  		g_l(\mu) = g^*_l + \sum_m c_m v^m_l \mu^{\lambda_m}\,,
  \end{equation}
  with coefficients $c_m$.
  
  The qualitative behaviour for $\mu\to\infty$ is thus governed by the eigenvalues $\lambda^m$ and the coefficients $c_m$. In particular,
  the fixed point is approached if and only if $c_m=0$ for all $\lambda^m>0$. A zero eigenvalue may be caused by a vanishing $\beta_l$, in
  which case the fixed point is reached for any value of this coupling in a region around the fixed point.
  
  The eigenvectors of the negative eigenvalues define a subspace containing the trajectories attracted by the fixed point. Trajectories with
  support outside of this subspace may get very close to the fixed point but are eventually repelled.
  
  We close this section by pointing out that the eigenvalues $\lambda^m$ are invariant under a change of basis, that is going to a differently
  defined set of renormalised couplings $\tilde{g}_l$. The change $g\to\tilde{g}$ amounts to a similarity transformation of the matrix $M$
  and thus preserves its spectrum. This can be seen as follows. The new couplings will be functions of the $g$s and satisfy renormalisation
  group equations
	\begin{equation}
  		\mu\frac{\rd}{\rd \mu} \tilde{g}_l(\mu) = \sum_m \frac{\partial \tilde{g}_l(g)}{\partial g_m} \beta_m(g)
  				 = \tilde{\beta}_l(\tilde{g}(\mu))\,.
  \end{equation}
  Thus $\beta$ transforms as a contravariant vector in coupling space
	\begin{equation}
  		\tilde{\beta}_l(\tilde{g}) = \sum_m \frac{\partial \tilde{g}_l(g)}{\partial g_m} \beta_m(g)\,.
  \end{equation}
  Differentiating on both sides with respect to $g_k$ we get
	\begin{equation}
  		\sum_m \frac{\partial \tilde{\beta}_l(\tilde{g})}{\partial \tilde{g}_m} \frac{\partial \tilde{g}_m(g)}{\partial g_k} 
  		= \sum_m \frac{\partial^2 \tilde{g}_l}{\partial g_m \partial g_k} \beta_m(g)
  			+ \sum_m \frac{\partial \tilde{g}_l(g)}{\partial g_m} \frac{\partial\beta_m(g)}{g_k}\,.
  \end{equation}
  At the fixed point $g^*$ the first term on the right hand side vanishes and we are left with the matrix equation
	\begin{equation}
  		\tilde{M}\, S = S\, M \,,
  \end{equation}
  with
	\begin{equation}
  		\tilde{M}_{lk} = \li(\frac{\partial \tilde{\beta}_l(\tilde{g})}{\partial \tilde{g}_k}\re)_{\tilde{g}=\tilde{g}(g^*)}\,,\quad
  		S_{lk} = \li(\frac{\partial \tilde{g}_l(g)}{\partial g_k}\re)_{g=g^*}\,.
  \end{equation}
  As long as $S$ is invertible this is a similarity transformation and the eigenvalues of $\tilde{M}$ are those of $M$.

\section{Discretisation}
\label{discretisation}

	The regulator	used in the computations of this thesis is the Euclidean lattice. If the (Minkowskian) time coordinate is
	Wick rotated to imaginary (Euclidean) time
	\begin{equation}
			x^E_0 = ix^M_0\,,
	\end{equation}
 	the imaginary unit in front of the Minkowski-space action in the path integral of a quantum field theory becomes a minus sign
	\begin{equation}
			\int \rD\; \e^{iS_M} \to \int \rD\; \e^{-S_E}\,.
	\end{equation}
	If the action is bounded from below (which is the case for physically relevant theories) the weight factor can be interpreted as a
	probability distribution for field configurations. Evidently there is a close connection between field theory and statistical physics if
	the weight is identified with the Boltzmann factor. The only subtle point at this stage is,
	one has to assure that the analytic continuation
	of the $n$-point Green's functions to imaginary time exists (see Section 1.3 in \cite{Montvay:1994cy} and
	references therein). The path integral becomes a mathematically well defined object, that is 
	a convergent multidimensional integral, by discretising Euclidean space-time on a finite lattice.
	This is the basis of non-perturbative Monte-Carlo techniques.
	
	From now on we work in $D=d+1$ dimensional space-time with Euclidean metric $\delta_{\mu\nu}$ and drop the superscript, that is $x_0$ refers
	to Euclidean time (Greek subscripts always run from $0$ to $d$). The Euclidean Dirac matrices satisfy anti-commutation relations
	\begin{equation}
			\{\gmu,\gnu\}=2\delta_{\mu\nu}\,,
	\end{equation}
	are all hermitian 
	\begin{equation}
			\gmu^\dagger = \gmu\,,
	\end{equation}	
	and are related to their Minkowskian counterparts as
	\begin{equation}
			\go = \go^M\,,\quad \gi = -i\gi^M\,.
	\end{equation}
	We are here dealing with theories in two and four dimensions and the definition of the Euclidean $\gfive$ differs by
	a factor due to demanding hermiticity
	\begin{equation}
			D=2:\; \gfive=i\go\gamma_1\,,\quad  D=4:\; \gfive = \go\gamma_1\gamma_2\gamma_3\,.
	\end{equation}
	Explicit representations of the Dirac matrices are 	given in the Appendix \ref{gamma}.
	
	Although the ultimate goal is to consider field theories on a lattice with boundaries we start here with the more common
  hypercubic lattice with either infinite extension or periodic boundary conditions. The sites of the lattice are labelled
  by $x_\mu=an_\mu$ with integer $n_\mu$ and lattice spacing $a$, which is the same in all directions. On a finite lattice
  the coordinates are restricted to $0\le x_\mu < L$ giving a total number of lattice
  sites $V/a^{D}=L^{D}/a^{D}$.
  
  Continuum space-time integrals are on the lattice replaced by sums over all lattice sites
  \begin{equation}
    	\int \rd^D x \to\; a^D \sum_{x}\,.
	\end{equation}
  The lattice spacing $a$ is the minimal distance in the system and thus introduces an ultraviolet
  cut-off. The momenta can be restricted to the first Brillouin zone
  \begin{equation}\label{BrillouinZone}
    	-\frac{\pi}{a} < \pmu \le \frac{\pi}{a}\,.
	\end{equation}
	On an infinite lattice continuum momentum integrals are cut-off
	\begin{equation}\label{BrillouinInt}
    	\int \frac{\rd^{D} p}{(2\pi)^D}\; \to\; \int_{-\pi/a}^{+\pi/a} \frac{\rd^D p}{(2\pi)^D}\,.
	\end{equation}
	On a finite lattice the allowed momenta are a discrete set in the range \eqref{BrillouinZone}. For
	periodic boundary	conditions and integer $n_\mu$ the $V/a^{D}$ allowed momenta are
	\begin{equation}\label{discrete_momenta}
  		\pmu = \frac{2\pi n_\mu}{L}\,, \quad n_\mu = -L/2<n_\mu\le L/2\,,
	\end{equation}
	and the momentum integrals also become momentum sums
	\begin{equation}\label{BrillouinSum}
  		\int_{-\pi/a}^{+\pi/a} \frac{\rd^D p}{(2\pi)^D}\; \to\; \frac{1}{V} \sum_{p}\,.
	\end{equation}
	
	The discretisation of integrals was straightforward. However, continuum differential operators have
	infinitely many 
	valid lattice representations. We introduce here the simplest possibilities which will serve as
	building blocks for
	more difficult choices. We consider lattice fields $\psi(x)$ defined at the sites of the lattice.
	On a finite lattice we have to specify boundary conditions. We choose general periodic boundary
	conditions
  \begin{equation}\label{PBconditions1}
    	\psi(x+L\hat{\mu}) = \e^{ia\theta_\mu}\psi(x)\,,\quad -\pi< \theta_\mu \le \pi\,,
	\end{equation}
	parametrised by the phases $\theta_\mu$.
	The forward and backward finite difference operators are
	\begin{align}
	    \dmu \psi(x) & =\tfrac{1}{a}[\psi(x+a\hat{\mu})-\psi(x)]\,,\\
	    \dmub \psi(x) & = \tfrac{1}{a}[\psi(x) - \psi(x-a\hat{\mu})]\,,
	\end{align}
	where $\hat{\mu}$ is a unit vector in $\mu$-direction.
  
  There is a different way of incorporating such general boundary conditions. One
  takes the lattice fields as periodic 
  \begin{equation}\label{PBconditions}
    	\psi(x+L\hat{\mu}) = \psi(x)\,,
	\end{equation}
	and defines the forward and backward finite difference operators as
	\begin{align}
	    \dmu \psi(x) & =\tfrac{1}{a}[\lambda_\mu\psi(x+a\hat{\mu})-\psi(x)]\,,\\
	    \dmub \psi(x) & = \tfrac{1}{a}[\psi(x) - \lambda^{-1}_\mu\psi(x-a\hat{\mu})]\,.
	\end{align}
	The phase factors $\lambda_\mu$ depend on $\theta_\mu$
  \begin{equation}
		    \lambda_\mu = \e^{ia\theta_\mu/L}\,, \quad -\pi< \theta_\mu \le \pi\,.
  \end{equation}
  The two notations are connected by an Abelian gauge transformation.
  A non-zero $\theta_\mu$ introduces a ``momentum'' that is
  not restricted to the values of \eqref{discrete_momenta}. The construction with the
  phase factors $\lambda_\mu$ in the finite differences is computationally easier and therefore
  adopted here. On infinite and periodic
  lattices the forward and backward finite difference operators obey
  \begin{equation}
    	(\dmu)^\dagger = -\dmub\,.
	\end{equation}

	Other important lattice operators are the anti-hermitian averaged finite difference operator
  \begin{equation}\label{averaged_difference}
    	 \md_\mu \psi(x)= \tfrac{1}{2}(\dmu + \dmub)\psi(x) = \tfrac{1}{2a}[\lambda_\mu\psi(x+a\hat{\mu}) - \lambda^{-1}_\mu\psi(x-a\hat{\mu})]\,,
	\end{equation}
	and the hermitian lattice Laplace operator
  \begin{equation}\label{lattice_laplace}
    	 \dmu\dmub \psi(x) = \dmu\dmub \psi(x) = \tfrac{1}{a^2}[\lambda_\mu\psi(x+a\hat{\mu}) + \lambda^{-1}_\mu\psi(x-a\hat{\mu}) - 2\psi(x)]\,.
	\end{equation}
	
	As already mentioned there is some freedom in discretising differential operators and therefore the action of a given field theory. This is
	due to the smaller symmetry on the lattice. In particular Lorentz invariance is broken and an infinite number of (in this case
	irrelevant from the point of view of renormalisation) terms can appear in the lattice action. In this way infinitely many different lattice actions
	for the same continuum theory are possible. However, they are expected to be equal in the continuum limit, that is when the
	cut-off is removed. One says the different lattice actions fall into the same universality class characterised by the target
	 continuum theory.
	 
	In numerical Monte-Carlo computations one is interested in lattice actions that balance between complexity (numerical cost)
	and the rate at which the continuum limit is approached (systematic error). Lattice perturbation theory is an essential tool
	to provide analytic understanding of these lattice artefacts.

\section{Continuum limit and lattice artefacts}
\label{continuum_limit}

	The renormalisation group function $\beta(g_R)$  describes the variation of the renormalised coupling $g_R$ with the cut-off at fixed
	bare coupling
	\begin{equation}
  		\beta(g_R) = \mu\frac{\partial g_R(g_0,a\mu)}{\partial \mu} \Bigg|_{g_0} = a\mu\frac{\partial g_R(g_0,a\mu)}{\partial a\mu} \Bigg|_{g_0}\,.
  \end{equation}
	Now we can ask how the bare coupling has to be varied with the cut-off for fixed renormalised coupling. The lattice beta-function $\beta_{\text{LAT}}(g_0)$
	is defined through
	\begin{equation}
  		a\frac{\rd}{\rd a}g_R = \li\{a\frac{\partial }{\partial a} -  \beta_{\text{LAT}}(g_0) \frac{\partial }{\partial g_0} \re\} g_R(g_0,a\mu) = 0\,.
  \end{equation}
	\begin{equation}
  		\beta_{\text{LAT}}(g_0) = -a\frac{\partial g_0(g_R,a\mu)}{\partial a} \Bigg|_{g_R}\,.
  \end{equation}
	Since 
	\begin{equation}
  		a\frac{\partial }{\partial a} g_R(g_0,a\mu) = \mu\frac{\partial }{\partial \mu} g_R(g_0,a\mu) = \beta(g_R)\,,
  \end{equation}
	the two beta-function are related in the following way
	\begin{equation}
  		\beta_{\text{LAT}}(g_0) \frac{\partial g_R(g_0,a\mu)}{\partial g_0} = \beta(g_R)\,.
  \end{equation}
	And because at lowest order in perturbation theory the both couplings are equal, $g_R = g_0 + \rO(g_0^2)$, we know from Section \ref{asymptotic_freedom},
	 that
	the first two coefficients of the beta-functions are identical. In particular, this means that the bare coupling vanishes in the limit $a\to0$.
	
	As mentioned at the end of Section \ref{MIndependentRenScheme} renormalised lattice Green's functions have a finite continuum
	limit $a\to 0$ and differ from this limit by terms of order $a$ 
	\begin{equation}\label{renormGamma2}
  		\Gamma_R(p;g_R,m_R,\mu,a\mu) = \Gamma_R(p;g_R,m_R,\mu) + \rO\li(a(\ln a)^k\re)\,,
  \end{equation}
  at $k$-loop order in perturbation theory \cite{Symanzik:1979ph}. It is widely believed that this behaviour also holds beyond perturbation theory and
  non-perturbative Monte Carlo data (naturally only available at finite $a$) is extrapolated to the continuum accordingly.
  Since perturbative computations are the only analytic tool to learn about the size of these {\it lattice artefacts} such computations
  are essential,  especially if new methods are used. For example, the amplitude of the lattice artefacts can be very different for different
  lattice actions.
	 
	In order to obtain numbers in lattice perturbation theory one almost always is forced to evaluate some lattice or momentum sums
	numerically. Consider a quantity $P(a/L)$ that is a sum of lattice diagrams at 2-loop order, where we suppress any dependence on the external lines.
	We supose that $P$ is dimensionless. If it is not, it can be made
	so by appropriate factors of $a$. Futhermore we suppose that it is finite at $a/L=0$, which always can be achieved by multiplication with 
	appropriate factors of $a/L$. Then $P$ has an expansion in	$a/L$
	\begin{equation}\label{cl:expansion}
  		P(a/L) = \sum_{n=0}^\infty\; \li[ r_n + s_n\, \ln(a/L)+ t_n\, \ln^2(a/L)\re]\, (a/L)^{n}\,.
  \end{equation}
	The generalisation to abitrary loop order should be obvious.
	
	In practice, $P(a/L)$ is computed at several values of $a/L$.	 In order to extract the coefficients of the expansion \ref{cl:expansion} we use the method
	described in Appendix D of Ref. \cite{Bode:1999sm}. In this way it is possible to reliably determine the systematic uncertainties that are inevitable involved in
	such a 	numerical analysis of	lattice Feynman diagrams.

\chapter{Chiral symmetry on the lattice}
\label{chiral_symmetry}

	The Euclidean Lagrangian of $N$ free fermions is
	\begin{equation}\label{FreeLagrangian}
			\mathcal{L} = \psibar(x)\, (\gmu\dmu + m)\, \psi(x)\,,
	\end{equation}
	where the fermion fields carry suppressed Dirac and flavour indices.
	 The flavour indices, labelling the $N$ fermions, are contracted by a unit matrix
	in flavour space.
	For $m=0$ the theory is invariant under a global $U(N)\times U(N)$ flavour symmetry which can be decomposed into $U(1)_V\times U(1)_A$
	 transformations
	\begin{equation}\label{U1trafo}
			\psi \to \e^{i\omega_V + i\gfive \omega_A}\, \psi\,, \quad \psibar \to  \psibar\, \e^{-i\omega_V + i\gfive \omega_A}\,,
	\end{equation}
	acting equally on all flavours
	and chiral $SU(N)\times SU(N)$ transformations
	\begin{equation}\label{SUNtrafo}
			\psi \to \e^{i \theta_V^a \lambda^a + i\gfive \theta_A^a \lambda^a}\, \psi\,, \quad
			 \psibar \to  \psibar\, \e^{-i \theta_V^a \lambda^a + i\gfive \theta_A^a \lambda^a}\,,
	\end{equation}
	where the generators of the $SU(N)$ algebra $\lambda^a$ act on the flavour indices.
	 \footnote{The generators $\lambda_a$ are	normalised to obey $\Tr[\lambda^a\,\lambda^b]=2\delta_{ab}$.
	  See Appendix \ref{ap:SUNGenerators} for more details.}
	The subscripts $V$ and $A$ refer to the associated Noether currents (see next section), which are of Lorentz vector and axial-vector type:	
	\begin{equation}\label{iso_singlet}
			V_\mu = \psibar \gmu \psi\,,\quad A_\mu = \psibar \gmu \gfive \psi\,,
	\end{equation}
	and
	\begin{equation}\label{iso_vector}
			V_\mu^a = \psibar \gmu \lambda^a \psi\,,\quad A_\mu^a = \psibar \gmu \gfive \lambda^a \psi\,.
	\end{equation}
	We use the same letters for the $SU(N)$ singlet \eqref{iso_singlet} and vector currents \eqref{iso_vector} but indicate 
	the difference by an additional superscript for the vector ones.
	
	We can introduce a new set of generators
	\begin{equation}\label{LRgernerators}
			t_L^a = (1-\gfive)\lambda^a\,,\quad  t_R^a = (1+\gfive)\lambda^a\,,
	\end{equation}
	with commutation relations
	\begin{align}\label{LRcommutators}
			[t_L^a,t_L^b] & = 2 i f^{abc} t_L^c\,,\\
			[t_R^a,t_R^b] & = 2 i f^{abc} t_R^c\,,\\
			[t_L^a,t_R^b] & = 0\,.
	\end{align}
	The $t_L$s and $t_R$s obviously form two closed subalgebras. Thus, as indicated by the notation, chiral $SU(N)\times SU(N)$
	is the direct sum of two $SU(N)$ subgroups
	acting independently on the left- and right-handed components of the fermion field
	\begin{equation}\label{LeftRight}
			\psi_L = \tfrac{1}{2}(1-\gfive)\psi\,,\quad \psi_R = \tfrac{1}{2}(1+\gfive)\psi\,.
	\end{equation}
		
	A general mass term $\psibar m\psi$ with $m=\text{diag}(m_1,\dots,m_N)$  in \eqref{FreeLagrangian} breaks all of these symmetries but
	multiplication with an $U(1)_V$ phase. $N$ mass degenerated fermions ($m\propto 1$) lift this to an $U(N)_V$
	since also the generators $\lambda^a$	remain unbroken.
	
	In QCD chiral symmetry plays a key role in understanding the mass spectrum of the light mesons. The pions, for example, are seen
	as the Goldstone bosons \cite{Goldstone:1961eq,Goldstone:1962es}
	associated with the spontaneously  axial generators  of $SU(2)\times SU(2)$
	\cite{Nambu:1961tp}.
	The QCD
	Lagrangian with only the two light $u$ and $d$ quarks has this symmetry in the massless limit, which is a good approximation because
	$m_{u,d}$ is much smaller than a typical hadron mass scale. Indeed isospin symmetry and quark number conservation, $U(1)_V$ and $SU(N)_V$ 
	respectively, are experimentally confirmed to high precision \cite{Yao:2006px}.
	The axial generators $\gfive\lambda^a$ are spontaneously broken in the
	quantum theory by a non-vanishing quark condensate leading to three massless Goldstone bosons: the pions. The small but non-zero mass
	of the pions is due to the fact that $SU(2)\times SU(2)$ is only an approximate symmetry. 
	
	So far, nothing has been said about the axial $U(1)_A$, i.e. continuous chiral phase transformations $\psi\to e^{i\omega_A\gfive}\psi$.
	In QCD this symmetry is also broken, but it must be in a different way since there is no light iso-singlet state expected from the Goldstone
	 theorem. The effect can be traced back to the topological structure of the vacuum
	gauge field \cite{Luscher:1998pe} and also explains the suppression of the electromagnetic decay of
	the neutral pion \cite{Peskin:1995ev}.	
	However, this chiral or axial {\it anomaly} is tightly connected to gauge symmetry.
	The two-dimensional fermion model considered in this thesis has no gauge symmetry and therefore the
	$U(1)_A$ is not anomalous.
	
	As already stated above understanding the structure of the low energy regime of QCD needs understanding of chiral symmetry and how it is
	violated. Since the QCD coupling is large (cf. Section \ref{asymptotic_freedom}) at low energies non-perturbative methods
	such as numerical lattice QCD are needed. The discretisations used today handle chiral symmetry quite differently. We 
	will discuss two different lattice Dirac operators and present their chiral properties. 
  But before that we introduce the currents associated with the symmetries and derive operator 
	identities from infinitesimal variable transformations in the path integral.
	
\section{Continuum Ward identities}
\label{continuum_wi}
	
	Here we derive operator identities, so called Ward identities. They are conveniently derived by variable transformations in the
	path integral for the expectation value of operators, where the new variables are connected to the old ones by infinitesimal
	{\it local} transformations
	\begin{equation}\label{infinitesimal_trafo}
			\psi \to \psi + \delta\psi\,,\qquad \psibar \to \psibar + \delta\psibar\,.
	\end{equation}	
  
	The expectation value of a local operator $\Op$, that is a field composed of fermion fields and their derivatives
	evaluated at the same space-time point, is
	\begin{equation}\label{PathIntegral}
			\ev{\Op} = \frac{1}{Z}\, \int \rD\psibar\rD\psi\; \Op\; \e^{-S}\,,
	\end{equation}
	with the Euclidean action
	\begin{equation}\label{FreeAction}
			S = \int\rd^D x\; \mathcal{L} = \int\rd^D x\; \psibar(x)\, (\gmu\dmu + m)\,\psi(x) \,.
	\end{equation}
	
	A linear change of the Grassmann valued fermion fields in \eqref{PathIntegral} not only effects the operator and the action but also the
	integration measure by introducing a non-trivial Jacobian
	\begin{equation}
			\psi' = A\psi\,,\qquad \psibar' = \bar{A}\psibar\,,
	\end{equation}
	\begin{equation}
			\rD\psibar\rD\psi = \det A \det \bar{A} \cdot \rD\psibar'\rD\psi' = J \cdot \rD\psibar'\rD\psi'\,.
	\end{equation}	
	For simplicity we assume the infinitesimal transformations \eqref{infinitesimal_trafo} can be represented by
	\begin{equation}
			A = 1 + \omega X\,,\quad \bar{A} = 1 + \omega \bar{X}\,,
	\end{equation}
	where $\omega$ is an infinitesimal function of space-time evaluated at the point of the field and $X\,,\bar{X}$ are matrices acting
	on Dirac and flavour indices. Then we can use $\det(1 + \omega X)=1+\omega\Tr[X] + \rO(\omega^2)$
	to write $J = 1 + \delta J + \rO(\omega^2)$.
	
	Let $\delta S$ and $\delta \Op$ be the change in the action and the operator for a given change of fermion fields. Performing an infinitesimal
	change of variables in the path integral yields 
	\begin{align}
			\ev{\Op} & = \frac{1}{Z}\, \int \rD\psibar\rD\psi\; \Op\; \e^{-S}\,,\\
							&  = \frac{1}{Z}\, \int \rD\psibar\rD\psi\; (1+\delta J + \rO(\omega^2))\, (\Op+\delta\Op)\; \e^{-S} (1-\delta S)\,,\\
							&  = \ev{\Op} + \ev{\delta \Op} - \ev{\Op\, \delta S} + \ev{\Op\,\delta J} + \rO(\omega^2)\,.
	\end{align}
	Thus we derived the general identity
	\begin{equation}\label{Op_identity}
			\ev{\delta \Op} - \ev{\Op\, \delta S} + \ev{\Op\,\delta J} = 0\,.
	\end{equation}
	Let us restrict $\omega(x)$ to a region $\mathcal{R}$, that is $\omega(x)=0$ for $x\notin \mathcal{R}$.
	If we choose an operator $\Op_\text{ext}$ defined outside the region $\mathcal{R}$ where the variable transformation is performed then
	$\delta \Op_\text{ext}=0$. Likewise the change in the action is supported only in $\mathcal{R}$	
	\begin{equation}
			\delta S = \int_{\mathcal{R}}\rd^D x\; \delta\mathcal{L}\,.
	\end{equation}
	Eq. \eqref{Op_identity} further simplifies to
	\begin{equation}\label{Op_identity2}
			 \ev{\Op_\text{ext}\, \delta S} = \ev{\Op_\text{ext}\,\delta J}\,.
	\end{equation}

	Now we promote the global transformations of \eqref{U1trafo} and \eqref{SUNtrafo} to local but infinitesimal transformations
	like
	\begin{align}
      a) & \quad \phantom{s1}u(1)_V \text{:}\quad  \delta\psi = i \omega_V \psi\,, & \delta\psibar & = -i \omega_V\psibar\\
      b) & \quad \phantom{s1}u(1)_A \text{:}\quad   \delta\psi = 
      			i \omega_A \gfive\psi\,, & \delta\psibar & = i \omega_A \psibar\gfive \label{U1Aomega}\\
      c) & \quad su(N)_V \text{:}\quad  \delta\psi = 
      			i \omega_V^a \lambda^a \psi\,,& \delta\psibar & = -i \omega_V^a \psibar \lambda^a\\
      d) & \quad su(N)_A \text{:}\quad \delta\psi = 
      			i \omega_A^a \lambda^a \gfive\psi\,,& \delta\psibar & = i \omega_A^a \psibar\gfive \lambda^a \label{SUNAomega}\,,
	\end{align}
  where the $\omega$s are now infinitesimal functions of space-time evaluated at the point of the field.

	For the $SU(N)$ vector transformations $c)$ and $d)$ the Jacobian is unity ($\delta J =0$) because
	the $SU(N)$ generators are traceless. The change in the action is	
	\begin{align}
			\delta_V S & = \int_{\mathcal{R}}\rd^D x\; \omega_V^a \li[-\dmu V_\mu^a(x) - \psibar(x)[\lambda^a,m]\psi(x) \re]\,,\\
			& \text{and} \\ 
			\delta_A S & = \int_{\mathcal{R}}\rd^D x\; \omega_A^a \li[-\dmu A_\mu^a(x) + 2m P^a(x) \re]\,,
	\end{align}
	respectively.
	Then the identity \eqref{Op_identity2} becomes
	\begin{equation}\label{iso_vector_vector1}
			 \ev{\dmu V_\mu^a(x)\,\Op_\text{ext}} = \ev{\psibar(x)[m,\lambda^a]\psi(x)\,\Op_\text{ext}}
	\end{equation}
	and 
	\begin{equation}\label{iso_vector_axial_vector1}
			 \ev{\dmu A_\mu^a(x)\,\Op_\text{ext}} = \ev{\psibar(x)\gfive\{\lambda^a,m\}\psi(x)\,\Op_\text{ext}}\,.
	\end{equation}
	In deriving these identities one has to make use of the arbitrariness of the exact definition of $\omega(x)$.
	For mass degenerate fermions one finds the $SU(N)$ vector current conservation
	\begin{equation}\label{iso_vector_vector}
			 \ev{\dmu V_\mu^a(x)\,\Op_\text{ext}} = 0
	\end{equation}
	and the {\it partially conserved axial current} (PCAC)
	\begin{equation}\label{iso_vector_axial_vector}
			 \ev{\dmu A_\mu^a(x)\,\Op_\text{ext}} = 2m\ev{P^a(x)\,\Op_\text{ext}}\,,
	\end{equation}
	with the flavour vector pseudo-scalar density
	\begin{equation}\label{iso_vector_ps_density}
	 		 P^a(x) = \psibar(x)\gfive\lambda^a\psi(x)\,.
	\end{equation}
	Although we considered here free fermions without gauge interactions
	the calculation in QCD yields the same identities (see \cite{Luscher:1998pe} for example).	
	
	For the $U(1)$ transformations $a)$ and $b)$ the change in the action is 
	\begin{align}
			\delta_V S & = \int_{\mathcal{R}}\rd^D x\; \omega_V \li[-\dmu V_\mu(x)\re]\,,\\
			& \text{and} \\ 
			\delta_A S & = \int_{\mathcal{R}}\rd^D x\; \omega_A \li[-\dmu A_\mu(x) + 2m P(x) \re]\,.
	\end{align}
	The corresponding operator identities are the fermion number conservation
	\begin{equation}\label{singlet_vector}
			 \ev{\dmu V_\mu(x)\,\Op_\text{ext}} = 0
	\end{equation}
	and the singlet PCAC relation
	\begin{equation}\label{singlet_axial_vector}
			 \ev{\dmu A_\mu(x)\,\Op_\text{ext}} = 2m\ev{P(x)\,\Op_\text{ext}}\,,
	\end{equation}
	with the singlet pseudo-scalar density
	\begin{equation}\label{singlet_ps_density}
	 		 P(x) = \psibar(x)\gfive\psi(x)\,.
	\end{equation}
	
	As already mentioned in QCD the $U(1)_A$ symmetry is anomalous and $A_\mu(x)$ is not conserved in the quantum theory even for
	vanishing masses. Along the presented steps, also in QCD we would have arrived at \eqref{singlet_axial_vector}, in contradiction to
	the anomaly. This is because the treatment here was very formal. That is, we performed variable transformations in an integral
	that is not well defined in first place. If QCD is regularised first \eqref{singlet_axial_vector} is changed and the anomaly is
	recovered. Using Ginsparg-Wilson fermions the anomaly can be computed easily \cite{Luscher:1998pq}.

\section{Lattice Ward identities}
\label{lattice_wi}
	
	The steps that led us to the continuum PCAC relations can be applied on the lattice \cite{Bochicchio:1985xa}.
	The expressions for lattice currents are then multilocal, in the sense that they involve more then one lattice site, and additional terms can appear that
	are allowed by the lattice symmetries.
	However, the difference	between multilocal and local\footnote{Here locality refers to operators composed of fields that are all taken at the same
	space-time point. It may not be confused with the locality of a differential operator like the Dirac operator.} becomes irrelevant in the continuum limit.
	Futhermore the continuum PCAC relations are part of the definition of the theory and thus, at finite lattice spacing and for properly renormalised
	operators they hold up to cutoff effects.
	 
	In perturbation theory on expects terms of $\rO(a\ln(a)^l)$ in $l$-loop order. In the case of massless fermions,
	for example, eq. \eqref{iso_vector_axial_vector} implies the lattice version
	\begin{equation}\label{latticePCAC}
			 \ev{\mdmu \ren{A_\mu^a(x)}\,\ren{\Op_\text{ext}}} = \rO(a\ln(a)^l)\,.
	\end{equation}
	On the lattice we also use the local currents and densities. Thus the continuum definitions \eqref{iso_vector}, \eqref{iso_singlet}, 
	\eqref{iso_vector_ps_density}, \eqref{singlet_ps_density} transfer to lattice. Although the axial-vector and vector currents
	are conserved in the massless theory they receive a finite renormalistion on the lattice 
	\begin{equation}\label{renAxialCurrent}
			 \ren{A_\mu^a(x)} = Z_A A_\mu^a(x)\quad \text{and}\quad \ren{A_\mu(x)} = Z_A A_\mu(x)\,,
	\end{equation}
	\begin{equation}\label{renVectorCurrent}
			 \ren{V_\mu^a(x)} = Z_V V_\mu^a(x)\quad \text{and}\quad \ren{V_\mu(x)} = Z_V V_\mu(x)\,,
	\end{equation}
	with
	\begin{equation}\label{ZCurrent}
			 Z_I = 1 + Z_I^{(1)}\,g^2 + \dots\,.
	\end{equation}

\section{Nielsen-Ninomiya theorem}
\label{nielsen_ninomiya_theorem}

	Before discussing the two lattice Dirac operators used in this thesis we briefly describe the broader context of
	lattice fermions and chiral symmetry. As we will see chiral symmetry in the lattice regularisation is tightly connected
	to other desirable properties of the operator $D=\gmu D_\mu$ in the massless action
  \begin{equation}\label{LPT:massless_action}
      S = a^{D} \sum_{x}\, \psibar(x)\,D\psi(x)\,.
  \end{equation}
  The kernel $D(x-y)$ and the Fourier transform $\tD(p)$ of the Dirac operator are defined through%
  \begin{equation}\label{LPT:dirac_kernel}
      D\psi(x) = a^D\sum_{y}\, D(x-y)\psi(y)\,, \quad D(x-y)=\int_{-\pi/a}^{\pi/a} \frac{\rd^D p}{(2\pi)^D}\,\e^{ip(x-y)}\tD(p)\,.
  \end{equation}
	
	To give an example consider the
	simplest lattice Dirac operator one could think of 
  \begin{equation}\label{naive_dirac}
      D_{\text{naive}} = \tfrac{1}{2}\{\gmu(\dmub+\dmu)\}\,,
  \end{equation}
  that is the averaged finite difference \eqref{averaged_difference} contracted with the gamma-matrices.
  Indeed, $D_{\text{naive}}$ anti-commutes with $\gfive$ and thus \eqref{LPT:massless_action} would be invariant under the 
  chiral transformations \eqref{U1Aomega} and \eqref{SUNAomega}. But it turns out that this Dirac operator leads
   to a proliferation of fermion
  species. The operator in Fourier space
  \begin{equation}\label{naive_dirac2}
      \tD_{\text{naive}}(p) = \tfrac{1}{a}\sum_\mu\, \sin(a\pmu)\,, \quad -\pi/a < \pmu \le \pi/a\,,
  \end{equation}
  has 4 and 16 zeros in two and four dimensions respectively. Thus $\tD^{-1}_{\text{naive}}$ is not only propagating
  a single physical fermion (the zero at $\pmu=0$), but also the so called {\it doubler modes}.

	This is not an accident. The appearance of the doublers is in accordance with the Nielsen-Ninomiya theorem
	 \cite{Nielsen:1981hk,Nielsen:1980rz,Nielsen:1981xu,Friedan:1982nk}.
	It can be stated in the form:\footnote{The theorem was originally proven for fermions on two- and four-dimensional lattices
	 \cite{Nielsen:1981xu}.}%
	\begin{theorem}\label{NN_theorem}
		Any massless lattice Dirac operator cannot satisfy the following four properties simultaneously:
		\begin{enumerate}
			\item[a)] $a D(x)$ is local in the sense that it is bounded by $C\e^{-\gamma |x-y|/a}$
			\item[b)] $\tD(p)=i\gmu p_\mu + \rO(ap^2)\,$ for $\,|p|\ll \pi/a$ 
			\item[c)] $\tD(p)$ is invertible at all non-zero momenta
			\item[d)] $\gfive D - D \gfive = 0$
		\end{enumerate}
	\end{theorem}
	In the bound in a) $C$ and $\gamma>0$ are constants  that do not depend on $a$. With respect to the universality of the 
	continuum limit a Dirac operator with such exponentially small tails is certainly as good as an ultra-local operator
	with only nearest neighbour interactions. The property b) is the right continuum behaviour at small momentum. Doublers
	are excluded by c) and chiral symmetry in the form of \eqref{U1Aomega} and \eqref{SUNAomega} is guaranteed by d).
	
	Any fermion discretisation in two or four dimensions has to abandon at least one of these desirable properties. In the
	example given in this section one finds a bunch of doubler modes, i.e. c) is violated. In the free theory this may not be a
	problem,
	but in the interacting theory they would contribute in the loop corrections.
	
	In the following two sections we discuss the Wilson and the Ginsparg-Wilson fermions, both abandoning d), but in a very different way
	and with very different consequences.

\section{Wilson fermions}
\label{wilson_fermions}
  
	In the view of the last section Wilson's approach \cite{Wilson:1975id} to remove the doublers is to sacrifice chiral symmetry by adding
	an irrelevant term to \eqref{naive_dirac}. The Wilson-Dirac operator
	is given by\footnote{Often there is an additional parameter $0<r\le 1$ multiplying the lattice Laplace operator: $- a r\dmub\dmu$.
	Throughout this thesis we set $r=1$.}
  \begin{equation}\label{wilson_dirac}
      \Dw = \tfrac{1}{2}\{\gmu(\dmub+\dmu) - a\dmub\dmu\}\,,
  \end{equation}
  and corresponding massive Dirac operator is simply
  \begin{equation}\label{wilson_dirac_m}
      D_m = \Dw + m_0\,.
  \end{equation}    
  Looking at the Fourier transform
  \begin{equation}\label{wilson_dirac_fourier}
      \widetilde{D}_\text{W}(p) = \tfrac{1}{a}\sum_\mu\{ \gmu \sin(a p_\mu) + \tfrac{2}{a}\sin^2(a p_\mu/2) \}\,,
  \end{equation}
  we see that the doublers receive a mass of the order of the cut-off and hence are strongly suppressed at finite
  $a$ and eventually disappear in the continuum limit.
  
  The price to pay is that the lattice Laplace operator $a\dmub\dmu$ explicitly breaks chiral symmetry at finite lattice 
  spacing. As a consequence a vanishing bare mass does not automatically imply a conserved axial current as in the
  continuum \eqref{iso_vector_axial_vector}. But this operator identity has to hold for the renormalised quantities also on the
  lattice up to scaling violations
	\begin{equation}\label{latticePCAC2}
			 \ev{\li[\mdmu \ren{A_\mu^a(x)} - 2m_R\ren{P(x)}\re]\,\ren{\Op_\text{ext}}} = \rO(a\ln(a)^l)\,.
	\end{equation}
	In fact, together with \eqref{renMass}
	\begin{equation}\label{renMass2}
			 	m_R = m_q\,Z_m(g_0,a\mu)\,, \quad m_q=m_0-m_c\,,
	\end{equation}	
	eq. \eqref{latticePCAC2} serves as definition of the {\it critical mass} $m_c$ \cite{Bochicchio:1985xa},
	that is, the value of the bare mass at which the axial current is conserved up to scaling violations
	\begin{equation}\label{critical_mass}
			 	m_0 = m_c(g)\quad \text{such that}\quad  \ev{\mdmu \ren{A_\mu^a(x)}\,\ren{\Op_\text{ext}}} = \rO(a\ln(a)^l)\,.
	\end{equation}	
	In the two-dimensional Gross-Neveu model the global chiral $U(1)$ symmetry is not anomalous and thus a critical mass
	for the conservation of the singlet axial current can be defined
	\begin{equation}\label{critical_mass2}
			 	m_0 = m_c(g)\quad \text{such that}\quad  \ev{\mdmu \ren{A_\mu(x)}\,\ren{\Op_\text{ext}}} = \rO(a\ln(a)^l)\,.
	\end{equation}	
	In this thesis we employ the latter definition. As indicated the critical mass can be computed in perturbation theory
	as a power series in the coupling constant
	\begin{equation}\label{critical_mass_exp}
			 	m_c(g) = m_c^{(0)} + m_c^{(1)} g^2 + m_c^{(2)} g^4 + \rO(g^6)\,.
	\end{equation}
	
	Having mass dimension one, $m_c$ will cancel the linear divergence spoiling the PCAC relation
	(\eqref{iso_vector_axial_vector} or \eqref{singlet_axial_vector}) due to the explicitly broken chiral symmetry. However,
	there may also be a term of order $\rO(1)$. In QCD such a term is not present because there is no chiral symmetry breaking operator
	of mass dimension four. Therefore it is enough to tune $m_c$. But in two dimensions there are chiral symmetry breaking
	four fermion interactions and one has to tune a dimensionless parameter in addition to $m_c$ (see Section \ref{CGN:wilson_fermions}).

\section{Ginsparg-Wilson fermions}
\label{gw_fermions}

	From the Nielsen-Ninomiya theorem (page \pageref{NN_theorem}) one may conclude that it is not possible to construct a meaningful 
	lattice theory of fermions with chiral symmetry. However, the way out was discovered shortly after the original paper
	by Nielsen and Ninomiya. Studying block-spin renormalisation in lattice QCD, Ginsparg and Wilson \cite{Ginsparg:1981bj}
	found the relation\footnote{In Ref. \cite{Ginsparg:1981bj} the authors actually derive a more general relation, but
	 \eqref{LPT:gw_relation} is the most spread version and is used solely
	in this thesis.}
	\begin{equation}\label{LPT:gw_relation}
			\gfive\,D + D\,\gfive = a\,D\gfive D\,,
	\end{equation}
	for the lattice Dirac operator. But it was not recognised at the time and remained unappreciated until Hasenfratz
	\cite{Hasenfratz:1998jp}
	``rediscovered'' it in the context of the perfect action. Indeed, the Dirac operator of the perfect action obeys the Ginsparg-Wilson
	relation \eqref{LPT:gw_relation}.
	
	In the following Neuberger \cite{Neuberger:1997fp} found an explicit solution to \eqref{LPT:gw_relation} 
	and the dimensional reduced	Dirac operator of the domain wall fermion was recognised to obey the Ginsparg-Wilson relation
	\cite{Kikukawa:1999sy}.
	
	Before turning to the solution presented by Neuberger we point out that the importance of \eqref{LPT:gw_relation} is due to the
	fact that it allows for an exact chiral symmetry \cite{Luscher:1998pq}.
	Consider the infinitesimal transformation of the fermion fields%
	\footnote{Note that the fermion fields can be transformed independently in Euclidean space.}
	\begin{equation}\label{LPT:gw_chiral_symmetry}
			\psi \to \psi + \epsilon\gfive(1-aD)\psi\,,\quad \psibar \to \psibar + \epsilon\psibar\gfive\,.
	\end{equation}
	This is an exact symmetry of the action \eqref{LPT:massless_action}
	\begin{equation}
			\delta S  = a^D\sum_x\; \epsilon \psibar(x) \li(D\gfive(1-aD) + \gfive D \re)\psi(x) \weq{LPT:gw_relation} 0\,.
	\end{equation}
	In the continuum limit \eqref{LPT:gw_chiral_symmetry} become the familiar chiral phase transformations \eqref{U1Aomega}.
	Chiral flavour transformations are defined analogously by replacing $\epsilon\gfive \to \epsilon^a\gfive\lambda^a$ in
	\eqref{LPT:gw_chiral_symmetry}.
	
	As a consequence on the lattice one gets \eqref{latticePCAC}, and similar for the singlet axial current, for free. Of course
	there is a exactly conserved current corresponding to the symmetry \eqref{LPT:gw_chiral_symmetry}, but it is 
	complicated in structure and not ultra-local.
	
	The Neuberger-Dirac operator \cite{Neuberger:1997fp} is given by the expression
	\begin{equation}\label{LPT:neuberger_op}
			D = \frac{1}{\bar{a}}\,\li\{ 1- A\,(A^\dagger A)^{-1/2} \re\}\,, \quad \bar{a} = \frac{a}{1+s}\,,
	\end{equation}
  \begin{equation}
			A = 1 + s -a\Dw\,,
	\end{equation}
	with the Wilson Dirac operator $\Dw$ as introduced in Section \ref{wilson_dirac} and a parameter $0\le s\le 1/2$
	that allows for some optimisation. For any $A$ obeying $A\gfive = \gfive A^\dagger$ this operator satisfies
	\eqref{LPT:gw_relation} with $a$ replaced	by $\bar{a}$. That is certainly the case for $\Dw$.
	More properties of the operator \eqref{LPT:neuberger_op} are discussed in \cite{Niedermayer:1998bi}.
	The best choice for the massive Dirac operator, for example, is
  \begin{equation}
			D_m = (1-\tfrac{1}{2}\bar{a}m)\,D + m\,.
	\end{equation}

\chapter{The Schrödinger functional}
\label{schroendinger_functional}

		Nonperturbative lattice QCD is set up to deliver predictions from first principles without uncontrolled approximations, mostly in the form
		of MC simulations. One ingredient on the way from numbers to physics is renormalisation. In order to stay on the road of first principles,
		also the renormalisation has to be done non-perturbatively. Here the Schrödinger functional (SF) has proved to be a powerful framework.
		
		The Schrödinger functional in a quantum field theory is the transition amplitude between field configurations at time zero and some later
		time. In an Euclidean prescription it can be represented by a functional integral
    \begin{equation}
        Z = \int \,  \e^{-S} \,,
    \end{equation}		
		where the fields are defined on a space-time manifold with boundaries and obey Dirichlet boundary conditions.
		
		There exists a reasonable amount of literature about the Schrödinger functional in lattice QCD and its merits. One of the motivations to study the SF was,
		that it provides a infrared cut-off $\rO(1/T)$ to the theory \cite{Sint:1993un}, where $T\propto L$ is the time extension of the lattice. Since it is
		a finite size regularisation scheme, the size of the system $L$ is a natural scale in the theory. By employing finite size recursion techniques, it is thus 
		possible to connect the low energy regime of the light mesons with the high energy regime of perturbative QCD.	 In this way the running of the
		strong coupling and the fundamental 	parameters of QCD can be computed \cite{DellaMorte:2004bc,DellaMorte:2005kg,Sommer:2006wx}.
		 We can not give a 	
		thorough introduction to these topics here. Instead we refer the reader to the Les Houches lectures by Lüscher \cite{Luscher:1998pe}, which give an
		overview about the techniques. In the present chapter we discuss the renormalisability of the SF and on its lattice 
		representation. 

    In the continuum formulation of the Schrödinger functional the fermion fields are defined in the time interval $x_0=[0,T]$. At the 
    boundaries they are subject to the conditions
    \begin{equation}\label{bc_1}
        P_+ \psi(x) = \psibar(x) P_- = 0 \quad \text{at} \quad x_0=0\,,
    \end{equation}
    \begin{equation}\label{bc_2}
        P_- \psi(x) = \psibar(x) P_+ = 0 \quad \text{at} \quad x_0=T\,.
    \end{equation}
    The projectors occurring in these equations are defined as
    \begin{equation}
        P_\pm=\tfrac{1}{2}(1\pm\go)\,.
    \end{equation}
    Note that these boundary conditions are invariant under space rotations, parity, time reflections and charge conjugation,
		 but not under chiral transformations \eqref{U1trafo} and \eqref{SUNtrafo}.

		In this chapter we concentrate on bilinear fermionic actions
    \begin{equation}\label{bilinear_fermion_action}
        S_\mathrm{F} = \int_0^T \rd\,x_0\, \int \rd^d\,\vx\; \psibar(x)\, D\, \psi(x) \,,
    \end{equation}
    \begin{equation}
        D = \gmu \dmu + {\cal A} + m_0\,,
    \end{equation}
    where $d=1$ or $d=3$ and ${\cal A}(x)$ is the sum of bosonic fields mediating the interactions.
    In the case of QCD obviously ${\cal A}=\gmu A_\mu(x)$, with the gauge field $A_\mu(x)$.
    In the case of the GN the four fermion interaction terms
    can be replaced by bosonic auxiliary fields in the functional integral. The fermion action is then also 
    of the form \eqref{bilinear_fermion_action} (cf. Section \ref{auxiliary_fields}).   
		The gauge field or the auxiliary fields will play a spectator role most of the time. However, desirable properties like renormalisability
		and locality (in the lattice formulation) of the theory may depend on their details.
		
		In the standard formulation of the SF inhomogeneous boundary conditions are adopted and the boundary values are used as sources for the
		fermion fields at the boundary \cite{Sint:1993un}.
		With the homogeneous boundary conditions of Eqs. (\ref{bc_1}, \ref{bc_2}) the boundary fermion fields can be
		defined through the non-zero Dirac components 
    \begin{equation}\label{boundary_sources1}
        \zeta(\vx) = P_- \psi(x)\,,\quad  \zetabar(\vx) = \psibar(x) P_+ \quad \text{at} \quad x_0=0\,,
    \end{equation}
    \begin{equation}\label{boundary_sources2}
        \zeta'(\vx) = P_+ \psi(x)\,,\quad  \zetabar'(\vx) = \psibar(x) P_- \quad \text{at} \quad x_0=T\,.
    \end{equation}
    This is the convention used in \cite{Luscher:2006df}.
    
    The boundary conditions of the SF may cause additional divergences. In quantum field theory one has learned to deal with divergences.
    They are absorbed into a ``normalisation'' of the fields and parameter of the theory. If
    one gets away with a redefinition of a finite number of parameters, the theory is called renormalisable. It is then not obvious, that the
    theory with the boundaries stays finite. 
      
		Symanzik argues in \cite{Symanzik:1981wd} (see also \cite{Luscher:1985iu} for an introduction) 
		that the SF of any renormalisable quantum field theory can be rendered finite by adding a finite number of boundary counterterms.
		These are local polynomials in the fields and their derivatives, integrated over the boundary. They are restricted by the symmetries of
		the theory and have mass dimension $d$ or less.
	  
	  The argument is based on explicit calculations in scalar $\phi^4$ theory, where one has to add two new boundary counterterms $\phi^2$ and
		$\phi\partial_0\phi$. The expectation could be shown to hold up to 2-loop of perturbation theory in SU(N) Yang-Mills theory 
		\cite{Luscher:1992an,Narayanan:1995ex} and in QCD \cite{Sint:1995rb,Bode:1999sm}.
		In the former case the symmetries forbid any boundary counterterms and in the later 
		they can be absorbed in a multiplicative renormalisation of the quark fields at the boundary . Although a proof is still missing, there is little doubt that the
		SF of QCD is renormalisable, given the success of the method in non-perturbative computations \cite{Sommer:2006sj}.
		
		Discretising the Schrödinger functional of a given quantum field theory involves then two questions. Boundary conditions like Eqs.
		(\ref{bc_1}, \ref{bc_2}) make only sense when imposed on smooth functions, but the lattice is discrete by definition and the lattice
		fields are only defined at the sites of the lattice. So the question 
		arises how the continuum boundary conditions are represented on the lattice. Furthermore the lattice breaks some continuum
		symmetries, like continuous rotations, translations and chiral transformations.
		This may give rise not only to new terms in the bulk of the lattice, but also to additional terms at the boundaries.

		In Section \ref{boundaries} we address the first question with an heuristic argument to make plausible, that the Schrödinger
		functional boundary conditions arise naturally in the continuum limit and need no special adjustments of the lattice action. Then
		in Section \ref{wilson_sf} we get explicit and present formulae for the well studied lattice SF with Wilson fermions.
		We present a rather recent proposal \cite{Luscher:2006df} for an overlap Dirac operator in the SF in Section \ref{gw_sf}.
		There is another proposal for a SF with Ginsparg-Wilson fermions by Taniguchi \cite{Taniguchi:2004gf}. There the boundary conditions are realised 
		through an orbifold projection. However, there are some technical difficulties, i.e the fermion determinant has a phase and fermions masses
		can not be introduced straightforwardly. Therefore	we consider only the proposal of Ref. \cite{Luscher:2006df}.
		At the end of this Chapter we come back to the question of boundary counter terms.

\section{Lattices with boundaries}
\label{boundaries}
		
		
		Discretising a field theory involves some freedom in the definition of the lattice action. For example the differential operator
		$\dmu$ can be replaced by the forward finite difference $a\dmu \psi(x) = \psi(x+a\mu) - \psi(x)$ or by the averaged finite difference
		$a\md_\mu \psi(x) = \psi(x+a\mu) - \psi(x-a\mu)$. That both choices lead to the same continuum theory is assured by the local nature of
		the resulting actions. In the continuum limit the relevant length scale of the theory becomes much larger than the lattice spacing $a$ and
		the microscopic (length scale $a$) details become irrelevant. One says the two actions belong to the same universality class.
		
		These classes are characterised by global properties like dimensionality and symmetries. In statistical mechanics universal behaviour
		occurs in the vicinity of the critical lines and there the concept of universality has been extended to systems with boundaries
		(see \cite{Diehl:1996kd} for a review). This means for the discretisation of field theories that boundary conditions imposed in the
		continuum theory, together with
		the dimensionality and the symmetries of the theory, define an universality class.		
		In general requiring locality, symmetries and power-counting reduce the number of possible boundary conditions and therefore the 
		number of universality classes.
	
		The line of argument followed here is borrowed from Section 3 in Ref. \cite{Luscher:2006df} and we also start with the simple example of
		a free scalar field.
		
\subsection{Free scalar field}
		
		We consider a free scalar field in the half-space $x_0\le0$. We do not specify the boundary condition at $x_0=0$, but rather explore where
		we are lead to in the continuum limit starting from different lattice actions. The lattice fields $\phi(x)$ are defined at the sites of
		a hypercubic lattice with spacing $a$.
		Only the fields at $x_0=a,2a,3a,\dots$ are dynamical degrees of freedom and are integrated over 
		in the functional integral. A possible lattice action reads
    \begin{equation}\label{free_scalar_action}
        S = a^4\,\sum_{x_0\ge a}\, \sum_{\vx}\, \tfrac{1}{2}\{\dmu\phi(x)\dmu\phi(x) + m^2\phi(x)^2\}\,,
    \end{equation}
		where $\dmu$ is the forward difference operator in $\mu$ direction. Note that the action depends only on the dynamical degrees of freedom.
		
		In order to calculate the propagator we write Eq. \eqref{free_scalar_action} as a quadratic form in $\phi$
    \begin{equation}\label{free_scalar_action2}
        S = a^4\,\sum_{x_0\ge a}\, \sum_{\vx}\, \tfrac{1}{2}\phi(x)\{-\dmub\dmu + m^2 + P\}\phi(x)
    \end{equation}
    with the backward difference operator $\dmub$. Replacing the symmetric nearest neighbour interaction of Eq. \eqref{free_scalar_action}
    by the one in Eq. \eqref{free_scalar_action2} produces some mismatched terms near the boundary. This is cured by the
    inclusion of the boundary term
    \begin{equation}
        P = -\frac{1}{a}\,\delta_{x_0,a}\, \partial_0^\ast \,.
    \end{equation}
    The defining equation of the propagator $G(x,y)$ follows directly from the action \eqref{free_scalar_action2}
    \begin{equation}\label{def_propagator}
        (-\dmub\dmu + m^2 + P)\, G(x,y) = a^{-4}\delta_{x,y}\,,\quad		x_0\,,y_0 \ge a\,.
    \end{equation}
    Due to translation invariance in the space directions the propagator can be calculated in a time-momentum representation
    \begin{equation}
        G(x,y) = \int_{-\pi/a}^{\pi/a}\frac{\rd^3 \vp}{(2\pi)^3}\, \e^{i\vp(\vx-\vy)}\, \widetilde{G}(x_0,y_0,\vp)  \,.
    \end{equation}
    This is done by first determining the eigenfunctions of $-\dmub\dmu + m^2$ that are annihilated by $P$
    \begin{equation}\label{eigenfunctions}
        \e^{i\vp \vx}\, \cos\li((x_0-\tfrac{a}{2})p_0\re)\,.
    \end{equation}
    It is then easy to write down the propagator
    \begin{equation}\label{propagator_tilde}
        \widetilde{G}(x_0,y_0,\vp) = \int_{0}^{2\pi/a}\frac{\rd p_0}{\pi}\,
        						\frac{1}{m^2 + \phat^2}\, \cos\li((x_0-\tfrac{a}{2})p_0\re)\cos\li((y_0-\tfrac{a}{2})p_0\re) \,.
    \end{equation}
    In the continuum limit the integral can be done and the result is
    \begin{equation}
        G_{\text{cont}}(x,y) = \int\frac{\rd^3 \vp}{(2\pi)^3}\, 
        						\frac{\e^{i\vp(\vx-\vy)}}{2\epsilon(\vp)}\, \li(\e^{-\epsilon(\vp)|x_0-y_0|}+\e^{-\epsilon(\vp)(x_0+y_0)}\re)\,,
    \end{equation}
    \begin{equation}
        \epsilon(\vp)=\sqrt{m^2 + \hat{\vp}^2}\,.
    \end{equation}
   	Examining the behaviour at the boundary one infers that the propagator satisfies Neumann boundary conditions in the continuum
   	limit
    \begin{equation}
        \partial_0\,G_{\text{cont}}(x,y)\big|_{x_0=0} = 0\,.
    \end{equation}
    
    Now we slightly modify the action by adding a boundary term
    \begin{equation}\label{free_scalar_action3}
        S \to S + a^3\sum_\vx\, \frac{c}{2a}\, \phi(x)^2\Big|_{x_0=a}\,.
    \end{equation}
    Note that the powers of $a$ are such that $c>0$ is dimensionless.
    The propagator is still given by \eqref{def_propagator} but now $P$ has an additional term
    \begin{equation}
        P = \frac{1}{a}\,\delta_{x_0,a}\, \li(-\partial_0^\ast + \frac{c}{a}\re)\,.
    \end{equation}
    The eigenfunctions that are annihilated by $P$ are a bit more complicated
    \begin{equation}\label{eigenfunctions_c}
        \e^{i\vp \vx}\, \sin\li((x_0-a)p_0+\varphi(p_0)\re)\,,
    \end{equation}
    \begin{equation}
        \varphi(p_0) = \arctan \left( {\frac {\sin \left( p_0 \right) }{\cos \left( p_0 \right) + c-1}} \right) \,.
    \end{equation}
		The propagator is similar to \eqref{propagator_tilde} but with the eigenfunctions \eqref{eigenfunctions_c}.
		Again the integral can be performed in the continuum limit and the result is
    \begin{equation}
        G^c_{\text{cont}}(x,y) = \int\frac{\rd^3 \vp}{(2\pi)^3}\, 
        						\frac{\e^{i\vp(\vx-\vy)}}{2\epsilon(\vp)}\, \li(\e^{-\epsilon(\vp)|x_0-y_0|}-\e^{-\epsilon(\vp)(x_0+y_0)}\re)\,.
    \end{equation}
    This propagator satisfies Dirichlet boundary conditions
    \begin{equation}
        G^c_{\text{cont}}(x,y)\big|_{x_0=0} = 0\,.
    \end{equation}
    
    A small change in the action led to a totally different class of boundary conditions. This can be made a bit more transparent
    by looking at the field equation
    \begin{equation}
        \ev{\eta(x)\phi(y)} = a^{-4}\delta_{x,y}\,, \quad \eta(x) = \frac{\delta S}{\delta \phi(x)}\,.
    \end{equation}
    From the action in the form \eqref{free_scalar_action3} we derive
    \begin{equation}
        \eta(x) = \{-\dmub\dmu + m^2\}\,\phi(x)\,,\quad x_0>a\,.
    \end{equation}
   	Thus in the bulk of the lattice we find the Klein-Gordon equation. But at the boundary $x_0=a$ we find
    \begin{equation}
        \eta(x) = \frac{c}{a^2}\,\phi(x) - \frac{1}{a}\,\partial_0\,\phi(x) + \{-\dkb\dk + m^2\}\,\phi(x)\,.
    \end{equation}
   	The negative powers of $a$ in front of the first two terms are due to the fact that they are supported only at the 
   	boundary. Taking the limit $a\to 0$ the first term dominates for all $c>0$. Thus in the continuum limit
   	the field equation at the boundary implies Dirichlet boundary conditions at $x_0=0$. In the somehow special case
   	$c=0$ the second term dominates and the field equation implies Neumann boundary conditions.
   	
\subsection{The SF universality classes}
\label{sf_universality}
		
		In the example of the last section no boundary conditions were imposed onto the lattice fields. The boundary conditions were encoded
		in the lattice action and emerged when
		we took the limit $a\to 0$. But we found two distinct classes of boundary conditions. One that is generic, in the sense that it is
		found for a wide range of actions (all $c>0$). And one that is sensitive to small perturbations of the action.
		
		Clearly a free scalar field is a trivial example. In the case of interacting theories or more complicated actions the analysis of the
		field equations will not be so transparent. But boundary conditions that respect locality will always be of the form
    \begin{equation}
        \mathcal{O}(x)\big|_{x_0=0}=0\,,
    \end{equation}
    where $\mathcal{O}(x)$ is a linear combination of local fields and their derivatives with the appropriate symmetry properties.
    
    Inspired by the above example it is plausible that the generic boundary conditions, that are stable under perturbations
    of the lattice action, are those imposed on the fields with the lowest dimension. All other possible boundary conditions will require
    some tuning of the lattice action, unless there are symmetries that protect them.

		We want to apply this argument now to the case of fermions in two (GN) and four dimensions (QCD). In both cases 
		the fermions are represented on the lattice by Dirac spinors $\psi(x)$ at each lattice site. Both theories are asymptotically free,
		assuring that the scaling dimension of local fields is equal to their engineering dimension (cf. Section \ref{renormalisation}).
		Thus the fields of lowest dimension are the fermion fields themselves and the generic boundary conditions are of the from
    \begin{equation}
        B\,\psi(x)\big|_{x_0=0}=0\,,
    \end{equation}
		where $B$ is a constant matrix with Dirac, flavour and, in the case of QCD, colour indices. Boundary conditions at a later time $x_0=T$
		are then linked to those at $x_0=0$ by time reflection symmetry and the conditions for the anti-fermion field are given by charge
		conjugation symmetry. 
		
		The possible boundary conditions are further restricted by the lattice symmetries. If the lattice theory is invariant under gauge and 
		flavour transformations, cubic rotations and parity, so should be the boundary conditions. And finally the matrix $B$ cannot have full rank
		since the Dirac equation is a first order differential equation and the two conditions at $x_0=0$ and $x_0=T$ would imply
		a vanishing fermion propagator. Up to a constant factor there are only two matrices satisfying all these properties
    \begin{equation}
        B=P_- \quad \text{and} \quad B=P_+\,.
    \end{equation}
    
    The homogeneous Dirichlet boundary conditions of the Schrödinger functional are therefore the generic ones and do not need any fine
    tuning or particular adjustment of the lattice action. There are two universality classes of lattice theories which differ by the sign
     in the boundary condition
    \begin{equation}
        P_\pm\,\psi(x)\big|_{x_0=0}=0\,.
    \end{equation}
		Since the two are connected by a finite chiral transformation, the difference matters only for non-vanishing fermion masses. In this case
		the sign can be determined by inspection of the free propagator.
		
\section{Free Wilson fermions}
\label{wilson_sf}

		In the continuum the Schrödinger functional can be represented by a functional integral
    \begin{equation}
        Z = \int \rD\psi\rD\psibar \,  \e^{-S_\text{F}[\psibar,\psi]} \,,
    \end{equation}
		where the fields obey Dirichlet boundary conditions \eqsref{bc_1}{bc_2} and the action may be given by 
		\eqref{bilinear_fermion_action}.
		For simplicity we omit any interaction (gauge or four fermion) and introduce a $d+1$
		dimensional lattice with spacing $a$ and label the sites by integer multiples $x_\mu/a \in \text{Z}\,,\; \mu=0,\dots,d$.
		The space directions are taken to be periodic with length $L$
    \begin{equation}
       \psi(x+L\hat{k}) = \psi(x)\,,
    \end{equation}
    where $\hat{k}$ is an unit length vector in direction $k=1,\dots,d$.
    The fermion fields at times $x_0=a,2a,\dots,T-a$ are the dynamical degrees of freedom (the fields that are integrated over in the
    functional integral).
    It is convenient to assume that the fermion fields are defined at all other values of $x_0$ as well,
    but that they are zero there. The free lattice action takes then the familiar form
    \begin{equation}\label{SF:free_action}
        S_0 = a^{D} \sum_x\, \psibar(x)\,D_m\,\psi(x)\,,
    \end{equation}
    where $D_m$ is some discretisation of the massive Dirac operator.

\subsection{Dirac operator and propagator}
\label{sf_wilson_dirac}
		
		In the present section we consider
    \begin{equation}\label{massive_wilson_dirac}
        D_m = \Dw + m_0\,,
    \end{equation}
    where the Wilson Dirac operator is defined as (cf. Section \ref{wilson_fermions})
    \begin{equation}\label{wilson_dirac2}
        \Dw = \tfrac{1}{2}\{\gmu(\dmub+\dmu) - a\dmub\dmu\}\,,
    \end{equation}
    in the range $0<x_0<T$. At all other times the target field $\chi=D_m\psi$ is set
	  to zero\footnote{The Dirac operator maps the space of fermion fields that are defined at all $x_0$, but are zero at $x_0<a$ and $x_0>T-a$, into itself. }
    \begin{equation}\label{wilson_dirac3}
        D_m\psi(x)\Big|_{x_0\le 0} = 0 = D_m\psi(x)\Big|_{x_0\ge T}\,.
    \end{equation}  
    Note that we set $r=1$ and that we introduce factors $\e^{ia\theta_k/L}$ in the spacial lattice difference operators
    (which is equivalent with periodic boundary conditions with a phase, see Section \ref{discretisation}).
   	For $\theta=0$ this is the Dirac operator introduced in \cite{Sint:1993un}.
    
    Thus the Dirac operator can be considered as a linear mapping in the space of fermion fields that vanish at the boundaries.
    The propagator $S(x,y)$ is defined through
    \begin{equation}\label{propagator}
        \li(\tfrac{1}{2}\{\gmu(\dmub+\dmu) - a\dmub\dmu\} + m_0\re)\, S(x,y) = \frac{1}{a^{d+1}}\,\delta_{x,y}\,,\quad 0<x,y<T\,,
    \end{equation}  
    with boundary values
    \begin{equation}
        P_+\,S(x,y)\Big|_{x_0=0}=P_-\,S(x,y)\Big|_{x_0=T}=0\,.
    \end{equation}
    Since the operator $\gfive D_m$ is hermitian the propagator has the property
    \begin{equation}\label{sf:g5hermiticity}
        S(x,y)^\dagger=\gfive\,S(y,x)\,\gfive\,.
    \end{equation}
    
		In the free theory it is possible to derive an explicit expression for the propagator in a time-momentum representation
		\cite{Luscher:1996vw}. An elegant form is
    \begin{equation}\label{FreePropagator1}
        S(x,y) = (\Dw^\dagger + m_0)\, G(x,y)\,, \quad 0<x_0,y_0<T\,,
    \end{equation}
    where $G(x,y)$ is defined through
    \begin{eqnarray}
        G(x,y) = &  & \mspace{-27.0mu} L^{-d}\,\sum_\vp \left\{ -2i \po^\+_0 A(\vp^\+)R(\pp)\right\}^{-1}\,\e^{i\vp(\vx-\vy)}\nonumber\\
               &   & \mspace{-60.0mu} \times \Big\{ (M(\pp) - i\po^\+_0)\,\e^{-\omega(\vp)|x_0-y_0|} + (M(\pp)
               	 + i\po^\+_0)\,\e^{-\omega(\vp)(2T-|x_0-y_0|)}\nonumber\\
               &   & \mspace{-60.0mu} - (M(\pp) + i\go\po^\+_0)\,\e^{-\omega(\vp)(x_0+y_0)} - (M(\pp)
               	 - i\go\po^\+_0)\,\e^{-\omega(\vp)(2T-x_0-y_0)}\Big\}\,.\nonumber\\
               &   & \mspace{-60.0mu} \label{FreePropagator2}
    \end{eqnarray}
    All undefined functions and notations in this expression are introduced in Appendix \ref{Def:FreeTheory}. In particular
    the momenta $p_\mu$ are given by eq. \ref{BoundaryCondition} and \ref{PosEnergy}.

\section{Ginsparg-Wilson fermions}
\label{gw_sf}

		The Wilson lattice Dirac operator violates chiral symmetry explicitly and leads to computational difficulties
		such as additive mass and multiplicative current renormalisation. Therefore it is desirable to have a lattice operator with
		better chiral properties. In Section \ref{nielsen_ninomiya_theorem} we learned that it is not possible to formulate a 
		lattice theory of fermions with a continuum like chiral symmetry.
		The way out in Section \ref{gw_fermions} was to ease the restriction to an exact continuum like chiral symmetry in favour
		of a lattice chiral symmetry and a lattice Dirac operator given as the solution to the Ginsparg-Wilson relation		
		\begin{equation}\label{gw_relation2}
				\gfive\,D + D\,\gfive = a\,D\gfive D\,.
		\end{equation}
		In the limit $a\to 0$ this relation becomes the known anti\-com\-mutation relation and the lattice chiral transformations
		become the continuum ones.
		
		In the presence of the boundaries the same approach immediately leads to inconsistencies. As stated in the introduction in the continuum
		the SF boundary conditions break chiral symmetry. This can be seen considering the solution to the massless Dirac equation
		\begin{equation}
				D\,\psi(x) = 0\,,
		\end{equation}
		which has to obey the boundary conditions \eqsref{bc_1}{bc_2}. Note that the massless Dirac operator anti-commutes with $\gfive$.
		Therefore, given
		the propagator $D\,S(x,y)=\delta(x-y)$ the sum $\gfive\,S(x,y) + S(x,y)\,\gfive$ is a solution of the Dirac equation for all $y$.
		And since the propagator obeys the boundary conditions \eqsref{bc_1}{bc_2} this solution can also be obtain from 
		the boundary values at $x_0=0$ and $x_0=T$ (actually the non-zero components there)
		\begin{multline}\label{continuum_propagator}
				\gfive\,S(x,y) + S(x,y)\,\gfive =\\
							 \int_{x_0=0}\rd^d z\, S(x,z)\gfive S(z,y) + \int_{x_0=T}\rd^d z\, S(x,z)\gfive S(z,y)\,.
		\end{multline}
		Thus the continuum propagator anti-commutes with $\gfive$ up to boundary terms. In QCD with more than one massless quark this leads
		to a non-singlet chiral Ward identity with a unit mass term at the boundaries \cite{Luscher:2006df}.
		
		Now consider a lattice Dirac operator satisfying \eqref{gw_relation2} in the presence of the boundaries. Then the lattice propagator
		anti-commutes with $\gfive$
		\begin{equation}\label{gw_propagator}
				\gfive\,S(x,y) + S(x,y)\,\gfive = \frac{1}{a^d}\,\delta_{x,y}\gfive\,,
		\end{equation}
	  for any finite separation $x-y$ and any finite $a$.
	  Thus this theory can not have the right continuum limit, i.e. the one where the propagator obeys \eqref{continuum_propagator}.
	  
\subsection{Modified Neuberger-Dirac operator}
\label{SF:mod_neuberger_op}

	  As explained in Section \ref{sf_universality} the SF boundary conditions form a universality class and need no special
	  adjustment of the lattice action. This means in particular that there may be many possible lattice operators leading to the right continuum
	  limit. In Ref. \cite{Luscher:2006df} it is proposed to allow for an additional term on the right hand side of Eq. \eqref{gw_relation2}
		\begin{equation}\label{gw_relation3}
				\gfive\,D + D\,\gfive = a\,D\gfive D + \Delta_B\,,
		\end{equation}
	  where $\Delta_B$ is supported in the vicinity of the boundaries and decays exponentionally with the distance to them.
	  
	  The operator introduced in the same reference
		\begin{equation}\label{sf_neuberger}
				\Dgw=\frac{1}{\bar{a}}\li\{1-\tfrac{1}{2}(U+U^\sim)\re\}\,,
		\end{equation}
		\begin{equation}\label{sf_neuberger2}
				U=A\li(A^\dagger A + caP\re)^{-1/2}\,,\quad U^\sim=\gfive U^\dagger \gfive\,,\quad \bar{a}=\frac{a}{1+s}\,,
		\end{equation}
	  is a modification of the Neuberger-Dirac operator \eqref{LPT:neuberger_op}. Here $A$ is essentially the Wilson-Dirac operator in the presence of the
	  boundaries \eqsref{wilson_dirac2}{wilson_dirac3}
		\begin{equation}
				A = 1 + s - a\Dw\,.
		\end{equation}
	  The parameters $s$ and $c$ can be used to optimise numerical computations. The modification is due to the boundary operator
		\begin{equation}
				P\psi(x) = \frac{1}{a}\li\{\delta_{x_0,a}P_-\psi(x)\big|_{x_0=a} + \delta_{x_0,T-a}P_+\psi(x)\big|_{x_0=T-a}\re\}\,.
		\end{equation}
		
		The construction is such that $\Dgw$ inherits the transformation properties of the Wilson-Dirac operator under cubic rotations, parity,
		time-reflections and charge conjugation. And having the combination $U+U^\sim$ renders $\gfive \Dgw$ hermitian.
		Although we will be mainly concerned with massless fermions the extension to massive fermions is simple
		\cite{Niedermayer:1998bi} and given by
		\begin{equation}\label{sf_neuberger_massive}
				D_m = (1-\tfrac{1}{2}\bar{a}m_0)\Dgw + m_0\,.
		\end{equation}
		
		In Section \ref{SF:chiral_props} we will show that the operator \eqref{sf_neuberger} indeed obeys \eqref{gw_relation3} with $a$ replaced by
		$\bar{a}$.
		
\subsection{Free theory}
\label{SF:mod_neuberger_op_free}

		For our perturbative computation we need the Dirac operator and the fermion propagator in the free theory (no gauge field, no
		four fermion interactions). To obtain an explicit expression for the free propagator like in the Wilson case might be possible but is much
		more difficult. Instead it is computed numerically using well established methods \cite{Giusti:2002sm}.
		
		Nevertheless, the operator under the square root in \eqref{sf_neuberger2} can be worked out in the time-momentum representation and
		it is reassuring that for the interesting range of the parameters $s$ and $c$ its eigenvalues are positive. This can be seen as follows.
		In the free theory the operator under the square root explicitly reads
		\begin{equation}
				A^\dagger A + caP = (1+s)^2 + s a^2 \sum_\mu \dmub\dmu + \tfrac{1}{2}a^4\sum_{\mu<\nu} \dmub\dmu\db{\nu}\partial_\nu + (c-1)aP\,,
		\end{equation}
		and acts on the fermion fields in the presence of the boundaries (cf. Section \ref{sf_wilson_dirac}).
		It is hermitian and therefore has real eigenvalues. Due to translation invariance
		in space the spatial eigenfunctions are plane waves
		\begin{equation}
				w(x)=\frac{1}{L^d} \sum_\vp\, \e^{i\vp\vx}\,w_\vp(x_0)\,,
		\end{equation}
		with spatial momenta $p_k$ in the range \eqref{mom_range}. In the following we restrict ourself to a definite momentum $\vp$. Then
		the Dirac components of $w_\vp(x_0)$ are ordinary functions of $x_0$. Therefore we drop the subscript in the eigenvalue equation
		\begin{equation}\label{eigen_eq}
				\li( -q\,a^2\db{0}\partial_0 + m + (c-1)aP\re)\,w(x_0) = \lambda\, w(x_0)\,,
		\end{equation}
		where $q$ and $m$ are short hand for 
		\begin{equation}
				q=\tfrac{a^2}{2}\vphat^2 - s\,,\quad m = (1+s)^2 - sa^2\vphat^2 + \tfrac{a^4}{2}\sum_{k<l}\phat_k^2\phat_l^2\,.
		\end{equation}
		Because of the projectors in $P$, eqs. \eqref{eigen_eq} are two coupled equations for the plus and minus components in
		$w(x_0)=P_+\,w_+(x_0) + P_-\,w_-(x_0)$. In the special case $c=1$ they have to be identical and one easily finds the
		solutions
		\begin{equation}\label{c_one}
				\sin(p_0 x_0)\,, \quad p_0 = \frac{n\pi}{T}\,,\quad n=1,2,\dots,T/a-1\,,
		\end{equation}
		and the correspondent eigenvalues
		\begin{equation}\label{eigenvalues}
				\lambda = (1+s)^2 - sa^2\phat^2 + \tfrac{a^4}{2}\sum_{\mu<\nu}\phat_\mu^2\phat_\nu^2 = q\,\phat_0^2 + m\,.
		\end{equation}
		
		For arbitrary $c$ the solutions to \eqref{eigen_eq} are
		\begin{equation}\label{eigenfunctions_c2}
				P_-\sin\big(p_0 x_0 + b\big) + P_+\sin\big(p_0(T-x_0) + b\big)\,,
		\end{equation}
		\begin{equation}
				b=-\arctan\li(\frac{\sin(ap_0)}{\frac{q}{c-1}+\cos(ap_0)}\re)\,,
		\end{equation}
		and the allowed values of $p_0$ are given by the solutions of the equation
		\begin{equation}\label{p0_eq}
				\tan(p_0 T) = \frac{\sin(ap_0)}{\frac{q}{c-1}+\cos(ap_0)}\,.
		\end{equation}
		The solutions of this equation are all real for non-negative $q/(c-1)$. Since the eigenfunctions are odd functions of $p_0$ 
		it is sufficient to stick to the $T/a-1$ solutions in the interval $[0,\pi/a)$. In any case the eigenvalues \eqref{eigenvalues}
		are bounded from below by $(1-|s|)^2$. 
		
		For $q/(c-1)<0$ one finds a pure imaginary solution $p_0^*=ik$, $k>0$. This solution enters the eigenvalue \eqref{eigenvalues}
		through the product $q\,\phat_0^2$ and might lead to zero modes and negative eigenvalues. However, for $c\ge 1$ imaginary solutions
		exist only for $q<0$ and thus the product $q\,(\phat_0^*)^2$ is positive and the eigenvalues \eqref{eigenvalues}
		are as well bounded from below by $(1-|s|)^2$.
		
		Since the operator under the square root is bounded from below we can adopt the argument in \cite{Hernandez:1998et}, using 
		expansion in Legendre polynomials, to conclude that in the free theory the locality of the Dirac operator $\Dgw$ is guaranteed for all $|s|<1$ and $c\ge 1$.
		The eigenfunctions 
		\eqref{eigenfunctions_c2} can be orthonormalised and used to write down an analytical expression for the kernel $\Dgw(x,y)$ of the
		Dirac operator. But the evaluation of $\Dgw(x,y)$ in this way would be very expensive, since it involves a sum over momenta $p_0$
		which in turn are determined for each set of parameter values by the roots of \eqref{p0_eq}. 
		
		Even more desirable would be an explicit expression for the propagator. With the eigenfunctions of the operator under the square root
		at hand one would hope to find the eigenfunctions of the hermitian operator $\Dgw \Dgw^\dagger$ and write the propagator as
		\begin{equation}
				S(x,y) = \li[\Dgw^\dagger\frac{1}{\Dgw \Dgw^\dagger}\re](x,y)\,.
		\end{equation}
		But the eigenvalue problem of $\Dgw \Dgw^\dagger$ is much more complicated than the one treated above and we do not attempt to solve it
		here. Instead the propagator is computed numerically using established methods. This includes a polynomial
		approximation of the inverse square root in the Dirac operator and solving the linear equation
		\begin{equation}
				\Dgw\psi = \eta\,,
		\end{equation}
		for $\psi$ with the approximate Dirac operator and appropriate sources $\eta$. The precision can be controlled throughout all steps
		of the computation \cite{Giusti:2002sm}. In this way the whole propagator can be evaluated with a given accuracy. In practice we also
		Fourier transform to the space component to obtain the time-momentum representation (see Section \ref{FeynmanRules}). 
		
\subsection{Chiral properties}
\label{SF:chiral_props}

	\begin{figure}
		\centering
		\includegraphics[scale=0.7]{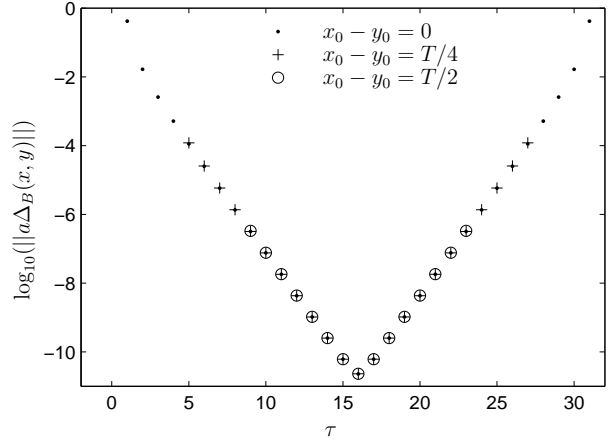}
		\caption{The deviation from the Ginsparg-Wilson relation $a\Delta_B$ in Eq. \eqref{gw_relation4} is 
			localised at the boundaries with tails that decrease exponentially with the distance from the
			boundaries. The plot is for a $16\times32$ lattice, $\theta=0.5$ and $x_1=y_1=L/2$.}
		\label{fig:DeltaB}
	\end{figure}
	\begin{figure}
		\centering
		\includegraphics[scale=0.7]{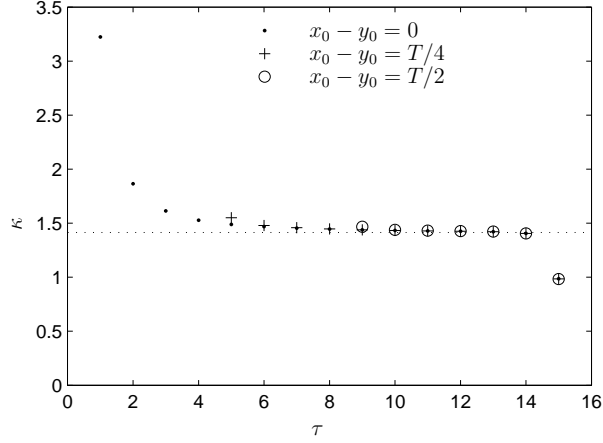}
		\caption{``Effective	mass'' plot for $a\Delta_B(x,y)$. For large distances from the boundary $a\Delta_B(x,y)$ is compatible with
		\eqref{DeltaB} with $\kappa\approx \sqrt{2}$ (dotted line). The plot is for the same parameters as Fig. \ref{fig:DeltaB}}
		\label{fig:kappa}
	\end{figure}
		As a check, it was tested numerically whether the operator \eqref{sf_neuberger} is a solution to
		\begin{equation}\label{gw_relation4}
				\gfive\,D + D\,\gfive = \bar{a}\,D\gfive D + \Delta_B\,,
		\end{equation}
		with
		\begin{equation}\label{DeltaB}
				 ||a\Delta_B(x,y)|| \approx \e^{-\kappa \cdot \tau/a} + \e^{-\kappa \cdot (T-\tau)/a}\,,\quad \kappa>0\,,
		\end{equation}
		for large distances $\tau$ and $T-\tau$ from the boundaries. For simplicity we define $\tau$ as the distance
		from the boundary at $x_0=0$ along the time direction as
		\begin{equation}\label{tau}
				 \tau = \frac{y_0 + x_0}{2}\,.
		\end{equation}
		In Fig. \ref{fig:DeltaB} $a\Delta_B(x,y)$ is plotted for a $16\times32$ lattice and $\theta=0.5$. The deviations from
		the Ginsparg-Wilson relation decay exponentially and from the ``effective	mass'' plot Fig. \ref{fig:kappa} the
		rate seems to approach $\kappa=\sqrt{2}$ for $\tau\to\infty$. It should be possible to compute $\Delta_B(x,y)$
		analytically to some extent, but we did not attempt to.

\section{The generating functional}
\label{SF:generating_functional}

    We want to compute expectation values of polynomials $\mathcal{O}$ in the fermion and anti-fermion bulk and boundary
    fields. A possible lattice representation of the boundary fields \eqsref{boundary_sources1}{boundary_sources2} is
		\footnote{In QCD one has to include link variables into this definition of the boundary fields.}
    \begin{equation}\label{boundary_lattice1}
        \zeta(\vx) = P_- \psi(x)\Big|_{x_0=a}\,,\quad  \zetabar(\vx) = \psibar(x)\Big|_{x_0=a} P_+\,,
    \end{equation}
    \begin{equation}\label{boundary_lattice2}
        \zeta'(\vx) = P_+ \psi(x)\Big|_{x_0=T-a}\,,\quad  \zetabar'(\vx) = \psibar(x)\Big|_{x_0=T-a} P_-\,.
    \end{equation}
    With this choice it is enough to introduce sources for the fermion and anti-fermion fields $\eta(x)$ and $\etabar(x)$ in the interior
    of the lattice $0<x_0<T$. The generating functional is then
    \begin{equation}
        Z[\etabar,\eta] = \int\rD\psi\rD\psibar\,\exp\li\{ -S_0[\psibar,\psi] + a^{d+1}\sum_x\, [\psibar(x)\eta(x) + \etabar(x)\psi(x)] \re\}\,.
    \end{equation}
    As said before only the fields at $0<x_0<T$ are integrated over in this functional integral.
    We can perform this integration and obtain
    \begin{equation}
        \ln Z[\etabar,\eta] =  a^{2(d+1)}\sum_{x,y}\, \etabar(x)\,S(x,y)\,\eta(y) + \text{const.}\,.
    \end{equation}
		Thus the generating functional is an exponential of a quadratic expression in the sources. Replacing the fields in $\mathcal{O}$
		by functional derivatives		
    \begin{equation}\label{SF:functional_derivatives}
        \psi(x) \to \frac{\delta}{\delta\etabar(x)}\,,  \qquad
        \psibar(x) \to -\frac{\delta}{\delta\eta(x)}\,,
    \end{equation}
    we may write the expectation value as
    \begin{equation}\label{sf:FreeEV}
        \<\mathcal{O}\z = \left\{ \frac{1}{Z}\, \mathcal{O}\, Z \right\}_{\etabar,\eta=0}\,.
    \end{equation}
		They are given as the sum of all Wick contractions. Below we list the basic contractions. 
    \begin{align}
    		[&\psi(x)\psibar(y)] = S(x,y)\,,\label{SF:contraction1} \\
        [&\psi(x)\zetabar(\vy)] = S(x;a,y_1)P_+\,, \\
        [&\psi(x)\zetabar'(\vy)] = S(x;T-a,y_1)P_-\,,\\
        [&\zeta(\vx)\psibar(y)] = P_-S(a,x_1;y)\,,\\
        [&\zeta'(\vx)\psibar(y)] = P_+S(T-a,x_1;y)\,,\\
        [& \zeta(\vx)\zetabar'(\vy)]  = P_-S(a,x_1;T-a,y_1)P_-\,, \\
        [& \zeta'(\vx)\zetabar(\vy)]  = P_+S(T-a,x_1;a,y_1)P_+\,, \\
        [& \zeta(\vx)\zetabar(\vy)]  =  P_-S(a,x_1;a,y_1)P_+\,, \\
        [& \zeta'(\vx)\zetabar'(\vy)]  =  P_+S(T-a,x_1;T-a,y_1)P_-\label{SF:contraction9}\,.
    \end{align}

\section{Boundary counter terms}
\label{boundary_counter_terms}

   The boundary conditions of the SF may give rise to new counter terms defined on the boundary, in the sense that the
   associated bare coefficients are needed to absorb infinities on the way to the continuum limit. Relevant operators or
   composite fields living at the boundary are those which have dimension $d$ or less. Thus objects like $\psibar \Gamma \psi$ with
   $\Gamma=\{1,\go,\gamma_1,\gfive\}$ can appear.

   The identity and $\go$ can be written as $P_++P_-$ and $P_+-P_-$ respectively. Hence the corresponding terms are proportional
   to the boundary fields (\ref{boundary_sources1},\ref{boundary_sources2}) in the continuum or (\ref{boundary_lattice1},\ref{boundary_lattice2})
	 on the lattice. The terms with $\gl$ and $\gfive$ violate parity
   and are therefore not present.
	 
	 Thus it is enough in the course of renormalisation to introduce a renormalisation factor $Z_\zeta$ for all boundary fields
    \begin{equation}
       	\zeta(\vx) \to Z_\zeta\,\zeta(\vx)\,, \quad 	\zetabar(\vx) \to Z_\zeta\,\zetabar(\vx)\,,
    \end{equation}
		and equivalently for $\zeta'$, $\zetabar'$.

\chapter{Self-coupled fermions in two dimensions}
\label{selfcoupled}

\section{Four fermion operators}
\label{four_fermion_op}

		In two dimensions fermion fields have mass dimension $1/2$ and a local four fermion operator
		\begin{equation}\label{most_general}
		    \psibar_{\alpha i} \psibar_{\beta j} \psi_{\gamma k}  \psi_{\delta l}\,,
		\end{equation}
		has a dimensionless coupling in the action. From the point of view of dimensional analysis such a interaction
		term is renormalisable. This means that if such a four fermion interaction is added to a theory that
		is renormalisable, it stays so. All new divergences can be absorbed into a redefinition of the four
		fermion coupling.
		
		In the literature one finds several two-dimensional fermion models with different symmetry and interaction content
		\cite{Thirring:1958in,Mitter:1974cy,Gross:1974jv}.
		This is because there are a lot of possible ways to contract the
		indices of four fermion fields. But in two dimensions there are also several relations between the possible contractions. 
		In the following we discuss this in some detail.
		
		The most general operator is an arbitrary contraction of the Dirac (Greek letters $\alpha,\beta,\dots$)
		and flavour (Latin letters $i,j,\dots$) indices in eq. \eqref{most_general}.
		Since we want a theory that is as similar to QCD as it can be in two dimensions, it certainly should be
		Lorentz invariant, even under parity and, in the massless case, have an $U(N)$ flavour symmetry.
		\footnote{This is meant in the continuum. At the end of the day we are interested in continuum QCD.}
		
		Lorentz invariance strongly constrains the possible contractions of Dirac indices. Since eq. \eqref{most_general}
		must be a Lorentz scalar it must be a product of two scalars, two pseudo-scalars, two vectors or two axial-vectors.
		Consider the case of two scalars\footnote{Repeated indices are summed over if not indicated otherwise.}
		\begin{equation}\label{two_scalar}
		    \psibar_{\alpha i} \psi_{\alpha j} \psibar_{\beta k}  \psi_{\beta l}\,.
		\end{equation}
		Invariance under $U(N)$ transformations
		\begin{equation}\label{UNtrafo}
			\psi_{\alpha i} \to U_{im}\, \psi_{\alpha m}\,, \quad 
			\psibar_{\alpha j} \to  \psibar_{\alpha n}\, U_{jn}^*\,,
		\end{equation}
		\begin{equation}
			\text{where} \quad U^\dagger U = U U^\dagger = 1\,,
		\end{equation}
		allows for the following two flavour contractions	
		\begin{equation}\label{two_scalar2}
		    \psibar_{\alpha i} \psi_{\alpha i} \psibar_{\beta j}  \psi_{\beta j}\,,\quad
		    \psibar_{\alpha i} \psi_{\alpha j} \psibar_{\beta j}  \psi_{\beta i}\,.
		\end{equation}
		The second contraction can be expanded in a basis of $N\times N$ matrices (Appendix \ref{Flavour})		
		\begin{equation}\label{flavour_mixing}
				\psibar_i \psi_j \psibar_j \psi_i =\frac{1}{N}(\psibar \psi)^2 + \frac{1}{2}\sum_a\, (\psibar \lambda^a\, \psi)^2 \,,
		\end{equation}
		where we suppressed all subscripts on the left hand side. The first term is proportional to the first one in \eqref{two_scalar2}
		and the sum is over the generators $\lambda^a$ of $SU(N)$, which act on the flavour indices of $\psibar$ and $\psi$.
		
		Therefore the most general four fermion operator consistent with the symmetries can be expanded in a basis of eight different
		contractions. One half of them are products of flavour singlet bilinear operators
		\begin{equation}\label{4FIdiagonal}
			\begin{array}{ll}
				O_{SS} =(\psibar\psi)^2\,, \\
				O_{PP} =(\psibar\gfive\psi)^2\,, \\
				O_{VV} =\sum_\mu(\psibar\gmu\psi)^2\,,\\
				O_{AA} =\sum_\mu(\psibar\gmu\gfive\psi)^2\,,
			\end{array}
		\end{equation}
		and the other half are products of flavour vector bilinear opeartors
		\begin{equation}\label{4FImixing}
			\begin{array}{ll}
				O'_{SS} =\sum_a(\psibar\lambda^a\psi)^2\,,\\
				O'_{PP} =\sum_a(\psibar \gfive\lambda^a\psi)^2\,,\\
				O'_{VV} =\sum_{\mu,a}(\psibar\gmu\lambda^a\psi)^2\,,\\
				O'_{AA} =\sum_{\mu,a}(\psibar\gmu\gfive\lambda^a\psi)^2\,.
			\end{array}
		\end{equation}
		
		But in two dimensions these operators are not independent. Suppose there is only a single fermion ($N=1$).  From the
		above list one would expect four different terms (no flavour vector operators for a single fermion). 
		But there are only four independent field components at each space-time point. Thus there is only one
		local four fermion operator.
		
		For $N>1$ the number of independent field components is no restriction. Nevertheless there are relations among 
		the operators above that reduce the number of really independent ones to three. Because of a peculiarity of the 
		$\gamma$-matrices in two dimensions ($\gmu\gfive=i\epsilon_{\mu\nu}\gnu$) there is no difference between
		vector and axial-vector and thus		
		\begin{equation}\label{dep12}
		    O_{AA} = -O_{VV}\,,\quad O'_{AA} = -O'_{VV}\,.
		\end{equation}
		
		More dependencies are due to Fierz identities. Fierz identities connect products of Dirac bilinear forms
		by rearranging the order of the Dirac spinors (see Appendix \ref{FT}). For 
		the two flavour contractions of \eqref{two_scalar2}, but with general Dirac structure, we find
		\begin{equation}\label{FI_general}
		    (\psibar_{i} \Gamma \psi_{j})\, (\psibar_{ j} \Gamma \psi_{ i}) 
		    		= -\frac{1}{4} \sum_I\, \tr(\Gamma_I\, \Gamma\, \Gamma_I\, \Gamma)
		    (\psibar_{i} \Gamma_I \psi_{ i})\, (\psibar_{ j} \Gamma_I \psi_{ j})\,,
		\end{equation}
		where $\Gamma,\Gamma_I \in \{\Gamma_S=1\,,\Gamma_P=\gfive\,,\Gamma_V=\go\,,\gl\}$.
		Using also \eqref{flavour_mixing} this yields three identities relating the flavour-singlet and the flavour-vector
		operators
		\begin{eqnarray}
		    O'_{SS} & = & -(1+2/N)\,O_{SS} - O_{PP} - O_{VV} \label{dep3}\\
		    O'_{PP} & = & -O_{SS} - (1+2/N)\,O_{PP} + O_{VV} \label{dep4}\\
		    O'_{VV} & = & -2\,O_{SS} + 2\,O_{PP} - 2/N\,O_{VV} \label{dep5}\,.
		\end{eqnarray}
		
		The number of independent operators has been reduced from eight to three. One possible choice would be
		\begin{equation}\label{terms2}
		    O_{SS}\,, O_{PP}\,, O_{VV}\,.
		\end{equation}
		But any other combination of three operators from \eqref{4FIdiagonal} and \eqref{4FImixing} is equally good.

\section{Chiral symmetry}
\label{CGN:chiral_symmetry}

		The eight operators of \eqref{4FIdiagonal} and \eqref{4FImixing}, of which only a set of three is independent, 
		are invariant under $U(N)$ transformations \eqref{UNtrafo} by construction. Since massless QCD at the classical level 
		has a global chiral $U(N)$ symmetry we investigate here the transformation properties of these operators under
		the infinitesimal transformations \eqref{U1Aomega} and \eqref{SUNAomega}. In Section \ref{continuum_wi} these were
		introduced as local transformations. But since they will act on operators that are local and contain no derivatives of
		fields the outcome is also applicable for the global case.
		
		For the above mentioned symmetry transformations  the change $\delta \Op$ of the operator $\Op$ linear in the
		infinitesimal parameter $\omega$ is
    \begin{equation}
        \Op \to \Op + \delta \Op + \rO(\omega^2)\,.
    \end{equation}
    If $\delta \Op$ vanishes the operator is invariant under finite transformations that can be build up from the
    infinitesimal ones.
		
		\paragraph{Axial $U(1)$ transformation}	
		Consider first the axial $U(1)$ transformation.
		Because of the $\gamma$-matrices in $O_{VV}$ and $O'_{VV}$ they are invariant like
		the kinetic term.
    \begin{equation}
        \delta \Ovv = 0\,,\quad \delta \Opvv = 0\,.
    \end{equation}
		But $\delta O_{SS}$ and $\delta O_{PP}$ do not vanish. Using the projectors $P_{R,L}=\tfrac{1}{2}(1\pm\gfive)$
		we write
    \begin{equation}
        O_{SS} = (\psibar\,(P_R+P_L)\, \psi)^2\,,\quad O_{PP} = (\psibar\, (P_R-P_L)\, \psi)^2\,.
    \end{equation}
    Now it is easy to see that
    \begin{eqnarray*}
        \delta O_{SS} & = & 2 i \omega_A (\psibar\, (P_R-P_L)\, \psi)\,(\psibar\,(P_R+P_L)\, \psi)\,,\\
        \delta O_{PP} & = & 2 i \omega_A (\psibar\, (P_R+P_L)\, \psi)\,(\psibar\,(P_R-P_L)\, \psi)\,.
    \end{eqnarray*}
    Therefore the difference of the two operators is invariant
    \begin{equation}
        \delta(O_{SS} - O_{PP}) = 0\,,
    \end{equation}
    and similar for the primed operators 
    \begin{equation}
        \delta(O'_{SS} - O'_{PP}) = 0\,.
    \end{equation}

		\paragraph{Axial $SU(N)$ transformation}	
    The axial $SU(N)$ transformation \eqref{SUNAomega} affects Dirac and flavour indices.
    However, $O_{VV}$ has a trivial flavour structure and it is easy to see that
    it is invariant
    \begin{equation}
        \delta O_{VV} = 0\,.
    \end{equation}
    Computing $\delta  O'_{VV}$ the commutator $[\lambda^a,\lambda^b]=2if^{abc}$ with the 
    structure constants $f^{abc}$ appear (cf. Appendix \ref{Flavour})
    \begin{eqnarray*}
        \delta O'_{VV} & = & 2 i \omega_A^a (\psibar\, \gfive\lambda^a\gmu\lambda^b\, \psi
                                 + \psibar\, \gmu\lambda^b\lambda^a\gfive\, \psi)\,(\psibar\,\gmu\lambda^b\, \psi)\\
                  & = & - 2 i \omega_A^a (\psibar\, \gmu\gfive[\lambda^a,\lambda^b]\, \psi)\,(\psibar\,\gmu\lambda^b\, \psi)\\
                  & = & 4 \omega_A^a f^{abc} (\psibar\, \gmu\gfive\lambda^c\, \psi)\,(\psibar\,\gmu\lambda^b\, \psi)\\
                  & = & 4 i \omega_A^a f^{abc} \big\{
                           (\psibar\, \gl\lambda^c\, \psi)\,(\psibar\,\go\lambda^b\, \psi)
                         - (\psibar\, \go\lambda^c\, \psi)\,(\psibar\,\gl\lambda^b\, \psi) \big\}\,.
    \end{eqnarray*}
    In the last step we used the definition of $\gfive$ and the Clifford algebra (cf. Appendix \ref{gamma}). Since
    the structure constants are totally anti-symmetric,
    the whole expression is symmetric in the indices $b$ and $c$ and hence does not
    vanish for general $\omega_A^a$
    \begin{equation}
        \delta O'_{VV} \neq 0\,.
    \end{equation}
    The result for $O_{VV}$ and $O'_{VV}$ can be used together with \eqsref{dep3}{dep5} to infer 
    \begin{equation}
        \delta(O_{SS} - O_{PP}) \neq 0\,, \quad \text{and} \quad \delta(O'_{SS} - O'_{PP}) \neq 0\,.
    \end{equation}

\section{Lattice chiral Gross-Neveu model}
\label{CGN:lattice_cgn}

		
		The continuum chiral Gross-Neveu model in Euclidean space-time is given by the action
    \begin{equation}\label{CGNaction1}
        S^c_{\text{CGN}} = \int \rd^2 x\,\li\{ \psibar\,\gmu\dmu\,\psi 
        			-\tfrac{1}{2} g^2 (O_{SS} - O_{PP}) - \tfrac{1}{2} g_V^2 O_{VV} \re\}\,.
    \end{equation}
    From the analysis of the last two sections we know that this is the most general action with chiral $U(1)\times U(1)$ and
    $SU(N)$ flavour symmetry. But we also learned that this form is not the only possibility. Using the identity \eqref{dep5}
    the action \eqref{CGNaction1} may equally be written as 
    \begin{equation}\label{CGNaction2}
        S^c_{\text{CGN}} = \int \rd^2 x\,\li\{ \psibar\,\gmu\dmu\,\psi 
        			+\tfrac{1}{4} g^2 O'_{VV} - \tfrac{1}{2} \delta_V^2 O_{VV} \re\}\,,
    \end{equation}
    \begin{equation}\label{couplings_relation}
    		\text{with}\quad \delta_V^2 = g_V^2 - g^2/N\,.
    \end{equation}
    The coupling $\delta_V^2$ is believed to have an exactly vanishing beta-function. This is connected
    to the decoupling of the $U(1)$ part of correlation functions that is derived by formal 
		manipulations of the continuum path integral \cite{Furuya:1982fh,Moreno:1987np}.
		The  vanishing of the beta-function has been proved up to two-loop perturbation theory \cite{Bondi:1989nq}.
    
    Passing now over to the lattice inevitably breaks part of the continuum symmetries. Generally this leads
    to more possibilities for the mixing of operators under renormalisation and to additonal parameters in the action.
    For example, an operator
    that is multiplicative renormalisable in the continuum may lose this property on the lattice.%
    \footnote{This also occurs in other regularisation schemes. In the two-loop computation of \cite{Bondi:1989nq} using
    dimensional regularization so called ``evanescent operators'' appear at an intermediate state of the calculation.}
    Being lattice artefacts\footnote{This is not true if there is an anomaly that
		breaks the symmetry.} one has to assure that these effects disappear in the continuum limit. As we will see, this may happen
    automatically, but in some cases additional parameters in the action have to be tuned.
    
    At first the theory is set up on a hypercubic lattice with periodic boundary conditions. The introduction of boundaries as in the
    Schrödinger functional involves additional problems that are addressed in a separate section (Section \ref{sf_cgn_model}).
    
    In the action we have to include terms that are forbidden by the continuum symmetries, but are allowed by the less restricting
    lattice symmetries. By dimensional analysis such operators with mass dimension $n$
    will have a factor $a^{n-D}$ in front (where $D=d+1$ is the dimension of space-time and should not be confused with the Dirac operator).
		Operators with $n>D$ can safely be neglected since 
    they automatically disappear for $a\to 0$.\footnote{The $n=D+1$ operators may be considered if one wants to achieve $\rO(a)$
    improvement.} But the operators with mass dimension $n\leq D$ are needed and lead to an additional renormalisation.
    
    Obviously (Euclidean) Lorentz symmetry $O(2)$ is broken on the lattice. The symmetry group of the hypercubic lattice, the
    hypercubic group, is a subgroup of $O(2)$ and contains rotations by $\pi/2$ and reflections. Luckily, the operators 
    allowed by the hypercubic group are the ones already present in \eqref{CGNaction1} or \eqref{CGNaction2} and operators
    with mass dimension $n>D$. So, ignoring all other symmetries for a moment, the effects of the broken Lorentz symmetry
    in the action automatically disappear in the continuum limit and we may use
    \begin{equation}\label{LatticeCGNaction1}
        S_{\text{CGN}} = a^2\sum_x\,\li\{ \psibar\,D\,\psi 
        			-\tfrac{1}{2} g^2 (O_{SS} - O_{PP}) - \tfrac{1}{2} g_V^2 O_{VV} \re\}\,,
    \end{equation}
    with the lattice Dirac operator $D$ as the lattice action. 
    The fermion fields $\psibar(x)$ and $\psi(x)$ are defined at the sites of a
    hypercubic lattice with periodic boundary conditions and lattice spacing $a$ (c.f. Section \ref{discretisation}). Note that
    there may be mixing amongst operators that cannot be neglected and that thus 
    has to be taken into account when calculating expectation values of such operators.
    
    Depending on the exact definition of $D$ there are further broken or modified symmetries that might lead to a modification
    of \eqref{LatticeCGNaction1}. The Wilson-Dirac operator explicitly breaks chiral symmetry and leads to an additive mass
    renormalisation. For staggered fermions the broken flavour symmetry causes new mixings. The lattice chiral symmetry associated with
    Ginsparg-Wilson fermions forbids additional operators of mass dimension $n\le D$ in the action.    
    In the next two sections we discuss Wilson and Ginsparg-Wilson fermions.

\subsection{Wilson fermions}
\label{CGN:wilson_fermions}

		As we have seen in Section \ref{wilson_fermions} the Wilson-Dirac operator explicitly breaks chiral symmetry and 
		a fine tuning of the bare mass is needed to restore chiral symmetry. The action should therefore contain
		all terms allowed by the remaining $U(N)$ flavour symmetry. 
		In Section \ref{four_fermion_op} a basis for $U(N)$ invariant four fermion operators was established. 
		Due to dependencies among the diverse operators it is enough to choose as set of three operators out of \eqref{4FIdiagonal}
		and \eqref{4FImixing}. In analogy to \eqref{LatticeCGNaction1} one possibility is
    \begin{multline}\label{LatticeCGNaction2}
        S_{\text{CGN,W}} = a^2\sum_x\,\Big\{ \psibar\,(\Dw + m_0)\,\psi \\
        			-\tfrac{1}{2} g^2 (O_{SS} - O_{PP}) - \tfrac{1}{2} \delta_P^2 O_{PP} - \tfrac{1}{2} g_V^2 O_{VV}  \Big\}\,,
    \end{multline}
		where we added a chiral symmetry breaking mass term and the coupling of $\Oss$ and 
		$\Opp$ are no longer related. 
		
		There are many different choices one could make. For example, one can invert eqs. \eqref{dep3}-\eqref{dep5} to obtain
    \begin{multline}\label{LatticeCGNaction3}
        S_{\text{CGN,W}} = a^2\sum_x\,\Big\{ \psibar\,(\Dw + m_0)\,\psi \\
        			-\tfrac{1}{2} \gp^2 (\Opss -\Oppp) - \tfrac{1}{2} \deltap_P^2 \Oppp - \tfrac{1}{2} \gp_V^2 \Opvv  \Big\}\,,
    \end{multline}
    where the primed couplings are related to the unprimed ones
		\begin{align}\label{SCF:translation1}
	    	\gp^2 & = \frac{N^2}{4(N^2-1)}\Big\{(2/N)\,g^2 - \delta_P^2 - 2g_V^2\Big\} \,,\\
	    	\deltap_P^2 & = \frac{N^2}{4(N^2-1)}\Big\{2(1/N-1)\,\delta_P^2\Big\} \,, \\
	    	\gp_V^2 & = \frac{N^2}{4(N^2-1)}\Big\{-2g^2 + \delta_P^2 + 2/N\,g_V^2\Big\}\label{SCF:translation2}\,.
		\end{align}
		The traceless $\lambda$-matrices in the four fermion operators in \eqref{LatticeCGNaction3} can help to reduce the number of
		non-zero diagrams in perturbative calculations. (This will be utilized in Section \ref{chiral_sym_restoration})
		Yet another version is obtained by decomposing the interaction as in \eqref{CGNaction2}
    \begin{equation}\label{LatticeCGNaction4}
        S_{\text{CGN,W}} = a^2\sum_x\,\li\{ \psibar\,(\Dw + m_0)\,\psi 
        			+\tfrac{1}{4} g^2 O'_{VV} - \tfrac{1}{2} \delta_P^2 O_{PP} - \tfrac{1}{2} \delta_V^2 O_{VV}  \re\}\,,
    \end{equation}
    \begin{equation}\label{couplings_relation3}
    		\text{with}\quad \delta_V^2 = g_V^2 - g^2/N\,.
    \end{equation}
    
    As pointed out in Section \ref{wilson_fermions} the bare mass $m_0$ has to be tuned to a non-zero value in order
    to restore chiral symmetry. In the same way $\delta_P^2$ has to be tuned since the value
     $\delta_P^2=0$ is not distinguished by a greater symmetry. A whole section (Section \ref{chiral_sym_restoration})
      is devoted to the restoration of chiral symmetry.

\subsection{Ginsparg-Wilson fermions}
\label{CGN:gw_fermions}

		A Dirac operator satisfying the Ginsparg-Wilson relation \eqref{LPT:gw_relation} implicates the lattice chiral symmetry
		\eqref{LPT:gw_chiral_symmetry}. This symmetry becomes the familiar chiral symmetry in the continuum limit and forbids
		a mass term. However, at finite lattice spacing the invariant four fermion operators are different from the continuum ones.
		
		The transformations \eqref{LPT:gw_chiral_symmetry} may be written as
		\begin{equation}\label{CGN:gw_chiral_symmetry}
				\psi \to \psi + \epsilon\gfiveh\psi\,,\quad \psibar \to \psibar + \epsilon\psibar\gfive\,,
		\end{equation}
		\begin{equation}
				\text{with}\quad \gfiveh = \gfive(1-aD)\quad \text{and}\quad \gfiveh^2 = 1\,.
		\end{equation}
		This looks like the usual chiral transformations except for the $\gfiveh$. Using \eqref{LPT:gw_chiral_symmetry} one
		proofs the identity
		\begin{equation}
				(1-\tfrac{a}{2}D)\gfiveh = \gfive(1-\tfrac{a}{2}D)\,.
		\end{equation}
		Which means that $(1-\tfrac{a}{2}D)\psi$ transforms under \eqref{CGN:gw_chiral_symmetry} like $\psi$ under
		\eqref{LPT:gw_chiral_symmetry}
		\begin{equation}\label{CGN:gw_chiral_symmetry2}
				(1-\tfrac{a}{2}D)\psi \stackrel{\text{\eqref{CGN:gw_chiral_symmetry}}}{\to}
				 (1-\tfrac{a}{2}D)\psi + \epsilon\gfive(1-\tfrac{a}{2}D)\psi\,.
		\end{equation}
		Therefore the chirally invariant operators of Section \ref{CGN:chiral_symmetry} are invariant under the lattice symmetry
		after replacing $\psi\to(1-\tfrac{a}{2}D)\psi$. In particular, if we define
		\begin{equation}\label{4FIdiagonal_gw}
			\begin{array}{ll}
				\hat{O}_{SS} =(\psibar(1-\tfrac{a}{2}D)\psi)^2\,, \\
				\hat{O}_{PP} =(\psibar\gfive(1-\tfrac{a}{2}D)\psi)^2\,, \\
				\hat{O}_{VV} =\sum_\mu(\psibar\gmu(1-\tfrac{a}{2}D)\psi)^2\,,
			\end{array}
		\end{equation}
		then the operators invariant under \ref{CGN:chiral_symmetry} are
    \begin{equation}
        \delta \Ohvv = 0\,,\quad \delta(\Ohss - \Ohpp) = 0\,.
    \end{equation}
		The treatment and result in the case of the operators which are products of flavour vector bilinear operators 
		\eqref{4FImixing} is equivalent. 		
		Note that the operators \eqref{4FIdiagonal_gw} are not ultra-local any more. Nevertheless the identities \eqref{dep12} and
		 \eqref{dep3}-\eqref{dep5} 	hold also for these operators. But in the derivation in Section \ref{four_fermion_op} 
		 one has to replaces 	$\psi\to\hat{\psi}=(1-\tfrac{a}{2}D)\psi$.
		 
		Now we gathered all 	prerequisites to write down the lattice action for Ginsparg-Wilson fermions in the form
    \begin{equation}\label{LatticeCGNaction_gw1}
        S_{\text{CGN,GW}} = a^2\sum_x\,\li\{ \psibar\,D\,\psi 
        			-\tfrac{1}{2} g^2 (\Ohss - \Ohpp) - \tfrac{1}{2} g_V^2 \Ohvv \re\}\,,
    \end{equation}
    or
    \begin{equation}\label{LatticeCGNaction_gw2}
        S_{\text{CGN,GW}} = a^2\sum_x\,\li\{ \psibar\,D\,\psi 
        			+\tfrac{1}{4} g^2 \Ohpvv - \tfrac{1}{2} \delta_V^2 \Ohvv \re\}\,,
    \end{equation}
    \begin{equation}\label{couplings_relation2}
    		\text{with}\quad \delta_V^2 = g_V^2 - g^2/N\,,
    \end{equation}
    where in both cases
    \begin{equation}
    		\gfive D + D \gfive = aD\gfive D\,.
    \end{equation}

\section{The discrete Gross-Neveu model}
\label{dgn}

		Note that for $g_V^2=0$ and $\delta_P^2=g^2$ \eqref{LatticeCGNaction2} becomes  the action of the discrete Gross-Neveu model	
    \begin{equation}\label{LatticeDGNaction}
        S_{\text{DGN}} = a^2\sum_x\,\li\{ \psibar\,(D + m_0)\,\psi 
        			-\tfrac{1}{2} g^2 O_{SS}  \re\}\,.
    \end{equation}
		The model is invariant under finite chiral transformations $\psi\to\psi\gfive$, $\psibar\to-\gfive\psibar$ and under a hidden $\rO(2N)$ symmetry.
		This invariance becomes evident in  the Majorana representation \cite{Korzec:2006hy}. Since its beta-function is well known in perturbation theory 
		\cite{Gracey:1991vy},	we will use this model as a cross-check of our calculation.

\section{Schrödinger functional of the CGN model}
\label{sf_cgn_model}

		The Schrödinger functional of free Wilson and Ginsparg-Wilson fermions in two dimensions reads (cf. Section \ref{schroendinger_functional})
    \begin{equation}\label{CGN:free_sf}
        Z_0 = \int \rD\psi\rD\psibar \,  \exp\li\{ -a^2\sum_x\, \psibar(x)\,D\,\psi(x) \re\} \,,
    \end{equation}
    where the Dirac operator $D$ in presence of the boundaries is given by the left hand side of
    \eqref{massive_wilson_dirac} and \eqref{sf_neuberger} respectively. The subscript of $Z_0$ is to indicate that it refers to the free theory.
    Remember that only the fermion fields at times $x_0=a,2a,\dots,T-a$ are
    integrated over in the functional integral and that they obey Dirichlet boundary conditions \eqsref{bc_1}{bc_2}.
    It is convenient to assume, as we do in \eqref{CGN:free_sf}, that the fermion fields are defined at all other values of $x_0$ as well,
    but that they are zero there.
    
    Since the four fermion operators have mass dimension two, they are irrelevant in the discussion of the naturalness of the
    Schrödinger functional  boundary conditions (Section \ref{sf_universality}) and the needed boundary counter terms
    (Section \ref{boundary_counter_terms}). Thus with the fermion fields defined as above the Schrödinger functional of the chiral Gross-Neveu
    model is
    \begin{equation}\label{CGN:cgn_sf}
        Z = \int \rD\psi\rD\psibar \,  \exp\li\{ -a^2\sum_x\, \li[\psibar(x)\,D\,\psi(x) + \tfrac{1}{2}\sum_I\, c_I\, O_{I}\re] \re\} \,,
    \end{equation}
    where the sum is over the four fermion operators multiplied by the corresponding coupling constant. This notation covers all the
    possible choices of operators presented in the Section \ref{CGN:lattice_cgn}.
        
\subsection{Generating functional}
\label{CGN:sec_generating_functional}

		In Section \ref{SF:generating_functional} we outlined how to compute expectation values of polynomials $\mathcal{O}$ in the fermion and
		anti-fermion bulk and boundary fields. A definition of the lattice boundary fields is given by \eqref{boundary_lattice1} and
		\eqref{boundary_lattice2}. The generating functional for the interacting theory is then 
    \begin{equation}\label{CGN:generating_functional}
        Z[\etabar,\eta] = \int \rD\psi\rD\psibar \, 
         \exp\li\{ -a^2\sum_x\,\li[ \psibar\,D\,\psi - \tfrac{1}{2}\sum_I\, c_I\, O_{I} - (\psibar\eta + \etabar\psi) \re] \re\} \,.
    \end{equation}
		Replacing in $\mathcal{O}$	the field operators by functional derivatives as defined in \eqref{SF:functional_derivatives}, 
		the expectation value can be written in a compact form
    \begin{equation}\label{CGN:expectaion_value}
        \ev{\mathcal{O}} = \left\{ \frac{1}{Z}\, \mathcal{O}\, Z[\etabar,\eta] \right\}_{\etabar,\eta=0} \,.
    \end{equation}
    
    For our perturbative expansion it is convenient to rewrite \eqref{CGN:expectaion_value}. If we also replace the fields 
    in the four fermion operators $O_I$ by functional derivatives the generating functional becomes
    \begin{equation}\label{CGN:generating_functional2}
        Z[\etabar,\eta] = \e^{\tfrac{a^2}{2}\sum_x\,c_I\, O_{I}}\, Z_0[\etabar,\eta]\,.
    \end{equation} 
    where $Z_0[\etabar,\eta]$ is the generating functional of the free theory (cf. Section \ref{SF:generating_functional}). After
    integration over the fermion fields it is given by
    \begin{equation}
        Z_0[\etabar,\eta] =  \exp\li\{a^{4}\sum_{x,y}\, \etabar(x)\,S(x,y)\,\eta(y)\re\}\,,
    \end{equation}
    where $S(x,y)$ is the propagator associated with the Dirac operator in \eqref{CGN:generating_functional}
    \begin{equation}\label{CGN:propagator}
        D\, S(x,y) = \frac{1}{a^{2}}\,\delta_{x,y}\,,\quad 0<x,y<T\,,
    \end{equation}  
    with boundary values
    \begin{equation}
        P_+\,S(x,y)\Big|_{x_0=0}=P_-\,S(x,y)\Big|_{x_0=T}=0\,.
    \end{equation}    
    Finally we use \eqref{CGN:generating_functional2} in \eqref{CGN:expectaion_value} to write
    \begin{equation}\label{CGN:expectaion_value2}
        \ev{\mathcal{O}} =
         \left\{ \frac{1}{Z}\, \mathcal{O}\,\e^{\tfrac{a^2}{2}\sum_x\,c_I\, O_{I}}\, Z_0[\etabar,\eta] \right\}_{\etabar,\eta=0} \,,
    \end{equation}
		
    Expanded in powers of the couplings the generating functional \eqref{CGN:generating_functional2} is the sum of all vacuum diagrams,
    that is, diagrams with no external lines. Clearly the factor $1/Z$ in
    \eqref{CGN:expectaion_value2} cancels all diagrams in the expansion of the expectation value, that contain vacuum diagrams as subdiagrams.
    Therefore we write the expansion of the  expectation value as
    \begin{equation}\label{CGN:expectaion_value4}
        \ev{\mathcal{O}} = \ev{\mathcal{O}}_0 + \tfrac{a^2}{2}\sum_x\,c_I\, \ev{\mathcal{O}\,O_{I}(x)}_0 
         + \tfrac{a^4}{8}\sum_{x,y}\,c_I\,c_J\, \ev{\mathcal{O}\,O_{I}(x)\, O_{J}(y)}_0 + \dots\,,
    \end{equation}
    where $\ev{\Op\,X}_0$, with $X$ a product of four fermion operators or unity, is defined as
    \begin{equation}\label{CGN:expectaion_value5}
        \ev{\mathcal{O}\, X}_0 = \left\{ \frac{1}{Z_0}\, \mathcal{O}\,X\, Z_0[\etabar,\eta] \right\}_{\etabar,\eta=0} 
        - \;\textrm{\parbox{.3\linewidth}{contractions with\\ vacuum subdiagrams}}\,.
    \end{equation}
    
    The terms in the expansion \eqref{CGN:expectaion_value4} are given by the sum of all Wick contractions of the fields in the 
    operator $\Op$ and the increasing number of four fermion operator insertions, that do not contain vacuum diagrams as subdiagrams.
    The basic contractions are given by \eqref{SF:contraction1}-\eqref{SF:contraction9}. The contractions are most conveniently represented
    by Feynman diagrams. The rules for the corresponding expressions are given in the next section.

\subsection{Feynman rules}
\label{FeynmanRules}

\begin{figure}
		\centering
		\includegraphics[scale=0.95]{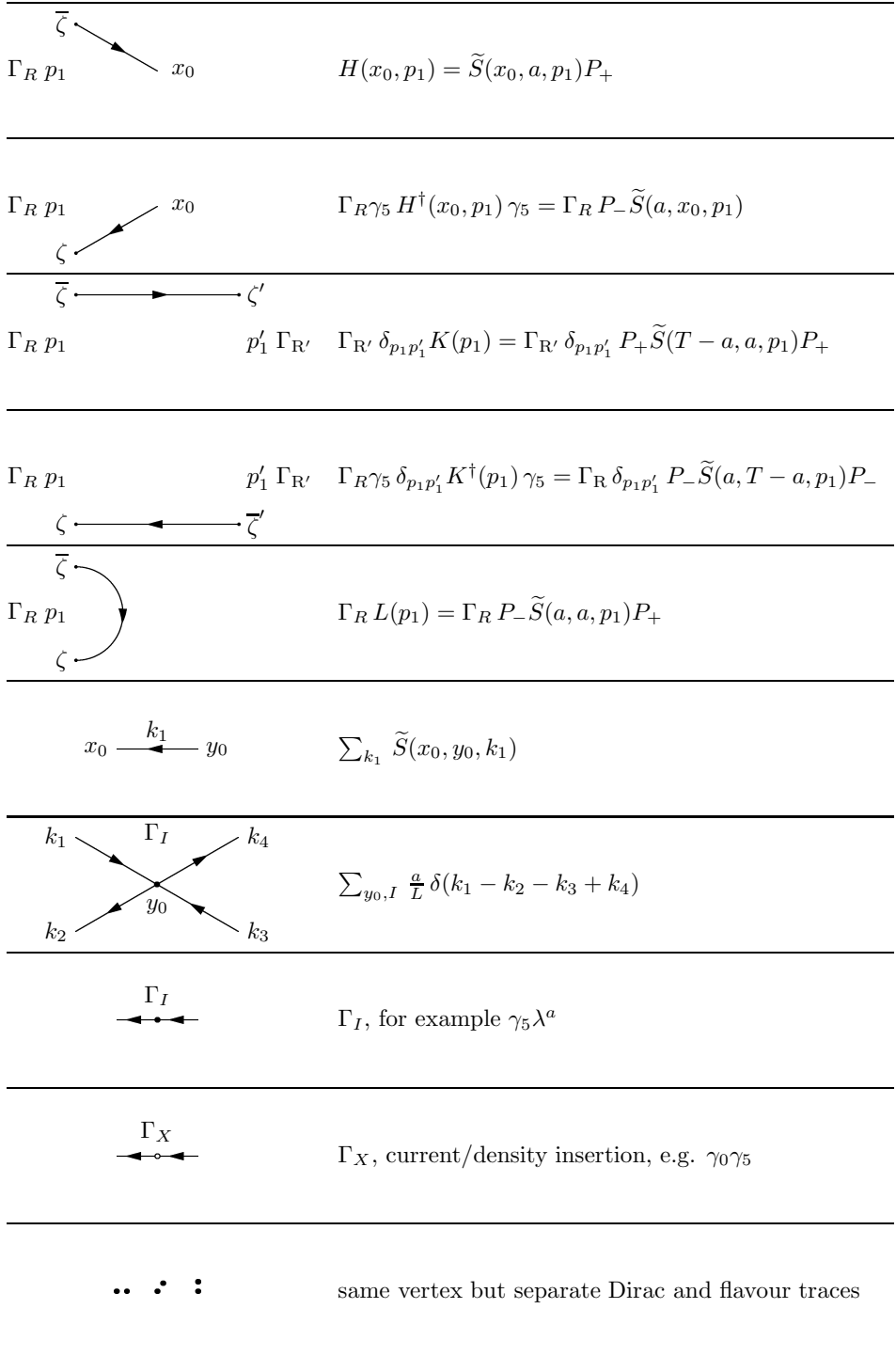}
        \caption{Feynman rules for Schrödinger functional of the chiral Gross-Neveu model.}
        \label{fig:FeynmanRules}
	\end{figure}

    Because translation invariance is broken in the time direction the Feynman rules in the Schr\"odinger functional
    are given in a half Fourier transformed space. That is, the
    half transformed Propagator $\Sp(x_0,y_0,p_1)$ is defined through
    \begin{equation}\label{halfFTProp}
        S(x,y)=\frac{1}{L}\,\sum_{p_1}\,\e^{i\p(x_1-y_1)}\,\Sp(x_0,y_0,p_1)\,.
    \end{equation}
    The rules are listed in Fig. \ref{fig:FeynmanRules}.
    In the case of Wilson fermions the propagators from the boundary to the interior $H(x_0,p_1)$, $H'(x_0,p_1)$ and from boundary to boundary
     $K(p_1)$, $K'(p_1)$ can be evaluated to some extent (see Appendix \ref{CF:analytical}).    
	 
	\paragraph{Statistical factors}
	Because the four fermion interactions consist of a product of two identical bilinear operators, each insertion of a four fermion interaction
	 can be connected to the rest of a diagram in two ways that lead to the same final contribution.
	In other words there are two different contractions giving the same diagram. This factor is always cancelled by the $1/2$ that comes
	with each insertion (see \eqref{CGN:expectaion_value5}).
	
	 The factor $1/n!$ from the Taylor expansion needs some more words.  At the $n$th order of 
	this expansion it multiplies terms with $n$ four fermion interactions. Their general strucure is
    \begin{equation}\label{polynomial}
        \li(\sum_{I=1}^F c_I\, O_I\re)^n = 
        	\sum_{m_1 + m_2 + \dots + m_F = n}\, \frac{n!}{m_1!\, m_2!\dots m_F!}\, O_{1}^{m_1}\, O_{2}^{m_2}\dots O_{F}^{m_F}\,,
    \end{equation}
  where $F$ is the number of different four fermion operators in the action ($F=3$ for Wilson, $F=2$ for Ginsparg-Wilson fermions).
	The number of equal contractions for a term of this sum due to interchange of vertices is $m_1!\, m_2!\dots m_F!$.
	Thus, taking also into account the overall factor $1/n!$ from the Taylor expansion of the exponential function (see
	 \eqref{CGN:expectaion_value5}) we have $(m_1!\, m_2!\dots m_F!)/n!$ which is exactly cancelled by the factor in \eq{polynomial}.
	Therefore none of these factors appears in the
	Feynman rules.
   
	\paragraph{Sign}
	The sign of a diagram is determined by the number of traces. Each trace comes with a minus sign. This gives the overall sign 
	\[(-1)^{\#\text{traces}}\,.\]


\chapter{Chiral symmetry restoration}
\label{chiral_sym_restoration}

	In this Section we use the chiral Ward identity to restore chiral symmetry in the Wilson discretisation
	at finite lattice spacing (up to $\rO(a)$). We show here that this can be achieved by a perturbative computation of the
	critical mass $m_c$ and a symmetric $\delta_{P,s}$. The value of the third coupling $g_V^2$ has not to be tuned. 

	On the lattice, using the Wilson discretisation, chiral symmetry is broken and nothing in general prevents
	$m_0$ to take a finite value or $\delta_P^2\neq 0$. However, as we pointed out in Section \ref{wilson_fermions}, the renormalised
	axial current $\ren{A_\mu}$ of properly defined fields is expect to obey
	\begin{equation}\label{dArenOa}
	        \<\ren{\Op}\,\md_\mu \ren{A_\mu}(x) \z = \rO(a)\,.
	\end{equation}
	This condition can be used to fix the bare parameters in perturbation theory \cite{Bochicchio:1985xa,Luscher:1996vw} (as well as in numerical simulations
	 \cite{Korzec:2005ed}). 	In the following section we discuss the strategy and the result in detail. 

\section{Correlation functions}\label{CorrelationFunctions}
	We define correlation functions in the Schödinger functional (SF) set up in order to utilise \eqref{dArenOa}.		
	The correlation functions
	\begin{equation}
		f_X(x_0) = -\frac{a^2}{2N}\,\sum_{y_1,z_1}\;
	  	 	\ev{\;\psibar(x)\,\Gamma_X\,\psi(x)\; \zetabar(y_1)\,\gfive\,\zeta(z_1)\;}\,,
	  	  \label{mc:fX}
	\end{equation}
	\begin{equation}
			 \Gamma_A=\go\gfive\,,\quad \Gamma_P=\gfive\,, \label{mc:fX2}
	\end{equation}
	to be considered are correlators of a zero momentum pseudo scalar boundary state built from the boundary fields $\zeta$, $\zetabar$ 
	(eq. \eqref{boundary_lattice1}) and
	insertions of the time component of the axial current ($f_A$) and the pseudo scalar density ($f_P$) respectively.
	The vacuum expectation value 	$\ev{\cdot}$ has been defined in Section \ref{CGN:sec_generating_functional}.
	
	As discussed in Section \ref{CGN:wilson_fermions} there is some ambiguity in the operators 
	of the four fermion interaction. Here we use the following form of the action
    \begin{multline}\label{PSR:LatticeCGNaction}
        S_{\text{CGN,W}} = a^2\sum_x\,\Big\{ \psibar\,(\Dw + m_0)\,\psi \\
        			-\tfrac{1}{2} \gp^2 (\Opss -\Oppp) - \tfrac{1}{2} \deltap_P^2 \Oppp - \tfrac{1}{2} \gp_V^2 \Opvv  \Big\}\,,
    \end{multline}
  for the actual computation. 
	Using \eqref{PSR:LatticeCGNaction} 
	in the perturbative expansion the computation gets simplified, because a number of contractions vanish due to 
	vanishing flavour traces. Results obtained for the primed couplings are then translated back into the more common unprimed couplings
	of \eqref{LatticeCGNaction2} using \eqref{SCF:translation1}-\eqref{SCF:translation2}.
  
	For small couplings the expectation value in \eqref{mc:fX} can be expanded as indicated in \eqref{CGN:expectaion_value4}.		
	Then the right hand side of \eqref{mc:fX} is a sum of an increasing number of insertions of the interactions but the expectation value
	taken with the free action $S_0$ only and all contractions with vacuum subdiagrams subtracted (cf. \eqref{CGN:expectaion_value5})
	\begin{equation}\label{fXepansion}
		f_X(x_0) = f_X^{(0)}(x_0) + \sum_{I}\, c_I\, f_{X,I}^{(1)}(x_0)	+ \sum_{I,J}\, c_I c_J\, f_{X,IJ}^{(2)}(x_0) + \rO(c^3)\,,
	\end{equation}
	where
	\begin{equation}\label{fXepansion2}
		c_I = \{\gp^2\,, \deltap_P^2\,, \gp_V^2\} \quad \text{and} \quad O_I = \{\Opss -\Oppp\,, \Oppp\,, \Opvv\}\,.
	\end{equation}
	\begin{figure}
		\centering
		\includegraphics[scale=0.7]{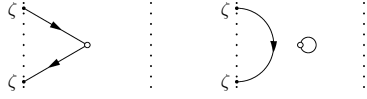}
		\caption{Tree level diagrams for $f_X$.}
		\label{fig:TreeLevelDiagramsForFX}
	\end{figure}
	\begin{figure}
		\centering
		\includegraphics[scale=0.7]{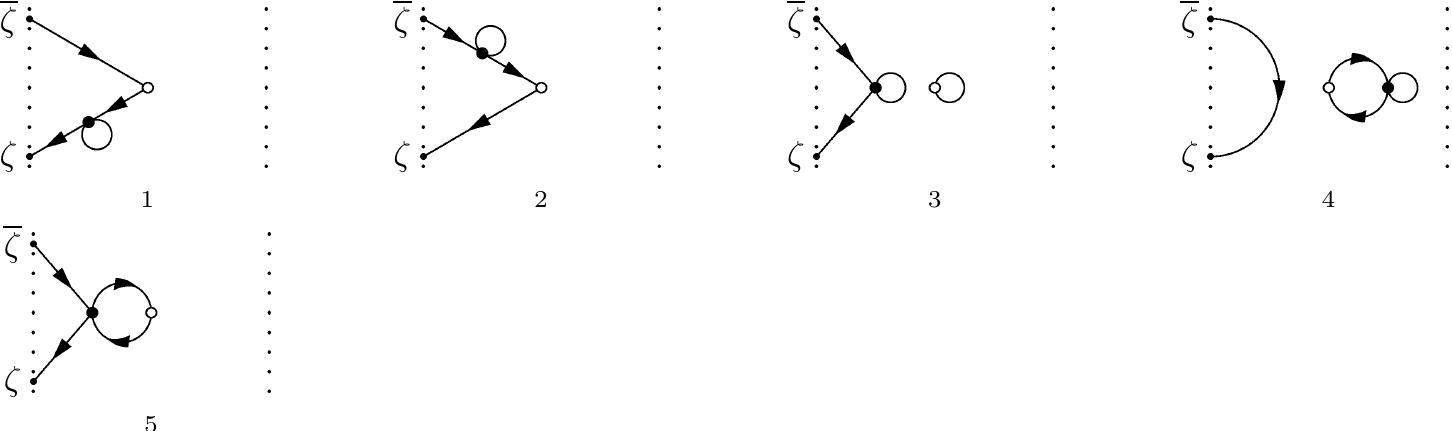}
		\caption{First order diagrams for $f_X$.}
		\label{fig:FirstOrderDiagramsForFX}
	\end{figure}
	The tree level amplitude 
	\begin{equation}\label{fXtree}
		f_X^{(0)}(x_0) = -\frac{a^2}{2N}\,\sum_{y_1,z_1}\; \ev{\;\psibar(x)\,\Gamma_X\,\psi(x)\; \zetabar(y_1)\,\gfive\,\zeta(z_1)\;}_0\,,
	\end{equation}
	is the sum of the two diagrams sketched in Fig. \ref{fig:TreeLevelDiagramsForFX}. In these diagrams the
	dotted lines represent the time slices $x_0=0$ and $x_0=T$. Plain lines are used for the fermion propagator and the small open circle in the
	middle symbolises the insertion of a current or density operator. Using the explicit form of the porpagator \eqref{FreePropagator2}
	the tree level amplitudes can be calculated analytically to some extent. The somewhat lengthy expressions are listed in Appendix \ref{boundary2interior}.

	The first order amplitudes	
	\begin{equation}\label{fX1loop}
		f_{X,I}^{(1)}(x_0) = -\frac{a^4}{4N}\,\sum_{y_1,z_1,u}\; 
				\ev{\;\psibar(x)\,\Gamma_X\,\psi(x)\; \zetabar(y_1)\,\gfive\,\zeta(z_1)\; O'_{I}(u) \;}_0\,,
	\end{equation}
 	are sums of five diagrams. Due to the $\gfive$-hermiticity of the Wilson-Dirac operator, and thus the
	fermion propagator, the diagrams 1 and 2 in Fig. \ref{fig:FirstOrderDiagramsForFX} are equal. In these diagrams the
	small filled circles (dots) represent the insertion of a four fermion interaction.
	Since the free propagator is diagonal in flavour space the flavour traces factorise from the rest of the computation. Differences in the
	resulting factors originate from different orders of the $\lambda$-matrices in the four fermion interaction and the number of separate traces.
	Only those combinations of the $\lambda$-matrices that can be reduced to the identity give a non-zero contribution.
	In the case of the first order diagrams there is only one possible order
	\begin{equation}\label{FirstOrderTrace}
		\lambda^a\lambda^a = \frac{2(N^2-1)}{N}\,\unity\,.
	\end{equation}
	The number of separate traces is one (diagrams 1,2,5) and two (diagrams 3,4). Together with the factor $1/N$ in the definition of the
	correlation function the first order amplitudes can be organised in powers of $N^p$ with $p=0,1$ and an overall factor
	 \eqref{FirstOrderTrace}. 

	\begin{figure}
		\centering
		\includegraphics[scale=0.7]{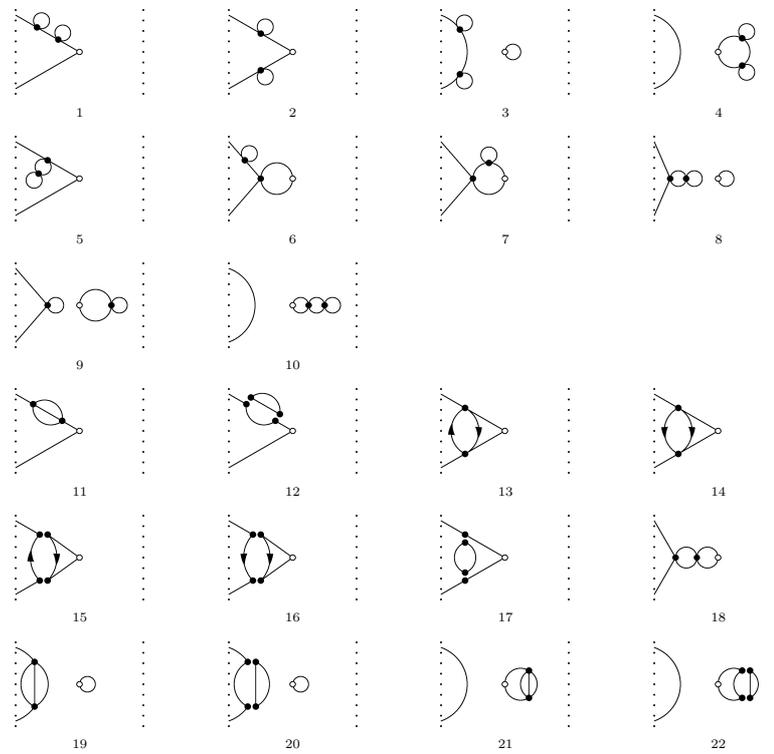}
		\caption{Second order diagrams for $f_X$.}
		\label{fig:SecondOrderDiagramsForFX}
	\end{figure}
	The second order amplitudes	
	\begin{equation}\label{fX2loop}
		f_{X,IJ}^{(2)}(x_0) = -\frac{a^6}{8N}\,\sum_{y_1,z_1,u,v}\; 
				\ev{\;\psibar(x)\,\Gamma_X\,\psi(x)\; \zetabar(y_1)\,\gfive\,\zeta(z_1)\; O'_{I}(u)\; O'_{J}(v) \;}_0\,,
	\end{equation}
 	are sums of the 22 diagrams depicted in Fig. \ref{fig:SecondOrderDiagramsForFX}. The diagrams 1, 5-7, 11 and 12 have to be counted
 	twice since when reflected at an horizontal line they give diagrams with the same numerical value but a different contraction (just like the 
 	first order diagrams 1 and 2 are equal).
	There are now four $\lambda$-matrices to combine producing factors
	\begin{align}
		\lambda^a\lambda^a\lambda^b\lambda^b & = \frac{4(N^2-1)^2}{N^2}\,\unity\,,\\
		\lambda^a\lambda^b\lambda^a\lambda^b & = -\frac{4(N^2-1)}{N^2}\,\unity\,,\\
		\lambda^a\lambda^b\,\otimes\, \lambda^a\lambda^b & = \frac{4(N^2-1)}{N^2}\,\unity\,\otimes\,\unity\,.	
	\end{align}
	The number of separate traces reaches from one to three. Together with the factor $1/N$ in the definition of the
	correlation function the second order amplitudes can be organised in powers of $N^p$ with $p=0,1,2,3$ and an overall factor $4(N^2-1)/N^2$.

\section{Strategy and result}
\label{CSR:strategy_result}

	We work with the Schrödinger functional of the Gross-Neveu model at fixed ratio $T/L=2$. The phase
  $\theta\equiv\theta_1$ characterising the spatial boundary conditions (cf. Section \ref{discretisation}) is a free parameter of this
  regularisation. The Ward identities are independent of $\theta$. Hence it provides a probe for the critical mass in the sense that
  \eqref{dArenOa} must hold for all $\theta$.
  
	We define the renormalised correlation function $\ren{f_A(x_0)}$ and its time derivative $h_A(\theta,x_0/L)$ as dimensionless quantities 
	\begin{equation}\label{hA}
		\ren{f_A(x_0)} = Z_A\, Z_\zeta^2\; f_A(x_0) \quad \text{and} \quad 
						h_A(\theta,x_0/L) = L\md_0\,\ren{f_A(x_0)}\,,
	\end{equation}
	where we introduced normalisation factors for the axial current and the boundary fields
	\begin{align}\label{CSR:norm_factors}
		Z_A = 1 +  c_I\, Z_{A,I}^{(1)} + \rO(c^2)\,,\quad	Z_\zeta = 1 +  c_I\, Z_{\zeta,I}^{(1)} + \rO(c^2)\,.
	\end{align}
	In general, because $\dmu A_\mu(x)$ can mix with $\tfrac{1}{a}\,P(x)$ we expect for $h_A$ an expansion in powers of $a/L$ starting with
	a linear divergence
	\begin{equation}\label{hAExpansion}
		h_A(\theta,x_0/L) = A_{-1}(\theta,x_0/L)\, L/a + A_{0}(\theta,x_0/L) + \rO(a/L)\,.
	\end{equation}	
	Eq. \eqref{dArenOa} enforces the coefficients of the divergence and the finite part to vanish
	\footnote{Be careful to not confuse the coefficient $A_0(\theta,x_0/L)$ with the time component of the axial vector current $A_0(x)$.}
	\begin{equation}\label{hAcond}
		A_{-1}(\theta,x_0/L) = A_{0}(\theta,x_0/L) = 0\,,\quad  \text{for all} \quad\theta\,,x_0/L\,,
	\end{equation}
	\begin{equation}\label{hAcond2}
		\text{at} \quad am_0=am_c\,\quad \text{and} \quad \deltap_P^2=\deltap_{P,s}^2\,.
	\end{equation}
	Thus we have a two-dimensional parameter space spanned by $\theta$ and $x_0/L$ for which these coefficients
	must vanish.
	
	For free fermions \eqref{dArenOa} is satisfied for $am_0=am_c=0$. In the interacting theory $m_c$ can be
	expanded
	\begin{equation}\label{mcExpansion}
	        am_c=am_{c,I}^{(1)}\,c_I + am_{c,IJ}^{(2)}\,c_I c_J + \rO(c^3)\,.
	\end{equation}
	As we will see also $\deltap_P^2$ is constrained by \eqref{dArenOa} and is given in terms of the other two couplings.
	
	The dimensionless $h_A$ just as well posses an expansion in powers of the couplings
	\begin{equation}\label{hAExpansion1}
		h_A(\theta,x_0/L) = h_A^{(0)} + \sum_I\, c_{I}\, h_{A,I}^{(1)} + \sum_{I,J}\,c_I c_J\,h_{A,IJ}^{(2)} + \rO(c^3)\,.
	\end{equation}
	Now the coefficients of this expansion, sums of lattice diagrams, may be expanded in powers of $a/L$. As already indicated
	 $h_A^{(0)}=h_0$ is at least $\rO(a^3/L^3)$ and therefore $am_c^{(0)}=0$.
	In the next two sections we compute the divergent and finite part of $h_{A,I}^{(1)}$ and $h_{A,IJ}^{(2)}$. Although there will be some
	analytic arguments to simplify the expressions, the final evaluation of the contributing diagrams is performed numerically as described in Section
	\ref{continuum_limit}.

\subsection{First order}

	Using expansions \eqref{fXepansion} and \eqref{mcExpansion} in \eqref{hAExpansion1} yields for the first order term
	\begin{equation}\label{hAfirstorder}
		\sum_I\, c_{I}\, h_{A,I}^{(1)} =
			 \sum_I\, c_{I}\, \Big\{  h_{1,I} + am_{c,I}^{(1)}\, h_2 + \li(Z_{A,I}^{(1)} + 2 Z_{\zeta,I}^{(1)}\re)\, h_0 \Big\}\,,
	\end{equation}
	with
	\begin{equation}
		h_0=L\md_0 f_A^{(0)}(x_0)\,, 
			\quad h_{1,I} = L\md_0 f_{A,I}^{(1)}(x_0)\,,
			 \quad h_2=\frac{\partial}{\partial am_0}h_0\,,
	\end{equation}
	all defined at $am_0=0$. The tree level amplitude $h_0$ vanishes identically for $\theta=0$ and is $\rO(a^3/L^3)$ for $\theta\neq 0$. Its
	derivative $h_2$ with respect to $am_0$ though diverges linearly with $L/a$.
	This implies at first order
	\begin{equation}\label{FirstOrderCoeffs}
		\sum_I\, c_{I}\,\li\{ h_{1,I} + am_{c,I}^{(1)}\, h_2\re\} = \rO(a/L)\,.
	\end{equation}
	
\paragraph{Linear divergence}
	The first order amplitudes $h_{1,I}=\sum_{d=1}^5\, h_{1,I}^d$ are the sums of the diagrams depicted in Fig.
	\ref{fig:FirstOrderDiagramsForFX}.
	Only diagrams one to four  contribute to the linear divergence due to the contact term induced by the bubble contraction (cf. Appendix 
	\ref{BubbleReduction}). The contribution of diagram five $h_{1,I}^5$ is $\rO(1)$ and thus is important for enforcing $A_{0}(\theta,x_0/L)=0$.
	
	For the linearly divergent diagrams one to four we find (with the help of \eqref{Insertion})
	\begin{multline}\label{BubbleDiagrams}
		\sum_{d=1}^4\, h_{1,I}^d = -\frac{a^3}{2N}\,\frac{2(N^2-1)}{N}\,L\md_0\; \sum_{y_1,z_1,u}\; \times\\
				 \ev{\;A_0(x)\; \zetabar(y_1)\,\gfive\,\zeta(z_1)\;
				 \psibar(u) \li(B(u_0) + F_I(u)\; a^2/L^2\, \theta_1 + \dots\re) \psi(u) \;}_0	\,.
	\end{multline}
	The factor $2(N^2-1)/N$ originates from the sum over the $\lambda$-matrices (cf. Eq. \eqref{FirstOrderTrace} and comment before).
	For fixing the critical mass we need only the leading term in \eqref{BubbleDiagrams}. Using \eqref{DiagonalPart} and \eqref{Dm0} we find
	\begin{equation}	
		\sum_{d=1}^4\, h_{1,I}^d = - \frac{2(N^2-1)}{N}\,B_1\, h_2  + \rO(1)\,.
	\end{equation}
	Thus the contribution of these diagrams is proportional to $h_2$ which also multiplies $am_{c,I}^{(1)}$  in \eqref{hAfirstorder}.
	Abbreviating $am_c^{(1)} = \sum_I\,am_{c,I}^{(1)}\,c_I$ the linearly divergent part at this order is given by
	\begin{equation}\label{FirstOrderA}
		A_{-1}(\theta,x_0/L) = \li( am_c^{(1)} - \li(\deltap_P^2 + 2\gp^2_{V}\re)\,\frac{2(N^2-1)}{N}\,B_1 \re)\,
						 C_{-1}(N,\theta,x_0/L)\,.
	\end{equation}
  Where $C_{-1}(N,\theta,x_0/L)$ is the coefficients of the linear divergences in $h_2$
	\begin{equation}\label{h2divergent}
		 h_2  = C_{-1}(N,\theta,x_0/L)\,\tfrac{L}{a} + \rO(1)\,,
	\end{equation}
  \begin{equation}
		 \text{and}\quad C_{-1}(N,\theta,x_0/L) = C_{-1,0}(\theta, x_0/L) + N\, C_{-1,1}(\theta, x_0/L)\,.
	\end{equation}
	The linear divergences in $h_2$ arise from the mass derivative of the two tree level diagrams in Fig.
	 \ref{fig:TreeLevelDiagramsForFX}.
	Diagram two receives an additional factor of $N$ because its evaluation involves two separate flavour traces
	(it has two closed fermion loops). These amplitudes can be determined numerically.
	The right hand side of \eqref{FirstOrderA} vanishes for all $\theta\,, x_0/L$ only if the critical mass is
	\begin{equation}\label{FirstOrderMass}
		am_c^{(1)} = \frac{2(N^2-1)}{N}\,B_1\,\li(\deltap_P^2 + 2\gp^2_{V}\re)\,.
	\end{equation}
	
\paragraph{Finite part}
	Summing up all $\rO(1)$ contributions of the five 
	diagrams and organising it in powers of $N$ we find that the terms proportional to $\gp^2$ and $\gp^2_V$ cancel. Explicitly we have
	\begin{equation}\label{FinitePart}
		A_{0}(\theta,x_0/L) =  
		\deltap_P^2 \frac{2(N^2-1)}{N}\, C_{0}(N,\theta,x_0/L)\,,
	\end{equation}
	with
	\begin{equation}
		C_{0}(N,\theta,x_0/L) = C_{0,0}(\theta, x_0/L) + N\, C_{0,1}(\theta, x_0/L)\,.
	\end{equation}
	The right hand side of \eqref{FinitePart} has to vanish for all $\theta\,,\,x_0/L$. Therefore chiral symmetry is restored for
	\begin{equation}\label{FirstOrderCouplings}
		\deltap_{P,s}^2 = 0\,,
	\end{equation}
	at first order of perturbation theory.

\paragraph{Summary}
	Translating Eqs. \eqref{FirstOrderMass} and \eqref{FirstOrderCouplings} back to the unprimed couplings using
	\eqref{SCF:translation1}-\eqref{SCF:translation2} yields
	\begin{align}
	    am_c^{(1)} & = - 0.3849001\times \li( 2N g^2 - \delta_P^2 - 2g_V^2\re)\,,\label{amc1}\\
	    \delta_{P,s}^2 & = \rO(g^4)\,.\label{gP2}
	\end{align}
	At this order we find no constraint on the vector coupling $g_V^2$. Or phrased in another way, the vector coupling has not to be tuned
	in order to restore chiral symmetry.

\subsection{Second order}
	
	The relevant terms  for the second order term in \eqref{hAExpansion1} are
	\begin{multline}\label{hAsecondorder}
		\sum_{I,J}\,c_I c_J\,h_{A,IJ}^{(2)} = \\
		\sum_{I,J}\,c_I c_J\, \Big\{ am_{c,IJ}^{(2)}\, h_2
			+   h_{1,IJ} + am_{c,I}^{(1)} h_{3,J} + am_{c,I}^{(1)} am_{c,J}^{(1)}\, h_4  + \rO(a/L) \Big\}\,,
	\end{multline}
	with
	\begin{equation}
		h_{1,IJ} = L\md_0 f_{A,IJ}^{(2)}(x_0)\,,
			 \quad h_{3,I} = \frac{\partial}{\partial am_0} h_{1,I}\,,
			 \quad h_4=\frac{1}{2}\frac{\partial^2}{\partial am_0^2} h_0\,.
	\end{equation}
	It is understood that the first order critical mass is set to the value found above and the constraint $\delta_{P,s}^2  = \rO(g^4)$
	is employed.
	Also terms like $(Z_{A,IJ}^{(2)} + 2 Z_{\zeta,IJ}^{(2)})\, h_0$ and $(Z_{A,I}^{(1)} + 2 Z_{\zeta,I}^{(1)})\, h_{A,J}^{(1)}$ 
	that are at least $\rO(a/L)$ and suppressed in \eqref{hAsecondorder}.

\paragraph{Divergences}
 	Recall the reduction of bubble diagrams like the diagrams 1-10 in Fig. \ref{fig:SecondOrderDiagramsForFX} to insertions of
 	the scalar density and the relation between the mass derivative of a expectation value and the insertion of the scalar density
 	outlined in Appendix \ref{BubbleReduction}. After a little thought it is clear that the derivatives $h_{3,J}$	and $h_4$
 	cancel all the power divergences in the second order amplitudes $h_{1,IJ}$ that are due to bubbles, i.e. the contribution of diagrams
 	1-10. That this is indeed the case was checked numerically.
 	The only other linearly divergent diagrams are number 11, 12 and 21, 22 in the same Figure. 
 	In order to keep the equations clear we define
	\begin{equation}
	 	am_c^{(2)} = \sum_{IJ}\,am_{c,IJ}^{(2)}\,c_I c_J\,,
	\end{equation}
	and the linear divergent part of the summed up diagrams 11, 12, 21 and 22
	\begin{equation}
		\sum_{d=11,12,21,22}\, h_{1,IJ}^d = C_{-1}^{IJ}(N,\theta,x_0/L)\,\tfrac{L}{a} + \rO(1)\,,
	\end{equation}
	\begin{equation}
		\text{with}\quad C_{-1}^{IJ}(N,\theta,x_0/L) = \sum_{n=0}^2\, N^n\,C_{-1,n}^{IJ}(\theta, x_0/L)\,.
	\end{equation}
 	The coefficient of the linear divergence at second order is then
 	\begin{equation}\label{SecondOrderA}
		A_{-1}(\theta,x_0/L) = 
	  am_{c}^{(2)}\, C_{-1} + \frac{4(N^2-1)}{N^2}\,\sum_{I,J}\, c_{I} c_{J}\, C_{-1}^{IJ}\,,
	\end{equation}
	where $C_{-1}=C_{-1}(N,\theta,x_0/L)$ is defined in \eqref{h2divergent}.
	
	Setting the right hand side of \eqref{SecondOrderA} zero leads to
	\begin{equation}
		am_c^{(2)} = -\frac{4(N^2-1)}{N^2}\,\sum_{I,J}\,c_I c_J\, \frac{C_{-1}^{IJ}(N,\theta,x_0/L)}{C_{-1}(N,\theta,x_0/L)}\,.
	\end{equation}
 	When one evaluates the ratios in the last expression numerically, one finds that the $\theta$ and $x_0/L$-dependence of the numerator
 	is cancelled by the denominator. In other words $C_{-1}^{IJ}(N,\theta,x_0/L)$ factorises
	\begin{equation}
		C_{-1}^{IJ}(N,\theta,x_0/L) = (D_{-1,0}^{IJ} + D_{-1,1}^{IJ}\, N) \cdot C_{-1}(N,\theta,x_0/L)\,.
	\end{equation}
	where $D_{-1,n}^{IJ}$ are computable constants.
 	Organising the terms in powers of $N$ we find at this order
	\begin{multline}\label{SecondOrderMass}
		am_c^{(2)} =  -\frac{4(N^2-1)}{N^2} \li\{ (D_{1} + N\, D_{2})\,( \gp^4 + \gp_V^4)
				  + 2(D_{2} + N\, D_{1})\,\gp^2\,\gp_V^2\re\}
	\end{multline}
 	with
	\begin{equation}
		D_1 = 0.01195(1)\,, \quad D_2 = 0.22870(1)\,.
	\end{equation}
	As mentioned the first order result $\deltap_{P,s}=\rO(g'^4)$ must be used here in order to cancel all divergences.
	
\paragraph{Finite part}
	The first order result allows for a second order contribution. Thus we have to include the first order finite part
 \eqref{FinitePart} at second order
	\begin{equation}\label{FinitePart2}
		A_{0}(\theta,x_0/L) = \deltap_{P,s}\, \frac{2(N^2-1)}{N}\, C_{0} + \frac{4(N^2-1)}{N^2}\,\sum_{I,J}\, c_I c_J\, C_{0}^{IJ} \,.
	\end{equation}
	The coefficients
	\begin{equation}
		C_{0}^{IJ}(N,\theta,x_0/L) = \sum_{n=0}^3\, N^n\,C_{0,n}^{IJ}(\theta, x_0/L)\,,
	\end{equation}
	receive contribution from all the 21 diagrams of Fig. \ref{fig:SecondOrderDiagramsForFX}, which all have a factor of $4(N^2-1)/N^2$ in
	common. $C_{0} = C_{0}(N,\theta,x_0/L)$ was introduced in \eqref{FinitePart} and is the finite part of the
	first order diagrams up to a common factor of $2(N^2-1)/N$. Setting \eqref{FinitePart2} zero yields
	\begin{equation}
		\deltap_{P,s} = \frac{2}{N} \sum_{I,J}\, c_I c_J\, \frac{C_{0}^{IJ}(N,\theta,x_0/L)}{C_{0}(N,\theta,x_0/L) }\,.
	\end{equation}
	Again we find that the $\theta$ and $x_0/L$-dependence of the numerator is cancelled by the one of the denominator, which means that chiral symmetry
	is really restored. Note that in
	the sum on the left hand side many terms vanish because of $\deltap_{P,s}=\rO(g^4)$.
	Organised in powers of $N$ the result is
	\begin{equation}\label{SecondOrderCouplings}
		\deltap_{P,s} = \frac{2 D_3}{N} \li[ N\,\gp^4 + \li(1+N-N^2\re)\,\gp_V^4 + 2\,\gp^2\,\gp_V^2 \re] + \rO(\gp^6)\,,
	\end{equation}
	with
	\begin{equation}
		D_3 = 0.6192(1)\,.
	\end{equation}

\paragraph{Summary}
	Translating Eqs. \eqref{SecondOrderMass} and \eqref{SecondOrderCouplings} back to the more common couplings $g^2$, $\delta_P^2$, $g_V^2$
	using \eqref{SCF:translation1}-\eqref{SCF:translation2} yields
	\begin{align}
	    am_c^{(2)} & = (D_{1} - N\, D_{2})\,( g^4 + g_V^4) + 2(D_{2} - N\, D_{1})\,g^2\,g_V^2 + \rO(g^6)\,,\label{amc2}\\
	    \delta_{P,s} & = D_3 \, \li[ N\,g^4 - 2\,g^2\,g_V^2 - g_V^4  \re] + \rO(g^6)\,.\label{gP22}
	\end{align}

\section{Conclusion}

	We demanded the chiral Ward identity \eqref{dArenOa} to hold up to $\rO(a)$ on the lattice for the renormalised operators. The computation
	is carried out in second order perturbation theory in the Schrödinger functional of the Gross-Neveu model with Wilson fermions.
	 The result is 	that the bare mass has to diverge in order to cancel a linear divergence and that the chiral symmetry breaking coupling
	$\delta_P^2$, although zero at first order, has to take a finite value at second order.
	
	Explicitly the result is
	\begin{equation}\label{mc_result}
	        am_c=am_{c}^{(1)} + am_{c}^{(2)} + \rO(g^6)\,.
	\end{equation}
	with
	\begin{align}
	    am_c^{(1)} & = - B_1\times \li( 2N g^2 - \delta_P^2 - 2g_V^2\re)\,,\label{amc12}\\
	    am_c^{(2)} & = (D_{1} - N\, D_{2})\,( g^4 + g_V^4) + 2(D_{2} - N\, D_{1})\,g^2\,g_V^2 + \rO(g^6)\,,\label{amc22}
	\end{align}
	and
	\begin{equation}\label{deltap_result}
		\delta_{P,s}^2 = D_3 \, \li(N\,g^4 - 2\,g^2\,g_V^2 - g_V^4  \re) + \rO(g^6)\,.
	\end{equation}
	We find no constraint on the vector coupling $g_V^2$. Or phrased in another way, the vector coupling has not to be tuned
	in order to restore chiral symmetry.
	
	In order to compare with the large $N$ result of \cite{Aoki:1985jj} we set and rescale
	\begin{equation}
		g^2=g_S^2/N\,,\quad  \delta_P^2=(g_P^2+g_S^2)/N\quad\text{and}\quad g_V^2\to g_V^2/N\,,
	\end{equation}
 	and take the $N\to\infty$ limit in \eqref{mc_result}	and \eqref{deltap_result}
	\begin{align}
	    am_c^{N\to\infty} & = -0.7698002\times g_S^2 + \rO(g^6)\,,\\
	    \li[g_P^2/g_S^2\re]_s^{N\to\infty} & = 1 - 0.6192(1) \times g_S^2 + \rO(g^4)\,.
	\end{align}
	For $am_c$ this is the complete large $N$ result. The authors neglect the vector four fermion interaction from the beginning, but their result is not changed if
	it is taken	into account. 
	In the large $N$ limit chiral symmetry is restored for $1/g_P^2=1/g_S^2 + 0.619$ which is reproduced by our result at second order.

\chapter{Renormalised coupling}
\label{renormalised_coupling}

	If the time extension $T$ of the Schrödinger functional of the massless chiral Gross-Neveu model is fixed to a multiple of the
	spatial extension $L$, say $T=2 L$, then all dimensionfull quantities will depend on $L$.
	Keeping the volume finite usually leads to systematic errors. But here the finite volume is utilised to probe the theory. That is
	renormalised quantities are defined at the scale $\mu = 1/L$. Therefore by definition there are no finite size effects in the
	Schrödinger functional.
	
	In view of a Monte-Carlo simulation of the model we aim here at the definition of renormalised couplings at zero renormalised mass 
	in terms of renormalised correlation functions of the boundary fields \eqref{boundary_lattice1}, \eqref{boundary_lattice2}.
	For each coupling $g_I^2$ in the action a
	combination $F_I$ of such correlation functions defines the corresponding renormalised coupling $\tilde{g}_I^2$
	such that it is equal to the bare coupling at leading order of perturbation theory up to corrections of order $a$
	\begin{equation}\label{RC:definition}
  		\tilde{g}_I^2 = F_I\quad \text{such that}\quad \tilde{g}_I^2 = g_I^2 + \rO(ag^2) + \rO(g^4)\,.
  \end{equation}
	In the continuum limit of asymptotically free theories this ensures $\tilde{g}_I^2 \eqw{a\to 0} g_I^2 + \rO(g^4)$.%
	\footnote{Throughout the thesis $\rO((g^2)^n)$ means all terms $\rO((g_{I_1}^2)^{n_1}\dots (g_{I_m}^2)^{n_m})$, $\sum_{i=1}^m n_i = n$.}
  In turn, as we have seen in Section \ref{asymptotic_freedom}, such a definition of the renormalised coupling  ensures
	the universality of the first two coefficients of the corresponding
  beta-function in the single coupling case. For multiple couplings the situation is more complicated (see Section
  \ref{multiple_coupling}).
  Nevertheless condition \eqref{RC:definition} is a possible choice also in that case.

	In the next Section the used correlation functions are discussed. In Section \ref{RC:perturbative_expansion}
	they are formally expanded in powers of the couplings exploiting the cancellation of some diagrams. The results for Wilson and
	Ginsparg-Wilson fermions are presented in Sections \ref{RC:wilson} and \ref{RC:gw} respectively. In the last Section (\ref{RC:Lambda}) we
	use the results to compute the ratio of the $\Lambda$-parameters.

\section{Correlation functions}		
		
	The two- and four-point correlators of the boundary fields are defined as
	\begin{equation}\label{2fermions}
  	f_2 = -\frac{a^2}{NL} \sum_{u_1 z_1}\, \ev{ \zetabar(u_1)\, \zeta'(z_1) }\,,
  \end{equation}
  and
  \begin{equation}\label{4fermions}
  	f_4 = -\frac{a^4}{2(N^2-1)L^2} \sum_{u_1 v_1 y_1 z_1}\, 
  					\ev{ \zetabar'(u_1)\, \gfive \lambda^a \zeta'(v_1)\; \zetabar(y_1)\, \gfive\,\lambda^a \zeta(z_1) }\,.
  \end{equation}
	We sum over the spatial direction to project on to zero momentum. The powers of the lattice spacing $a$ are chosen to render the
	correlation functions dimensionless.
	The remaining factors are to normalise the tree level amplitudes.
	
	The matrices $\Gamma_B$ and $\Gamma'_B$ contract the Dirac indices of the boundary fields. In two dimensions such a matrix can be expanded
	in the identity, $\gmu$ and $\gfive$. But only terms with $\Gamma_B$ and $\Gamma'_B$ equal to $\gl$ and/or $\gfive$ 
	have a non-vanishing contribution because of the projectors $P_\pm$
	at the boundary of the Schrödinger functional (apparent for example in the basic Wick contractions at the end of 
	Section \ref{SF:generating_functional} or the Feynman rules in Fig. \ref{fig:FeynmanRules}). For the same reason the contribution
	of the possible combinations of the two are all equal up to a factor of $i$ and the overall sign 
	($P_\pm \gfive P_\mp = \pm i P_\pm \gl P_\mp$). Therefore in the computation of $f_4$ we use for the Dirac structure on the boundaries
	$\Gamma'_B = \Gamma_B = \gfive$.
	
	For the flavour structure we choose flavour vector bilinears. In combination with four fermion interactions of the flavour scalar type
	\eqref{4FIdiagonal} this choice simplifies the perturbative computation. This is because diagrams which do not connect the two boundaries
	involve the flavour trace $\TRf[\lambda^a]$ and hence vanish. 
	
	The renormalised two- and four-point correlation functions are
	\begin{equation}\label{RC:ren_corr}
  		\ren{f_2} = Z_\zeta^2\,f_2 \quad \text{and} \quad \ren{f_4} = Z_\zeta^4\,f_4
  \end{equation}
  The normalisation factor for the boundary fields was already introduced in \eqref{CSR:norm_factors} in Section \ref{CSR:strategy_result}.
	
	In the ratio of appropriate powers of the renormalised quantities the unknown wave function normalisation is cancelled. Then it is
	enough to evaluate the unrenormalised correlators at zero renormalised mass
	\begin{equation}\label{RC:ratio}
  		R(\theta) = \frac{\ren{f_4}}{\ren{f_2}^2} - 1
  					= \frac{f_4}{\li(f_2\re)^2}  - 1 \quad \text{at}\quad m_{\mathrm{R}}=0\,.
  \end{equation}
 	For later convenience we subtract the free theory value at zero mass. As indicated the ratio $R=R(\theta)$
 	depends on the phase $\theta\equiv\theta_1$
 	parametrising the spatial boundary conditions. This dependence will allow us to find combinations $R(\theta)+b\cdot R(\theta')$
 	to define the renormalised couplings as in \eqref{RC:definition}.

\section{Perturbative expansion}
\label{RC:perturbative_expansion}

\begin{figure}
		\centering
		\includegraphics[scale=1]{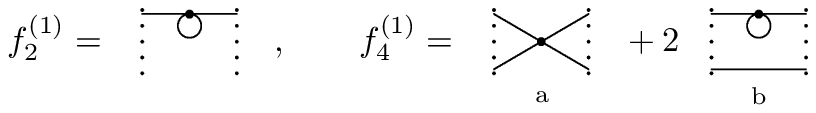}
		\caption{First order diagrams of $f_2$ and $f_4$. The bubble diagrams stand for the sum of the connected (one trace) disconnected (two traces) diagram.}
		\label{fig:FirstOrderDiagramsForf4f2}
	\end{figure}

\begin{figure}
		\centering
		\includegraphics[scale=1]{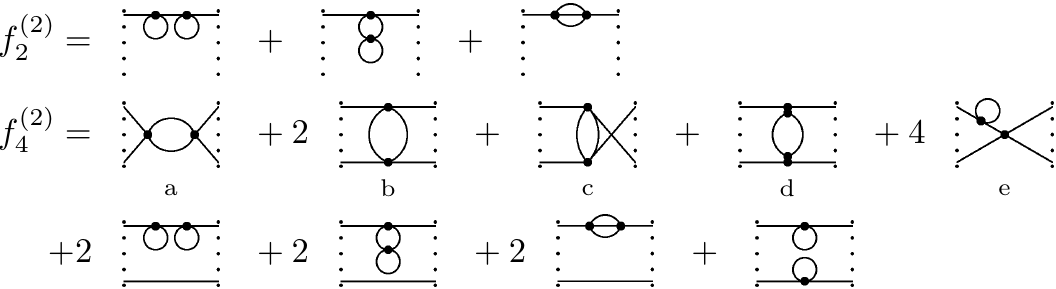}
		\caption{Second order diagrams of $f_2$ and $f_4$. The bubble diagrams stand for the sum of the connected (one trace) disconnected (two traces) diagram.}
		\label{fig:SecondOrderDiagramsForf4f2}
	\end{figure}

\begin{figure}
		\centering
		\includegraphics[scale=1]{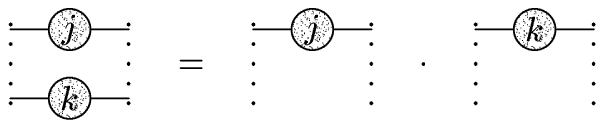}
		\caption{Diagrammatical representation of Lemma \ref{lemma:DiagramIdentity}.}
		\label{fig:dia_id}
	\end{figure}

	In order to expand $R_g$ in powers of the couplings $g_I^2$ we need the expansion of $f_4$ and $f_2$. We abbreviate
	$c_I=g_I^2$ and indicate that opposite to \eqref{fXepansion2} here $c_I$ refers to $g^2$, $\delta_P^2$ and $g_V^2$.
	\begin{equation}\label{f4exp}
  		f_4 = f_4^{(0)} + c_I\, f_{4,I}^{(1)} + c_I c_J\, f_{4,IJ}^{(2)} +\rO(c^3)\,,
  \end{equation}
  and $\li(f_2\re)^2$
	\begin{equation}\label{f2exp}
  		\li(f_2\re)^{2} =  \li(f_2^{(0)}\re)^2 + c_I\, 2 f_{2}^{(0)} f_{2,I}^{(1)}
  							 + c_I c_J \li\{ 2 f_2^{(0)} f_{2,IJ}^{(2)} + f_{2,I}^{(1)} f_{2,J}^{(1)} \re\} +\rO(c^3)\,.
  \end{equation}
  Multiplying \eqref{f4exp} with the inverse of \eqref{f2exp} yields
	\begin{multline}\label{ratioexp}
  		\frac{f_4}{\li(f_2\re)^2} = 1 + \frac{c_I}{\li(f_2^{(0)}\re)^2} \Bigg\{f_{4,I}^{(1)} - 2 f_{2,I}^{(1)} f_2^{(0)}\Bigg\}\\
  								 + \frac{c_I c_J}{\li(f_2^{(0)}\re)^2} \li\{ f_{4,IJ}^{(2)}- 2 f_{2,IJ}^{(2)} f_2^{(0)}
  								 						 + 3 f_{2,I}^{(1)} f_{2,J}^{(1)} - 2 \frac{ f_{2,I}^{(1)} f_{4,J}^{(1)}}{f_2^{(0)}} \re\} +\rO(c^3)\,.
  \end{multline}
  The first and second order diagrams of $f_2$ and $f_4$ are listed in Figs. \ref{fig:FirstOrderDiagramsForf4f2} and 
  \ref{fig:SecondOrderDiagramsForf4f2}. 
	Half of the diagrams of $f_4$ at first and second order can be shown to be identical to products of $f_2$ diagrams. To see this, 
	consider a diagram $f_{4,i}^{(n)}$ at order $n$ in the expansion of $f_4$, where the subscript $i$ labels the diagram. Then we define:
	\begin{definition} \label{def:DiagramIdentity}
		If a diagram $f_{4,i}^{(n)}$ at order $n$ in the expansion of $f_4$ can be cut horizontally
		without cutting through a fermion line or a vertex, then it is called reducible.		
	\end{definition}	
	In Appendix \ref{ProofLemma} we prove the following Lemma:
	\begin{lemma} \label{lemma:DiagramIdentity}
		If a diagram $f_{4,i}^{(n)}$ at order $n$ in the expansion of $f_4$ is reducible, then $f_{4,i}^{(n)}$
		can be written as a product $f_{2,j}^{(r)}\cdot f_{2,k}^{(s)}$ of two diagrams appearing in the expansion of $f_2$, with
		$n=r+s$. The amputated part of $f_{2,j}^{(r)}$ and $f_{2,k}^{(s)}$ is equal to the amputated part of the upper and the lower half of
		$f_{4,i}^{(n)}$, respectively.	
	\end{lemma}
	Diagrammatically this is depicted in Fig. \ref{fig:dia_id}.	
	As a consequence all products of $f_2$ diagrams in the expansion \eqref{ratioexp} are cancelled by corresponding $f_4$ diagrams. 
	Only $f_4$ diagrams that are not reducible are left
	\begin{multline}\label{ratioexp2}
  		\frac{f_4}{\li(f_2\re)^2} = 1 + \frac{c_I}{f_4^{(0)}}\; f_{4,I,\text{a}}^{(1)}\\
  								 + \frac{c_I c_J}{f_4^{(0)}} \li\{ 
  								 			f_{4,IJ,\text{a}}^{(2)} + 2 f_{4,IJ,\text{b}}^{(2)} + f_{4,IJ,\text{c}}^{(2)} +  f_{4,IJ,\text{d}}^{(2)} 
  								 			+ 4 f_{4,IJ,\text{e}}^{(2)}
  								 			 - 2 \frac{ f_{2,I}^{(1)}}{f_2^{(0)}}\, f_{4,J,\text{a}}^{(1)}  \re\} +\rO(c^3)\,.
  \end{multline}
  
  From this result one easily reads off the expansion of $R$
	\begin{equation}\label{Rgexp}
  		R(\theta) = R^{(1)}(\theta) + R^{(2)}(\theta) + \rO(c^3)\,,
  \end{equation}
  with
	\begin{equation}\label{Rgexp2}
  		R^{(1)}(\theta) = c_I\, R_I^{(1)}(\theta) \quad \text{and} \quad R^{(2)}(\theta) = c_I c_J\, R_{IJ}^{(2)}(\theta)\,.
  \end{equation}
  The first and second order terms are
	\begin{equation}\label{Rg1}
  		R_I^{(1)}(\theta) = \frac{1}{f_4^{(0)}}\; f_{4,I,\text{a}}^{(1)}\,,
  \end{equation}
	\begin{multline}\label{Rg2}
  		R_{IJ}^{(2)}(\theta) = \frac{1}{f_4^{(0)}} \Bigg\{ 
  								 			f_{4,IJ,\text{a}}^{(2)} + 2 f_{4,IJ,\text{b}}^{(2)} + f_{4,IJ,\text{c}}^{(2)} +  f_{4,IJ,\text{d}}^{(2)} 
  								 			+ 4 f_{4,IJ,\text{e}}^{(2)}
  								 			 - 2 \frac{ f_{2,I}^{(1)}}{f_2^{(0)}}\, f_{4,J,\text{a}}^{(1)}   \Bigg\}\,.
  \end{multline}
	We now proceed with the explicit computation for Wilson fermions.

\section{Wilson fermions}
\label{RC:wilson}
	
	The lattice action of the chiral Gross-Neveu model with Wilson fermions we use here is given by \eqref{LatticeCGNaction2}
  \begin{multline}\label{RC:LatticeCGNaction}
      S_{\text{CGN,W}} = a^2\sum_x\,\bigg\{ \psibar\,(\Dw + m_0)\,\psi \\
      			-\tfrac{1}{2} g^2 (O_{SS} - O_{PP}) - \tfrac{1}{2} \delta_P^2 O_{PP} - \tfrac{1}{2} g_V^2 O_{VV}  \bigg\}\,.
  \end{multline}	
	The renormalised mass vanishes if the bare mass is set to its critical value and the coupling of the 
	pseudo-scalar interaction to its symmetric value
	\begin{equation}\label{RC:ratio_w2}
  		m_0=m_c \quad \text{and} \quad \delta_P^2=\delta_{P,s}^2\,.
  \end{equation} 
  Since Lemma \ref{lemma:DiagramIdentity} holds for any $m_0$, it is also true for $m_0=m_c$. 
  In order to incorporate the expansion of $m_c$ in the expansion of $R$, two
  new terms have to be added at second order, corresponding to the derivative with respect to $am_0$ of the first order term in
  \eqref{ratioexp2} ($\partial_m = \partial/\partial am_0$)
	\begin{equation}\label{Rg1_w}
  		R_I^{(1)}(\theta) = \frac{1}{f_4^{(0)}}\; f_{4,I,\text{a}}^{(1)}\,,
  \end{equation}
	\begin{multline}\label{Rg2_w}
  		R_{IJ}^{(2)}(\theta) = \frac{1}{f_4^{(0)}} \Bigg\{ 
  								 			f_{4,IJ,\text{a}}^{(2)} + 2 f_{4,IJ,\text{b}}^{(2)} + f_{4,IJ,\text{c}}^{(2)} +  f_{4,IJ,\text{d}}^{(2)} 
  								 			+ 4 f_{4,IJ,\text{e}}^{(2)}\\
  								 		+ am_{c,I}^{(1)} \partial_m f_{4,J,\text{a}}^{(1)}
  							 - 2 \frac{ f_{2,I}^{(1)} + am_{c,I}^{(1)} \partial_m f_2^{(0)}}{f_2^{(0)}}\, f_{4,J,\text{a}}^{(1)} \Bigg\}\,,
  \end{multline}
	where all diagrams are evaluated at $m_0=0$. We do not set $\delta_P^2=\delta_{P,s}^2$ from the beginning, but keep it as a free
	parameter and only set it to its symmetric value at the end of the computation. This will allow us to compare the general result with
	computations in the discrete Gross-Neveu model (cf. Section \ref{dgn}).

	In the case of $R_I^{(1)}(\theta)$ we have to evaluate only one diagram which involves a sum over the time coordinate of the four fermion\
	interaction. This sum can be computed analytically to some extent (cf. Appendix \ref{FirstOrder}), i.e. the continuum limit can be extracted.
	Using \eqref{f41aResult} and \eqref{f40Result} 	we find
	\begin{equation}\label{Rg1Result}
  		R^{(1)}(\theta) =
  			\frac{T}{2L\, C(\theta)}\, \li\{2\,g^2 + \delta_P^2\, (A(\theta)-1) - 2\, g_V^2\, B(\theta) \re\} + \rO(a)\,.
  \end{equation}
	with
  \begin{align}
  		A(\theta) &= \frac{L}{2\theta T}\,\sinh(2\theta T/L) \stackrel{\theta\to 0}{\to} 1\\
  		B(\theta) &= \cosh(2\theta T/L) \stackrel{\theta\to 0}{\to} 1\\
  		C(\theta) &= \cosh^2(\theta T/L) \stackrel{\theta\to 0}{\to} 1\,.
  \end{align}
  
  In the case of $R_{IJ}^{(2)}(\theta)$ there are five non-reducible diagrams involving one momentum loop. The diagrams are evaluated
  numerically for a range of lattice sizes and several $\theta$ values and extract the finite
  and logarithmic divergent terms (cf. Section \ref{continuum_limit} and Appendix \ref{SecondOrder}). Using Eqs. (\ref{f42aResult}-\ref{f42eResult}) we find
	\begin{multline}\label{Rg2Result}
  		R^{(2)}(\theta) =
  			\frac{T}{2L\, C(\theta)}\, \frac{\ln(a/L)}{2\pi}\, \Big\{ -4\, g^4\,(N-B(\theta))
  					+ 2\,\delta_P^4\,(N-1)\li(A(\theta)-1\re)\\
  					  - 4\, g^2\,\delta_P^2\,\li((N-1)A(\theta)+B(\theta)-N\re) 
  					 + 8\, g_V^2\,\delta_P^2\, A(\theta) \Big\} + \dots\,,
  \end{multline}
	where the dots indicate the finite part that has been suppressed here.

\subsection{Discrete Gross-Neveu model}

	Our general four fermion interaction theory also contains the well studied Gross-Neveu model \cite{Gross:1974jv}. We call it here the discrete
	Gross-Neveu model for its discrete chiral symmetry $\psi\to\gfive\psi$, $\psibar\to-\psibar\gfive$ and in order to discriminate
	it from the chiral Gross-Neveu model.
	The discrete Gross-Neveu model has only the scalar four fermion interaction $g^2/2\,(\psibar\psi)^2$ (cf. Section \ref{dgn}). In our notation this amounts to
	setting
	\begin{equation}\label{RC:gn_settings}
  		\delta_P^2=g^2\quad \and \quad g_V^2=0\,.
  \end{equation}
	This model possesses a $\rO(2N)$ symmetry that allows no other interaction term.	 This strictly holds only for $\theta=0\,,\;\pi$. For all other values
	the boundary conditions break this symmetry. However, the local Ward identities associated with this symmetry will still hold and therefore 
	the ultra-violet divergences are expected to remain unchanged.

	The results of the last section can be used to calculate the one-loop beta-function, which then can be compared to previous results.
	
	With the settings \eqref{RC:gn_settings}, eqs. \eqref{Rg1Result} and \eqref{Rg2Result} simplify to
	\begin{equation}\label{Rg1ResultDis}
  		R^{(1)}_{\text{dgn}}(\theta) =
  			\frac{T}{2L\, C(\theta)}\,\li(A(\theta)+1+ \rO(a) \re) \, g^2 \,,
  \end{equation}
	\begin{equation}\label{Rg2ResultDis}
  		R^{(2)}_{\text{dgn}}(\theta) =
  			-\frac{T}{2L\, C(\theta)}\, \frac{\ln(a/L)}{\pi}\, (N-1)\, (A(\theta)+1) \, g^2 + \dots\,.
  \end{equation}
  A renormalised coupling can be defined by 
	\begin{equation}\label{grenGN}
			\tilde{g}_\text{dgn}^2 \equiv F_\text{dgn} = \frac{2L\, C(\theta)}{T(A(\theta)+1)}\, R_\text{dgn}(\theta)\,,
  \end{equation}
	with the expansion
	\begin{equation}\label{grenGN_exp}
			\tilde{g}_\text{dgn}^2 \eqw{a\to 0} g^2 -  g^4\,\li\{\frac{N-1}{\pi}\,\ln(a/L) - c_{\text{dgn}}\re\} + \rO(g^6)\,,
  \end{equation}
	with $c_{\text{dgn}} = c_{\text{dgn}}^{0}  + c_{\text{dgn}}^{1}\,N$.
	The corresponding beta-function \eqref{RGfunctions1} is ($\mu=1/L$)
	\begin{equation}\label{GNbeta}
  		\beta(\tilde{g}_\text{dgn}^2)=-\frac{N-1}{\pi}\,\tilde{g}_\text{dgn}^4  + \rO(\tilde{g}_\text{dgn}^6)\,.
  \end{equation}
	This result is in accordance with previous continuum calculations \cite{Gross:1974jv,Wetzel:1984nw,Bondi:1989nq,Gracey:1991vy}
	and recent lattice calculations \cite{Korzec:2006hy}.

\subsection{Chiral Gross-Neveu model}
\label{RC:cgn_w}
	
\begin{figure}
		\centering
		\includegraphics[scale=0.75]{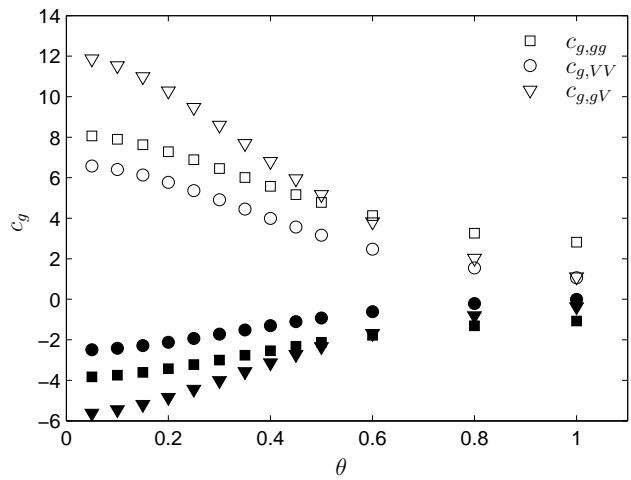}
		\caption{Dependence of the finite part of $\tilde{g}^2$ on $\theta$ at order $g^4$.}
		\label{fig:finite_cg}
	\end{figure}

\begin{figure}
		\centering
		\includegraphics[scale=0.75]{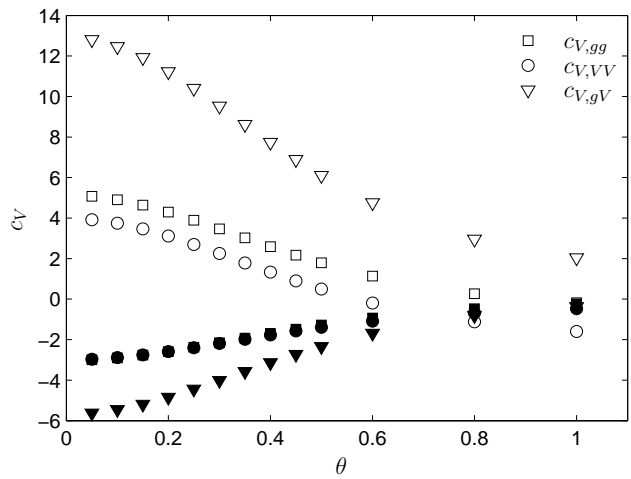}
		\caption{Dependence of the finite part of $\tilde{g}_V^2$ on $\theta$ at order $g^4$.}
		\label{fig:finite_cv}
	\end{figure}

	Now that we have confidence in our expansion, we set the pseudo-scalar coupling to its symmetric value
	$\delta_P^2=\delta_{P,s}^2$ to ensure continuous chiral symmetry
	at finite lattice spacing (up to $\rO(a)$, cf. Chapter \ref{chiral_sym_restoration})
	\begin{equation}\label{RC:deltap_sym}
		\delta_{P,s}^2 = D_3 \, \li(N\,g^4 - 2\,g^2\,g_V^2 - g_V^4  \re) + \rO(g^6)\,.
	\end{equation}
	Since the correction to zero is of order $g^4$, this enters as a finite term in $R^{(2)}(\theta)$. 
	Explicitly, in the chirally symmetric case Eqs. \eqref{Rg1Result} and
	 \eqref{Rg2Result} become
	\begin{equation}\label{Rg1ResultSym}
  		R_{\text{cgn}}^{(1)}(\theta) =
  			\frac{T}{L\, C(\theta)}\, \li\{ g^2 - g_V^2\, B(\theta) \re\} + \rO(a)\,,
  \end{equation}
	\begin{equation}\label{Rg2ResultSym}
  		R_{\text{cgn}}^{(2)}(\theta) =	-\frac{T}{L\, C(\theta)}\, \frac{\ln(a/L)}{\pi}\,(N-B(\theta))\,g^4 + \dots\,,
  \end{equation}
	where we suppressed the finite part. Thus renormalised couplings may be defined through
	\begin{align}\label{g_ren}
  		\tilde{g}^2  & \equiv F_g = \frac{L}{T(B(\theta)-1)}\,\li(B(\theta)\,R_{\text{cgn}}(0) - C(\theta)\,R_{\text{cgn}}(\theta)\re) \,,\\
			&\text{and}\nonumber\\
  		\tilde{g}_V^2  & \equiv F_V = \frac{L}{T(B(\theta)-1)}\,\li(R_{\text{cgn}}(0) - C(\theta)\,R_{\text{cgn}}(\theta)\re) \label{gV_ren}\,.
  \end{align}
  They obey the condition \eqref{RC:definition} as can be seen from their expansions
	\begin{align}\label{g_ren2}
  		\tilde{g}^2  & \eqw{a\to 0} g^2 -  g^4\,\frac{N}{\pi}\,\ln(a/L) + c_g + \rO(g^6)\,,\\
			&\text{and}\nonumber\\
  		\tilde{g}_V^2  & \eqw{a\to 0} g_V^2 - g^4\,\frac{1}{\pi}\,\ln(a/L) + c_V + \rO(g^6)\label{gV_ren2}\,,
  \end{align}
	where the finite parts at second order are given by
	\begin{align}\label{g_ren_finite}
  		c_g  & = c_{g,gg}\,g^4 +c_{g,VV}\,g_V^4 +c_{g,Vg}\,g_V^2 g^2 - \frac{D_3 D(\theta)}{2} \, \li(N\,g^4 - 2\,g^2\,g_V^2 - g_V^4  \re) \,,\\
			&\text{and}\nonumber\\
  		c_V  & =  c_{V,gg}\,g^4 +c_{V,VV}\,g_V^4 +c_{V,Vg}\,g_V^2 g^2  -\frac{D_3 D(\theta)}{2} \, \li(N\,g^4 - 2\,g^2\,g_V^2 - g_V^4  \re)
						 \label{gV_ren_finite}\,,
  \end{align}
	where $D(\theta)$ is the ratio
	\begin{equation}
  		D(\theta)=\frac{A(\theta)-1}{B(\theta)-1}\,.
  \end{equation}	 
	The coefficients $c_{I,JK} = c_{I,JK}^0 + c_{I,JK}^1\, N$ show a rather mild dependence on $\theta$. They have been determined numerically for several
	values of $\theta$ in the interval $[0,1]$. The result is shown in Fig. \ref{fig:finite_cg} and Fig. \ref{fig:finite_cv}, where open and filled symbols refer to
	$c_{I,JK}^0$ and  $c_{I,JK}^1$ respectively. The last term in \eqref{g_ren_finite} and \eqref{gV_ren_finite} originates from  $\delta_{P,s}$ $\delta_P^2$
	 \eqref{RC:deltap_sym}, which has to be included at this order.
	 
	It is straightforward to derive the associated beta-functions ($\mu=1/L$)
	\begin{align}\label{beta_g_ren2}
  		\beta_g(\tilde{g}^2)  & = -  \tilde{g}^4\,\frac{N}{\pi} + \rO(\tilde{g}^6)\,,\\
			&\text{and}\nonumber\\
  		\beta_V({\tilde{g}^2})  & = - \tilde{g}^4\,\frac{1}{\pi} + \rO(\tilde{g}^6)\label{beta_gV_ren2}\,.
  \end{align}
	Thus, from \eqref{beta_g_ren2} we see that the coupling $g^2$ is asymptotically free and from \eqref{gV_ren2} it is obvious that 
	the coupling $g_V^2$ receives  an additive renormalisation. The one-loop beta-functions derived here agree with the ones derived in the
	$\overline{\text{MS}}$ scheme in Ref. \cite{Bondi:1989nq}.

	As we have seen in Section \ref{CGN:lattice_cgn} there is some freedom in the choice of four fermion interactions due to the identities
	\eqref{dep3}-\eqref{dep5}.
	In particular, it is possible to find a combination of terms were one of the two beta-functions vanishes. This is also indicated by the formal continuum 
	argumentation in \cite{Furuya:1982fh,Moreno:1987np}. Indeed if we use \eqref{dep5} to rewrite \eqref{RC:LatticeCGNaction} as in \eqref{LatticeCGNaction4},
	the new coupling $\delta_V^2$ is related to the original ones by \eqref{couplings_relation3}
  \begin{equation}\label{RC:couplings_relation}
  		\delta_V^2 = g_V^2 - g^2/N\,.
  \end{equation}	 
	Then \eqref{g_ren} remains unchanged. But the renormalised coupling associated with $\delta_V^2$ is given by
	\begin{equation}
  		\tilde{\delta}_V^2  = \tilde{g}_V^2 - \tilde{g}^2/N = \delta_V^2 + c_V - c_g/N + \rO(g^6)\label{deltaV_ren}\,.
  \end{equation}
	There is no logarithmic divergence and therefore the corresponding beta-function vanishes at this order
	\begin{equation}
  		\beta_\delta(\tilde{\delta}_V^2)  = \rO(\tilde{g}^6)\label{beta_deltaV_ren}\,.
  \end{equation}

\section{Ginsparg-Wilson fermions}
\label{RC:gw}

		The free modified Neuberger Dirac operator $\Dgw$ defined in Section \ref{SF:mod_neuberger_op} obeys
		\begin{equation}\label{RC:gw_realtion}
				\gfive\,\Dgw + \Dgw\,\gfive = a\,\Dgw\gfive \Dgw + \Delta_B\,,
		\end{equation}
	  The parameters $c$ and $s$ are set to 1 and 0, respectively, in the following. The term
		$\Delta_B$ is supported in the vicinity of the boundaries and decays exponentionally with the distance to them (cf. Fig \ref{fig:DeltaB}).  
		In particular, the rate of the decay is constant if the distance is measured in lattice units. 
		Then transformation \eqref{CGN:gw_chiral_symmetry} is a symmetry of the 
		action and the associated Ward identities are expected to hold in the interior (well separated from the boundaries) of lattice with small corrections which
		vanish in the continuum limit. Therefore we can use an action with two four fermion interactions (cf. Section \ref{CGN:gw_fermions})
    \begin{equation}\label{RC:LatticeCGNaction_gw}
        S_{\text{CGN,GW}} = a^2\sum_x\,\li\{ \psibar\,\Dgw\,\psi 
        			-\tfrac{1}{2} g^2 (\Ohss - \Ohpp) - \tfrac{1}{2} g_V^2 \Ohvv \re\}\,,
    \end{equation}
		to compute the correlation functions \eqref{2fermions} and \eqref{4fermions} in the Schr\"odinger functional with Ginsparg-Wilson fermions. 
		The operators $\hat{O}_{I}$ differ from the operators $O_I$ by the substition $\psi\to\hat{\psi}=(1-\tfrac{a}{2}D)\psi$.
		
		The free propagator entering in the perturbative expansion is computed numerically as indicated in Section \ref{SF:mod_neuberger_op_free} for
		lattice sizes $L/a=4\,,5\,,\dots\,,48$ and $\theta=0\,,0.1\,,0.5\,,1$.
		
\subsection{Discrete Gross-Neveu model}
		As in the case of Wilson fermions we check if the beta-function of the discrete Gross-Neveu model is correctly reproduced. The renormalised 
		coupling was defined in \eqref{grenGN}. At finite lattice spacing one expects 
		\begin{equation}\label{grenGNgw}
					\tilde{g}_\text{dgn}^2 = g^2\, k_{\text{dgn}}(a/L) -  g^4\,\li\{\frac{N-1}{\pi}\,\ln(a/L) - c_{\text{dgn}}(a/L) \re\} + \rO(g^6)\,,
 		\end{equation}
		where $c_{\text{dgn}}(a/L) = c_{\text{dgn}}^{0}(a/L) + c_{\text{dgn}}^{1}(a/L)\, N$.
	\begin{figure}
		\centering
		\includegraphics[scale=0.75]{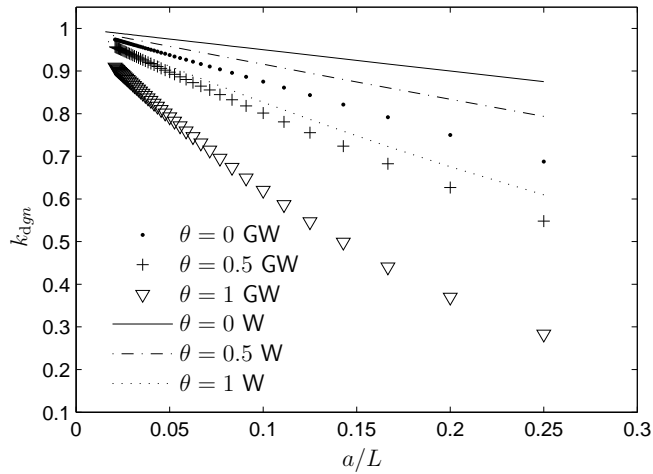}
		\caption{Cut-off dependence of the coefficient of the leading order term in the expansion of the renormalised coupling of the discrete Gross-Neveu model.
		 It is unity in the continuum limit, as it should be. For comparison we plot Ginsparg-Wilson and Wilson fermions.}
		\label{fig:leading_coeff_gdgn}
	\end{figure}
	\begin{figure}
		\centering
		\includegraphics[scale=0.75]{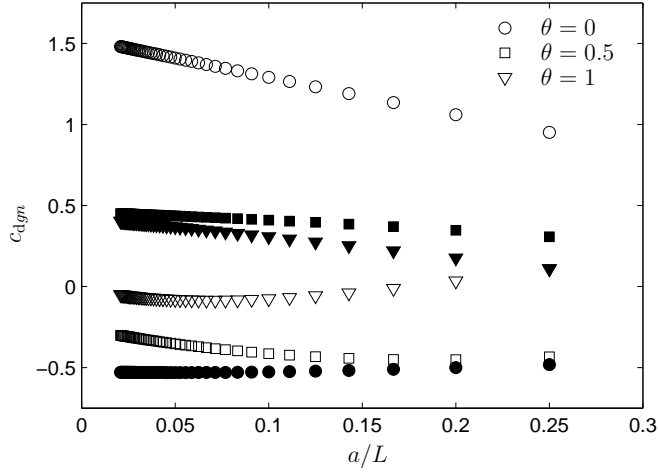}
		\caption{Subtracting the expected logarithmic divergence (as indicated in \eqref{grenGNgw}) from the second order diagrams one obtains the finite part
		 of $\tilde{g}_\text{dgn}^2$ at order $g^4$. We plot here the result for three different values of $\theta$. Open and filled symbols refer to
		 $c_{\text{dgn}}^{0}(a/L)$ and  $c_{\text{dgn}}^{1}(a/L)$ respectively.}
		\label{fig:finite_gdgn}
	\end{figure}	
	The coefficient of the leading order term $k_{\text{dgn}}(a/L)$	is expected to be unity in the continuum limit. In Fig. \ref{fig:leading_coeff_gdgn}
  its lattice spacing dependence is plotted for $\theta=0\,,0.1\,,0.5\,,1$. In any case it has the right continuum limit (the systematic error of the
	extrapolation is $\rO(10^{-5})$). In Fig \ref{fig:finite_gdgn} the finite part at second order is plotted for three $\theta$-values. These numbers are obtained
	by subtracting the expected logarithmic divergence from the second order diagrams (we also determined the coefficient of the logarithmic divergence
	directly with results compatible with $-(N-1)/\pi$ and a systematic error of $\rO(10^{-5})$). Note that the cut-off effects of the Ginsparg-Wilson
	fermions are roughly  twice as large as for the Wilson fermions.
	
	Thus the coupling of the lattice theory with the modified Neuberger operator \eqref{sf_neuberger} can be renormalised in the same way as with Wilson
	fermions. We now proceed with the chiral Gross-Neveu model.
	
\subsection{Chiral Gross-Neveu model}
	
	The renormalised couplings for the two interaction terms are defined in the same way as in the Wilson case, i.e. by implementing
	the $\theta$-dependence of the leading order term in the expansion of $R(\theta)$
	\begin{align}\label{g_ren_gw}
  		\tilde{g}^2  & \equiv F_g = \frac{L}{T(B(\theta)-1)}\,\li(B(\theta)\,R_{\text{cgn,gw}}(0) - C(\theta)\,R_{\text{cgn,gw}}(\theta)\re) \,,\\
			&\text{and}\nonumber\\
  		\tilde{g}_V  & \equiv F_V = \frac{L}{T(B(\theta)-1)}\,\li(R_{\text{cgn,gw}}(0) - C(\theta)\,R_{\text{cgn,gw}}(\theta)\re) \label{gV_ren_gw}\,.
  \end{align}
	The difference is that no additive mass renormalisation	is needed ($am_0=0$) and that the pseudo-scalar coupling vanishes exactly ($\delta_P^2=0$).
	
\begin{figure}
		\centering
		\includegraphics[scale=0.75]{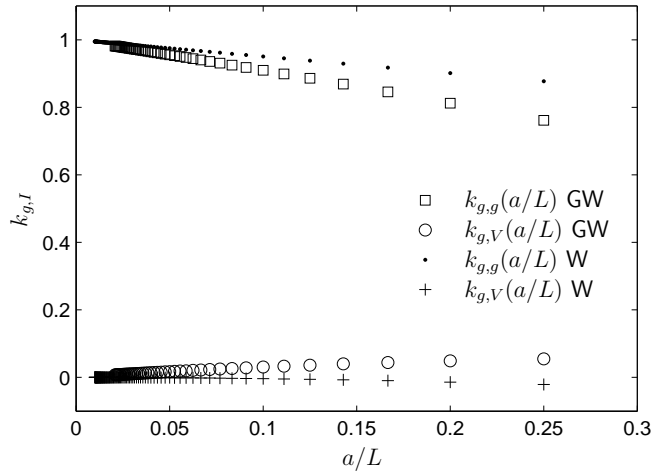}
		\caption{Cut-off dependence of the coefficients of the leading order terms in the expansion of the renormalised coupling $\tilde{g}^2$ of the chiral Gross-Neveu
		 model for $\theta=0.5$. For comparison we plot Ginsparg-Wilson and Wilson fermions.}
		\label{fig:leading_coeff_g}
	\end{figure}

\begin{figure}
		\centering
		\includegraphics[scale=0.75]{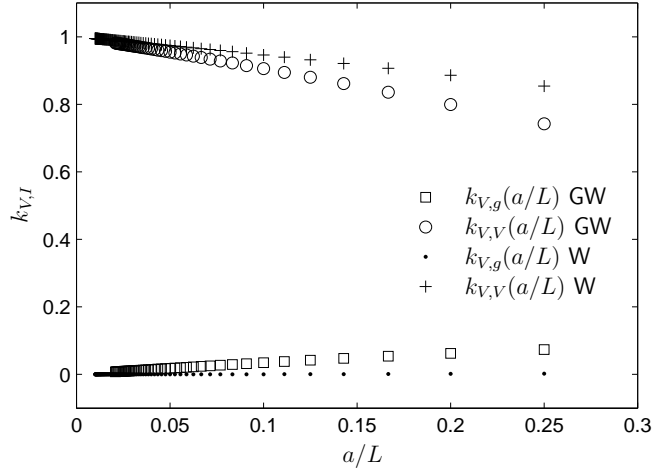}
		\caption{Cut-off dependence of the coefficients of the leading order terms in the expansion of the renormalised coupling $\tilde{g}_V^2$ of the chiral
		Gross-Neveu model for $\theta=0.5$. For comparison we plot Ginsparg-Wilson and Wilson fermions.}
		\label{fig:leading_coeff_gv}
	\end{figure}

	Since in the case of the modified Neuberger operator we have no analytic handle on the leading order, we first check whether the definitions
	above satisfy \eqref{RC:definition}.  To this end we expand
	\begin{align}\label{g_ren_gw2}
  		\tilde{g}^2  & =  g^2\, k_{g,g}(a/L) + g_V^2\,k_{g,V}(a/L) + \rO(g^4)\,,\\
			&\text{and}\nonumber\\
  		\tilde{g}_V^2  & = g^2\, k_{V,g}(a/L) + g_V^2\,k_{V,V}(a/L)  + \rO(g^4)\label{gV_ren_gw2}\,.
  \end{align}
	The coefficients were computed for lattice sizes $L/a=4\,,5\,,\dots\,, 48$. Fig. \ref{fig:leading_coeff_g} shows the coefficients of $\tilde{g}^2$ and
	Fig. \ref{fig:leading_coeff_gv} the coefficients of $\tilde{g}_V^2$. Their extrapolations to the continuum limit have the expected values within the
	systematic errors. For example, in the case $\theta=0.5$ the result is
	\begin{align}\label{cg_gw}
  		k_{g,g} & = 0.999995(17)\quad k_{g,V} = 0.000012(27)\,,\\
			&\text{and}\nonumber\\
  		k_{V,g} & = -0.000005(18)\quad k_{V,V} = 1.000011(28)\label{cV_gw}\,.
  \end{align}
	The results are similar for $\theta=0.1\,,1$. Again the cut-off effects of Ginsparg-Wilson fermions exceed the ones of Wilson fermions.
	
\begin{figure}
		\centering
		\includegraphics[scale=0.75]{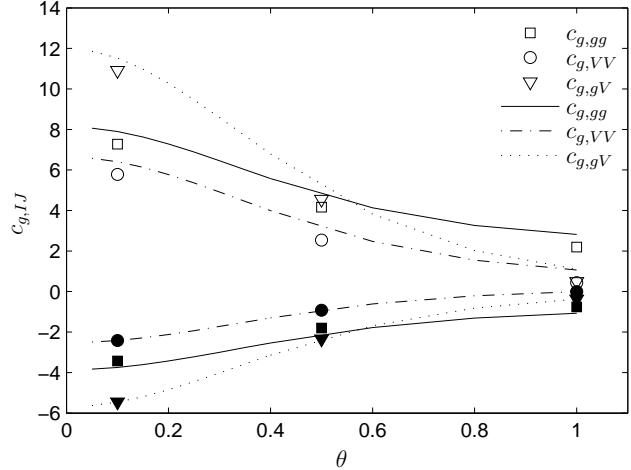}
		\caption{Dependence of the continuum extrapolation of the finite part of $\tilde{g}^2$ on $\theta$ at order $g^4$. The symbols are Ginsparg-Wilson and
		 curves are Wilson fermions. Open and filled symbols refer to $c_{I,JK}^{0}$ and  $c_{I,JK}^{1}$ respectively.}
		\label{fig:finite_cg_gw}
	\end{figure}

\begin{figure}
		\centering
		\includegraphics[scale=0.75]{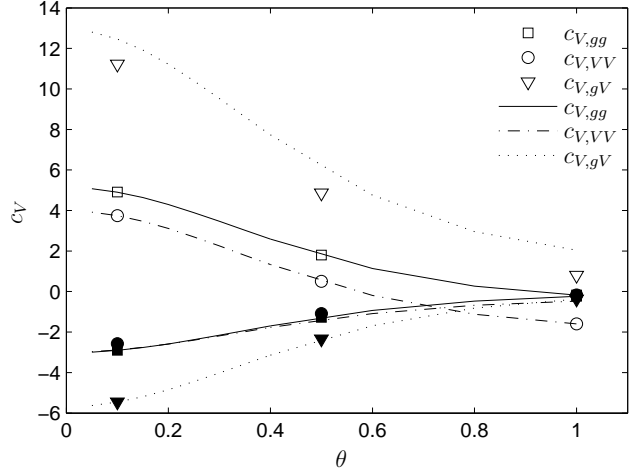}
		\caption{Dependence of the continuum extrapolation of the finite part of $\tilde{g}_V^2$ on $\theta$ at order $g^4$. The symbols are Ginsparg-Wilson and
		curves are Wilson fermions. Open and filled symbols refer to $c_{I,JK}^{0}$ and  $c_{I,JK}^{1}$ respectively.}
		\label{fig:finite_cv_gw}
	\end{figure}

	At next to leading order we expect to find the logarithmic divergences with coefficients $-N/\pi$ and $-1/\pi$, respectively, for $\tilde{g}^2$  and $\tilde{g}_V^2$.
	As in the case of the discrete Gross-Neveu model we first determined the coefficient of the divergence explicitly. After we were convinced that it has the right 
	value, we subtracted it from the sum of second order diagrams. The resulting amplitudes should have a finite or vanishing continuum limit. Expanding the
	renormalised couplings to second order
	\begin{align}\label{g_ren_gw3}
  		\tilde{g}^2  & =  g^2\, k_{g,g}(a/L) + g_V^2\,k_{g,V}(a/L)  -  g^4\,\frac{N}{\pi}\,\ln(a/L) + c_g(a/L) + \rO(g^6)\,,\\
			&\text{and}\nonumber\\
  		\tilde{g}_V^2  & = g^2\, k_{V,g}(a/L) + g_V^2\,k_{V,V}(a/L) - g^4\,\frac{1}{\pi}\,\ln(a/L) + c_V(a/L) + \rO(g^6)\label{gV_ren_gw3}\,,
  \end{align}
  these amplitudes are $c_g(a/L) $ and $c_V(a/L)$. They can be sorted by powers of the bare couplings
	\begin{align}\label{g_ren_finite_gw}
  		c_g  & = c_{g,gg}\,g^4 +c_{g,VV}\,g_V^4 +c_{g,Vg}\,g_V^2 g^2 \,,\\
			&\text{and}\nonumber\\
  		c_V  & =  c_{V,gg}\,g^4 +c_{V,VV}\,g_V^4 +c_{V,Vg}\,g_V^2 g^2   \label{gV_ren_finite_gw}\,,
  \end{align}
	and powers of $N$
	\begin{equation}
			c_{I,JK} = c_{I,JK}^0 + c_{I,JK}^1\, N\,.
	\end{equation} 
	In the continuum limit we do not expect to find the same values for these coefficients as in the Wilson case, because the finite part of the renormalised
	coupling is 	scheme dependent. However, as indicated above they should have a finite or vanishing continuum limit. Any uncancelled divergence would mean
	that additional terms are needed in the action to absorb them.
	\begin{figure}
		\centering
		\includegraphics[scale=0.75]{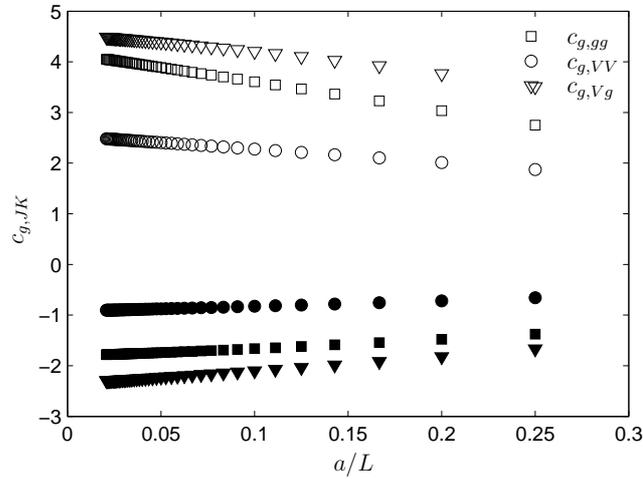}
		\caption{Cut-off dependence of the coefficients of the second order terms in the expansion of the renormalised coupling $\tilde{g}^2$
		of the chiral Gross-Neveu model 	for $\theta=0.1$. Open and filled symbols refer to $c_{I,JK}^{0}(a/L)$ and  $c_{I,JK}^{1}(a/L)$ respectively. }
		\label{fig:second_coeff_g}
	\end{figure}
	\begin{figure}
		\centering
		\includegraphics[scale=0.75]{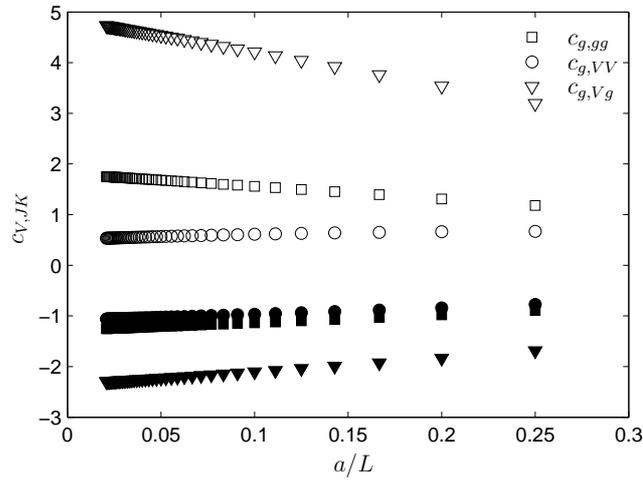}
		\caption{Cut-off dependence of the coefficients of the second order terms in the expansion of the renormalised coupling $\tilde{g}_V^2$ of the
		 chiral Gross-Neveu model 	for $\theta=0.1$. Open and filled symbols refer to $c_{I,JK}^{0}(a/L)$ and  $c_{I,JK}^{1}(a/L)$ respectively.}
		\label{fig:second_coeff_gv}
	\end{figure}	
  But as can be seen from Figs. \ref{fig:second_coeff_g} and \ref{fig:second_coeff_gv}
	there is no such divergence, all coefficients
	have a well defined continuum limit. The dependence of the values of the extrapolation on $\theta$
	is depicted in Figs. \ref{fig:finite_cg_gw}
	and \ref{fig:finite_cv_gw}. Although the amplitudes are different, the functional dependence is, as it should be, identical to the Wilson case.
	
	Since we find the same divergences the beta-functions associated with the renormalised couplings are the same as in the Wilson case. This, of course,
	comes as no suprise.
	If the theory with the modified Neuberger operator in the action is renormalisable and describes the same continuum theory,
	all coefficients of the beta-function must be equal since it is universal as a whole.

\section{Summary and ratio of Lambda parameters}
\label{RC:Lambda}

	In the last two Sections we calculated the renormalised couplings and beta-functions of the discrete and chiral Gross-Neveu model in one-loop perturbation theory
	for Wilson and Ginsparg-Wilson  fermions in the Schr\"odinger functional. Both models have a asymptotically free coupling. In the chiral Gross-Neveu model
	there is also a coupling that does not renormalise, i.e. its beta-function vanishes. The first coefficient of the beta-function is universal and we correctly
	reproduce the known values
	\begin{equation}\label{b0}
  		\text{DGN:}\quad b_0=-\frac{N-1}{\pi} \quad \quad \text{CGN:}\quad   b_0=-\frac{N}{\pi}\,.
  \end{equation}
	A nontrivial test of the equality of the theory with Wilson and with Ginsparg-Wilson fermions is now the computation of the ratio of the lattice
	$\Lambda$-pa\-ram\-e\-ters to show its independence of the angle $\theta$.
	
	\paragraph{Discrete Gross-Neveu model:}
	The continuum limit of the renormalised coupling of the discrete Gross-Neveu model calculated with Wilson fermions 
	\begin{equation}\label{grenGN_w}
			\tilde{g}^2 \eqw{a\to 0} g_\tw^2 + g_\tw^4\,\li\{b_0\,\ln(a/L) + c^\tw\re\} + \rO(g_\tw^6)\,,
  \end{equation}
	and with Ginsparg-Wilson fermions
	\begin{equation}\label{grenGN_gw}
			\tilde{g}^2 \eqw{a\to 0} g_\tgw^2 +  g_\tgw^4\,\li\{b_0\,\ln(a/L) + c^\tgw\re\} + \rO(g_\tgw^6)\,,
  \end{equation}
	must be equal. This allows to relate the bare couplings, e.g.
	\begin{equation}\label{gbareGN}
			g_\tgw^2 = g_\tw^2 + a_1\,g_\tw^4 + \rO(g_\tw^6)\,,\quad a_1 = c^\tw - c^\tgw\,.
  \end{equation}
	The finite parts $c_\tw^{\text{dgn}}$ and $c_\tgw^{\text{dgn}}$ are functions of $\theta$. Since it is a probe to the theory like external momenta,
	the relation between the bare parameters \eqref{gbareGN} can not depend on it. Therefore the difference $a_1$ has to be independent of $\theta$.
	This is equal to saying that the ratio of the lattice $\Lambda$-parameters
	\begin{equation}\label{ratioLambda}
  		\Lambda_{\text{LAT, gw}}/\Lambda_{\text{LAT, w}}  = \exp\li\{ \frac{a_1}{2b_0}\re\}\,.
  \end{equation}
	has to be independent of $\theta$. The lattice $\Lambda$-parameter is analogous to the $\Lambda$-parameter in \eqref{Lambda}, but with
	the renormalised coupling replaced with the bare coupling and the beta-function replaced with the lattice beta-function.
	 For the ratio of the lattice $\Lambda$-parameters in the discrete Gross-Neveu model at $\theta=0,0.5$ and $N=2,4$ we find
	\begin{align}
  		\Lambda_{\text{LAT, gw}}/\Lambda_{\text{LAT, w}}\Bigg|_{N=2,\;\theta=0\; [0.5]}  &= 0.987(7)\; [0.986(12)]\,, \\
			\Lambda_{\text{LAT, gw}}/\Lambda_{\text{LAT, w}}\Bigg|_{N=4,\;\theta=0\; [0.5]}  & = 0.712(12)\; [0.713(21)]\,.
  \end{align}

	\paragraph{Chiral Gross-Neveu model:}	 
	The chiral Gross-Neveu model has two coupling constants and the result for the renormalised couplings 
  can be given in the following form for Wilson fermions
	\begin{align}
  		\tilde{g}^2  & \eqw{a\to 0} g_\tw^2 -  g_\tw^4\,\frac{N}{\pi}\,\ln(a/L) + c^\tw_g + \rO(g_\tw^6)\,,\\
	  	\tilde{\delta}_V^2 & \eqw{a\to 0} \delta_{V,\tw}^2 + c^\tw_V - c^\tw_g/N + \rO(g_\tw^6)\,,	
	\end{align}
	and Ginsparg-Wislon fermions
	\begin{align}
  		\tilde{g}^2  & \eqw{a\to 0} g_\tgw^2 -  g_\tgw^4\,\frac{N}{\pi}\,\ln(a/L) + c^\tgw_g + \rO(g_\tgw^6)\,,\\
	  	\tilde{\delta}_V^2 & \eqw{a\to 0} \delta_{V,\tgw}^2 + c^\tgw_V - c^\tgw_g/N + \rO(g_\tgw^6)\,.		
	\end{align}
	These are couplings of \eqref{CGNaction2}. The finite parts are composed of three terms
	\begin{align}
  		c^i_g & = c^i_{g,gg}\,g_i^4 +c^i_{g,VV}\,g_{V,i}^4 +c^i_{g,Vg}\,g_{V,i}^2 g_i^2 \,,\\
  		c^i_V  & =  c^i_{V,gg}\,g_i^4 +c^i_{V,VV}\,g_{V,i}^4 +c^i_{V,Vg}\,g_{V,i}^2 g_i^2   \,,
	\end{align}
	for $i=\tw,\tgw$ (no sum).	The renormalised quantities must coincide.  This yields relations between the bare couplings
	\begin{align}
			g_\tgw^2 & = g_\tw^2 + a_g + \rO(g_\tw^6)\,,\quad a_g = c^\tw_g - c^\tgw_g\,,\\
			\delta_{V,\tgw}^2 & = \delta_{V,\tw}^2 + a_V + \rO(\delta_{V,\tw}^6)\,,\quad a_V = c^\tw_V - c^\tgw_V - a_g/N \label{bareWtoGW}\,.
	\end{align}
	Again the coefficients $a_g$ and $a_V$ have to be independent of $\theta$. That this is the case can be seen in Fig. \ref{fig:finite_cg_gw}
  and \ref{fig:finite_cv_gw}. There $a_g$ and $a_V$ are the differences between the curves and the corresponding symbols. It is finite, but constant.

  The beta-function of the vector-vector coupling $\tilde{\delta}_V$
  vanishes at one-loop \eqref{beta_deltaV_ren}.	This is in agreement with two-loop calculations \cite{Bondi:1989nq} and formal continuum arguments
	 \cite{Furuya:1982fh,Moreno:1987np}. 
	Furthermore the vector-vector coupling was not constrained by the restoration of chiral symmetry in the case of Wilson fermions (see Chapter
	\ref{chiral_sym_restoration}). Therefore no value of the bare coupling $\delta_V$ is distinguished. For each fixed value the theory is
	an one coupling theory. The ratio of the lattice $\Lambda$-parameters can then be calculated as in the discrete Gross-Neveu model
  (Eqs. \eqref{gbareGN} and \eqref{ratioLambda}).
	For convenience we take $\delta_{V,\tw}^2=0$. From \eqref{bareWtoGW} it follows that $\delta_{V,\tgw}^2=\rO(g_\tw^4)$ and the 
	coefficient $a_g$ simplifies to	
	\begin{equation}
  		a_g = a_1\,g_\tw^4\quad a_1=c^\tw_{g,gg} - c^\tgw_{g,gg}\,.
  \end{equation}	 
  We list the ratio for $\theta=0.1,0.5$ 
	\begin{align}
  		\Lambda_{\text{LAT,gw}}/\Lambda_{\text{LAT,w}}\Bigg|_{N=2,\theta=0.1\; [0.5]}  &= 0.9893(7)\; [0.9892(10)]\,, \\
  		\Lambda_{\text{LAT,gw}}/\Lambda_{\text{LAT,w}}\Bigg|_{N=4,\theta=0.1\; [0.5]}  &= 0.776(1)\;[0.776(2)]\,.
  \end{align}	 
	 The ratio of the $\Lambda$-parameters has been calculated in QCD on a  lattice with periodic boundary conditions in \cite{Alexandrou:1999wr}.
	  The values there are similar  to the ones obtained here.

\chapter{Conclusions}
\label{conclusions}

The main result of this work is the formulation of the Schr\"odinger functional (SF) for fermionic models of the Gross-Neveu type with
a finite number $N$ of fermion flavours.
In 1-loop lattice perturbation theory we showed that the theory is renormalisable with Wilson and Ginsparg-Wilson fermions. This is the first
check of the recently proposed Dirac operator \cite{Luscher:2006df} beyond the free theory.

In two dimensions four fermion interactions have dimensionless coupling constants. We discussed the symmetry properties
of the four fermion interaction terms and the relations among them. Due to its Abelian chiral symmetry the continuum chiral Gross-Neveu
(CGN) model has two independent interaction terms. A possible choice is $g^2/2\, ((\psibar\psi)^2 - (\psibar\gfive\psi)^2) + g_V^2/2\, (\psibar\gmu\psi)^2$.
Because Wilson fermions explicitly break chiral symmetry, the most general lattice action
 for the chiral Gross-Neveu model
has an independent four fermion interaction that breaks chiral invariance, for example $\delta_P^2/2\,(\psibar\gfive\psi)$, and a mass term. 
The model with three couplings becomes the discrete Gross-Neveu model (DGN), when two of the couplings are set appropriately.

Like in QCD chiral Ward identities can be used to define the critical mass. The Ward identity of the local axial current, strictly valid only in the continuum,
is demanded to hold on the lattice up to corrections of $\rO(a)$. The critical mass cancels the linear divergence that appears in the Ward identity because
of  operator mixing. However, for the restoration of chiral symmetry it is also necessary to tune the symmetry breaking coupling to its symmetry restoring value.
We computed the critical mass and the symmetric coupling up to second order in lattice perturbation theory. 	We find no constraint on the vector-vector
coupling $g_V^2$. Or phrased in another way, the vector-vector coupling has not to be tuned 	in order to restore chiral symmetry.
This is the first determination of this parameters at finite-$N$. The result is consistent with calculations in the large-$N$ limit.

Renormalised couplings for $g^2$ and $g_V^2$ are defined at vanishing renormalised mass. The natural choice are boundary-to-boundary correlation functions
with four external fermions. For Wilson fermions the non-trivial dependence of the first order diagrams on the phase $\theta$, which parametrises the spatial boundary conditions, could be
calculated analytically. This was crucial in order to define the renormalised couplings in such a way, that they are equal to 
the corresponding bare coupling at leading order perturbation theory. This definition has then been applied in the calculation with Ginsparg-Wilson fermions,
where one has no analytic handle on the diagrams.

The coefficient of the logarithmic divergence and the finite part at 1-loop is
computed numerically to a high precision using the method described in \cite{Bode:1999sm}. We correctly reproduced the first coefficient of 
the beta-function for the CGN and DGN. Both models have a asymptotically free coupling. Furthermore the couplings can be redefined such, that the vector-vector
coupling does not renormalise, i.e. its beta-function vanishes. This is consistent with 2-loop calculations and formal continuum arguments.
The finite part shows a mild dependence on $\theta$. The cut-off effects are clearly $\rO(a)$. 

The definition of the renormalised couplings is suitable for a computation in Monte Carlo simulations. The boundary-to-boundary correlation functions
are easily implemented and the $\theta$ dependence is strong enough to discriminate the couplings. Since the SF is a finite size regularisation scheme
the spatial extension provides a natural scale in the system and it is possible to define step scaling functions \cite{Luscher:1991wu} for the renormalised couplings. 
The renormalisation group invariant step scaling functions can then be used as benchmark observables for universality studies of different 
lattice actions.

We used the results of the computation with Wilson fermions to study a recently proposed modified Neuberger-Dirac operator \cite{Luscher:2006df}
 in 1-loop lattice perturbation theory. The operator is compatible with  the SF boundary conditions in the sense of Section \ref{sf_universality}.
The operator of the free theory is shown to be local (but not ultralocal)
 and to obey the Ginsparg-Wilson relation up to terms localised at the boundaries with exponentially decreasing
tails. Thus the lattice chiral symmetry associated with Ginsparg-Wilson fermions is a symmetry in the interior of the lattice and correlations of local
fields at physical distances from the boundaries obey the same Ward identities as they do on periodic lattices. 
After the substitution $\psi\to(1-\tfrac{a}{2}D)\psi$ the new four fermion operators transform under the lattice chiral symmetry in the same way as the old ones
under the continuum symmetry. Therefore $\delta_P^2=0$ is protected by the symmetry and no tuning is necessary. 

The modified Neuberger-Dirac operator \eqref{sf_neuberger} is rather complicated. The eigenfunctions of the operator under the square root were derived. We did
not succeed in deriving an analytic expression for the Dirac operator. But the Dirac operator and its inverse
can be computed numerically using standard techniques \cite{Giusti:2002sm}.

The computation of the previously  defined renormalised couplings was repeated with Ginsparg-Wilson fermions. For the coefficients of the logarithmic 
divergence we found perfect agreement with the Wilson result, thus proving the SF with Ginsparg-Wilson fermions to be renormalisable at the one loop level. As 
expected, the regularisation dependent finite part differs.
 The cut-off effects are $\rO(a)$ and larger than in the Wilson case. To achieve $\rO(a)$-improvement one would have to redefine the correlation 
 functions \cite{Capitani:1999uz} and to introduce four fermion interaction terms at the boundaries. 
 
 Finally we used the 1-loop calculation of the renormalised coupling with Wilson and Ginsparg-Wilson fermions to compute the ratio of corresponding
 $\Lambda$-parameters, which yields reasonable results.
 
 With this work the Schr\"odinger functional for Gross-Neveu models is well established. It can be used as a benchmark system for fermion actions.
 It should be possible to simulate with Wilson as well as Ginsparg-Wilson fermions. 



\appendix




\chapter{Notation}

\section{Definitions}\label{definitions}

\subsection{Dirac matrices}\label{gamma}
	The $\gamma$-matrices are defined through the Clifford algebra in Euclidean space
	\[\{\gmu^E,\gnu^E\}=2\delta_{\mu\nu}\,.\]
	Since the Pauli matrices
	\[\sigma_1= \left(\begin{array}{cc} 0&1\\1&0 \end{array}\right)\,,\quad
	  \sigma_2= \left(\begin{array}{cc} 0&-i\\i&0 \end{array}\right)\,,\quad
	  \sigma_3= \left(\begin{array}{cc} 1&0\\0&-1 \end{array}\right)\,,
	\]
	already have the right dimension and anticommutation properties,
	they can directly used to represent the Euclidean $\gamma$-matrices in $D=2$ dimensions. One possible choice is
	\[\go = \sigma_2\,,\]
	\[\gamma_1 = \sigma_1\,.\]
	and for hermitian $\gfive$
	\[\gfive \equiv i \go \gl = \sigma_3\,.\]
	In this representation the chiral projectors are diagonal. They are defined
	\begin{equation}\label{LRprojectors}
	    P_{R,L} = \frac{1}{2}(1\pm\gfive)\,,
	\end{equation}
	with the properties
	\begin{equation}
	    P_{R,L}^2 = P_{R,L}\,,\quad P_L P_R = P_R P_L = 0\,,
	\end{equation}
	\begin{equation}
	    \gfive P_{R,L} = \pm P_{R,L}\,,
	\end{equation}
	\begin{equation}
	    P_R + P_L = 1\,,\quad P_R - P_L = \gfive\,,
	\end{equation}
	
	We also need the projectors defined using $\go$
	\begin{equation}\label{PMprojectors}
	    P_{\pm} = \frac{1}{2}(1\pm\go)\,.
	\end{equation}
	These projectors become especially simple in another representation
	\[\go = \sigma_3\,,\]
	\[\gamma_1 = \sigma_1\,.\]
	\[\gfive \equiv i \go \gl = -\sigma_2\,.\]
	Then
	\begin{equation}
	    P_{+} = \left(\begin{array}{cc} 1&0\\0&0 \end{array}\right)\,,\quad P_{-} = \left(\begin{array}{cc} 0&0\\0&-1 \end{array}\right) \,.
	\end{equation}

\subsection{Generators of SU(N)}\label{ap:SUNGenerators}
	The Lie algebra $\mathrm{su}(N)$ of $SU(N)$ can be identified with the space of complex $N\times N$ matrices $M$ satisfying
	\begin{equation}
		M^\dagger=M\,, \quad \Tr[M]=0\,.
	\end{equation}	
	The generators of $SU(N)$ can be identified with a basis $\lambda^a$, $a=1,2,\dots,N^2-1$ in this space normalised to
	\begin{equation}
		\Tr[\lambda^a\, \lambda^b]=C\delta_{ab}\,, \quad C=2\,.
	\end{equation}	
	They obey commutation relation (repeated indices are summed over)
	\begin{equation}\label{comrelation}
		[\lambda^a,\lambda^b]=2 i f^{abc} \lambda^c\,,
	\end{equation}
	with the totally antisymmetric structure constants $f^{abc}$.
	
	Now every complex $N\times N$ matrix $X$ can be expanded in the complete basis
	\begin{equation}\label{ap:FlavourExp}
		\lambda^{A}=\li\{\sqrt{\tfrac{2}{N}}\,1,\lambda^a\re\}\,,\quad A=0,1,\dots,N^2-1\,.
	\end{equation}
	The expansion reads
	\begin{equation}
		X = \tfrac{1}{2}\sum_A \lambda^{A}\, \Tr[\lambda^{A}\,X]\,.
	\end{equation}
	Expanding in such a way the matrix $X^{(lk)}_{ij}=2\delta_{ik}\delta_{jl}$ one finds the identity
	\begin{equation}
		(\lambda^a)_{ij}(\lambda^a)_{lk} = 2\delta_{ik}\delta_{jl} - \tfrac{2}{N}\delta_{ij}\delta_{lk}\,.
	\end{equation}
	Thus the quadratic Casimir operator is
	\begin{equation}
		(\lambda^a\lambda^a)_{ij} = C_2\delta_{ij}\,, \quad C_2=\frac{2(N^2-1)}{N} \,.
	\end{equation}
	
	The following identities are usefull when evaluating flavour traces of Feynman diagrams
	\begin{align}
		\Tr[\lambda^a\lambda^a] & = N\,C_2\,,\\
		\Tr[\lambda^a\lambda^a\lambda^b\lambda^b] & = N\,(C_2)^2\,,\\
		(\lambda^a\lambda^b\lambda^a)_{ij} & = (C_2-2N)(\lambda^b)_{ij}\,,\\
		\Tr[\lambda^a\lambda^b\lambda^a\lambda^b] & = N\,C_2\, (C_2-2N)\,,\\
		\Tr[\lambda^a\lambda^b]\, \Tr[\lambda^a\lambda^b] & = C\,N\,C_2\,.	
	\end{align}

\subsection{Lattice notation}\label{ap:LatticeNotation}
	Here we define the lattice difference operators. Via the factor $\lambda_\mu$ we are able to define the theory with 
	general boundary conditions. (For details see \cite{Luscher:1996vw}.)
	\begin{eqnarray}
	    \dmu \psi(x) & = & \tfrac{1}{a}[\lambda_\mu\psi(x+a\hat{\mu})-\psi(x)]\label{BoundaryCondition}\\
	    \partial^*_\mu \psi(x) & = & \tfrac{1}{a}[\psi(x) - \lambda^{-1}_\mu\psi(x-a\hat{\mu})]\\
	    \lambda_\mu & = & \e^{ia\theta_\mu/L}\,, \quad \theta_0=0\,, \quad -\pi< \theta_1 \le \pi\,,
	\end{eqnarray}
	\begin{eqnarray}
	    p^{\scriptscriptstyle \pm}_\mu & = & p_\mu \pm \theta_\mu/L\\
	    \po_\mu & = & \tfrac{1}{a}\sin(ap_\mu)\\
	    \hat{p}_\mu & = & \tfrac{2}{a}\sin(ap_\mu/2)\,,
	\end{eqnarray}
	
	The left action of difference operators is defined as
	\begin{eqnarray}
	    \psibar(x)\moverset{\leftarrow}{\dmu} & = & \tfrac{1}{a}[\psibar(x+a\hat{\mu})\lambda^{-1}_\mu-\psibar(x)]\\
	    \psibar(x)\moverset{\leftarrow}{\partial}^*_\mu  & = & \tfrac{1}{a}[\psibar(x) - \psibar(x-a\hat{\mu})\lambda_\mu]\,.
	\end{eqnarray}
	They are related to the right difference operators
	\begin{eqnarray}
	    \sum_x \psibar(x)\moverset{\leftarrow}{\dmu} \psi(x) & = & -\sum_x \psibar(x)\partial^*_\mu \psi(x)\\
	    \sum_x \psibar(x)\moverset{\leftarrow}{\partial}^*_\mu \psi(x) & = & -\sum_x \psibar(x)\dmu \psi(x)\,.
	\end{eqnarray}
	and can be used to define the left action of the Dirac operator
	\begin{eqnarray}
	    \moverset{\leftarrow}{D} & = & \tfrac{1}{2}[(\moverset{\leftarrow}{\partial}^*_\mu 
	    				+ \moverset{\leftarrow}{\dmu})\gmu - a \moverset{\leftarrow}{\partial}^*_\mu\moverset{\leftarrow}{\dmu}]\\
	    \moverset{\leftarrow}{D}^\dagger & = & \tfrac{1}{2}[-(\moverset{\leftarrow}{\partial}^*_\mu 
	    				+ \moverset{\leftarrow}{\dmu})\gmu - a \moverset{\leftarrow}{\partial}^*_\mu\moverset{\leftarrow}{\dmu}]\,.
	\end{eqnarray}
	In particular, the action in the interior of the lattice can be written
	\begin{equation}
	    a^2 \sum_{x_0=a}^{T-a} \sum_{x_1=0}^{L-a}\, \psibar(x)\{D + m_0\}\psi(x) = 
	    a^2 \sum_{x_0=a}^{T-a} \sum_{x_1=0}^{L-a}\, \psibar(x)\{\moverset{\leftarrow}{D}^\dagger + m_0\}\psi(x)\,.
	\end{equation}

\section{Free theory}\label{Def:FreeTheory}

\subsection{Formulae}
	The positive energy plane wave solutions of the Dirac equation are
	\begin{equation}
	    \psi(x) = \e^{ipx}\,,\quad \mathrm{Im}\;p_0 > 0\,,
	\end{equation}
	with spatial momentum $p_1$ integer multiple of $2\pi/L$ in the range
	\begin{equation}\label{mom_range}
	    -\pi/a < p_1 \le \pi/a\,,
	\end{equation}
	The energy
	\begin{equation}\label{PosEnergy}
	    p_0 = p^\+_0 = i\omega(p^\+_1) \mod 2\pi/a\,,
	\end{equation}
	is constrained by Dirac equation giving
	\begin{equation}
	    (\po^\+)^2 + M(p^\+)^2=0\,,
	\end{equation}
	\begin{equation}
	    M(p) = m_0 + \tfrac{a}{2}\hat{p}^2\,,
	\end{equation}
	which defines $\omega(q_1)$
	\begin{equation}
	    \sinh\left[\tfrac{a}{2}\omega(q_1)\right] = \frac{a}{2}\left\{
	        \frac{\oover{q}_1^2 + (m_0 + \frac{a}{2}\hat{q}_1^2)^2}{1 + a(m_0 + \frac{a}{2}\hat{q}_1^2)} \right\}^{\frac{1}{2}}\,,
	\end{equation}
	This implies $\omega(\pp_1)\ge0$ for $m_0\ge0$.
	
	The following amplitudes appear in the free propagator
	\begin{equation}
	    A(q_1) = 1 + a(m_0 + \tfrac{a}{2}\hat{q}_1^2)\,,
	\end{equation}
	\begin{equation}
	    R(q) = M(q)\left\{1-\e^{-2\omega(q_1)T}\right\} - i\oover{q}_0\left\{1+\e^{-2\omega(q_1)T}\right\}\,.
	\end{equation}

\section{Four fermion operators}

%

%

\subsection{Fierz transformation}\label{FT}
	Fierz transformations connect products of Dirac bilinears by rearranging the order of the
	Dirac spinors. Thus some of these products are not independent.
	
	In two dimensions the set
	\[\Gamma^A=\{1,\go,\gamma_1,\gfive\}\]
	is normalised to
	\begin{equation}\label{normalization}
	    \tr(\Gamma^A \Gamma^B) = 2 \delta_{AB}
	\end{equation}
	and forms a complete basis for complex $2\times 2$ matrices $M_{\alpha\beta}$:
	\[M_{\alpha\beta} = \frac{1}{2}\sum_A\, \Gamma^A_{\alpha\beta}\, \tr(\Gamma^A M)\,.\]
	
	Considering four Dirac spinors $\psibar_1$, $\psi_2$, $\psibar_3$ and $\psi_4$ the general form of the Fierz identity is then
	\begin{equation}\label{fierztrafo}
	    (\psibar_1 \Gamma^A \psi_2)(\psibar_3 \Gamma^B \psi_4) =
	    \sum_{C,D} C^{AB}_{CD}(\psibar_1 \Gamma^C \psi_4)(\psibar_3 \Gamma^D \psi_2)\,.
	\end{equation}
	To fix the unknown coefficients $C^{AB}_{CD}$ consider the left-hand side with all Dirac indices explicit
	\begin{eqnarray*}
	    \psibar_{1\alpha} \Gamma^A_{\alpha\beta} \psi_{2\beta} \psibar_{3\lambda} \Gamma^B_{\lambda\sigma} \psi_{4\sigma}
	     & = & -\Gamma^A_{\alpha\beta} \Gamma^B_{\lambda\sigma} \psibar_{1\alpha} \psi_{4\sigma} \psibar_{3\lambda} \psi_{2\beta}\\
	     & = & -\Gamma^A_{\alpha\beta} \Gamma^B_{\lambda\sigma} M_{\sigma\alpha} {M'}_{\beta\lambda}\\
	     & = & -\frac{1}{4} \sum_{C,D} \Gamma^A_{\alpha\beta} \Gamma^B_{\lambda\sigma} \Gamma^C_{\sigma\alpha} \Gamma^D_{\beta\lambda}
	               \tr(\Gamma^C M)\tr(\Gamma^D M')\\
	     & = & -\frac{1}{4} \sum_{C,D} \tr(\Gamma^C \Gamma^A \Gamma^D \Gamma^B) (\psibar_1 \Gamma^C \psi_4)(\psibar_3 \Gamma^D \psi_2)
	\end{eqnarray*}
	where $M_{\sigma\alpha}=\psibar_{1\alpha} \psi_{4\sigma}$ and ${M'}_{\beta\lambda}=\psibar_{3\lambda} \psi_{2\beta}$
	are $2\times 2$ matrices, that are expanded in the third line in the basis given above. In the last line we used
	\[\tr(\Gamma^C M) = \Gamma^C_{\rho\nu} \psibar_{1\rho} \psi_{4\nu} = \psibar_1 \Gamma^C \psi_4\]
	and similar for $\tr(\Gamma^D M')$. The overall sign is due to the fact that the fields anticommute. Thus the coefficients are determined
	through
	\begin{eqnarray}
	    C^{AB}_{CD} & = & -\frac{1}{4}\,\tr(\Gamma^C \Gamma^A \Gamma^D \Gamma^A)\,.
	\end{eqnarray}
	
	We are mainly interested in the coefficients for $\Gamma^{A}=\Gamma^{B}$.
	One finds 
	\[C^{AA}_{CD} \propto \delta_{CD}\quad \text{for all}\quad \Gamma_C\,,\Gamma_D\,.\]
	The 16 nonzero coefficients $C^{AA}_{CC}$ are collected in \tab{tab:FierzCoefficients}.
	
    \begin{table*}
    \begin{center}
        \begin{tabular}{|c|c|c|}\hline
            $\Gamma^A$ & $\Gamma^C$ & $C^{AA}_{CC}$ \\\hline
            $1$        & $1,\go,\gamma_1,\gfive$ & $-\frac{1}{2}$ \\\hline
            $\gmu$     & $1$ & $-\frac{1}{2}$ \\
                       & $\gfive$ & $\frac{1}{2}$ \\
                   & $\gnu$ & $(-1)^{\delta_{\mu\nu}}\frac{1}{2}$ \\\hline
            $\gfive$   & $1,\gfive$ & $-\frac{1}{2}$ \\
                   & $\gnu$ & $\frac{1}{2}$ \\\hline
        \end{tabular}
    \end{center}
        \caption{Some coefficients of the Fierz transformation \eq{fierztrafo}.}
        \label{tab:FierzCoefficients}
    \end{table*}

	For the subtraction of $\Gamma^{A}=\Gamma^{B}=1$ and $\Gamma^{A}=\Gamma^{B}=\gfive$ only the terms with $\gmu$ survive
	\begin{eqnarray*}
	    (\psibar_1 \psi_1)(\psibar_2 \psi_2) - (\psibar_1\, \gamma_5\, \psi_1)(\psibar_2\, \gamma_5\, \psi_2)
	     & = & -\sum_{\mu} (\psibar_1\, \gamma_\mu\, \psi_2)\, (\psibar_2\, \gamma_\mu\, \psi_1)\,.
	\end{eqnarray*}
	In the case of $\Gamma^A=\gmu$ the terms with $\Gamma^C=\gnu$ cancel due to the implicit sum over $\mu$ and
	the fact that we have just two $\gamma$-matrices:
	\begin{eqnarray*}
	    (\psibar_1 \gmu \psi_1)(\psibar_2 \gmu \psi_2) & = &
	     -(\psibar_1 \psi_2)(\psibar_2 \psi_1) + (\psibar_1\, \gamma_5\, \psi_2)(\psibar_2\, \gamma_5\, \psi_1)\,.
	\end{eqnarray*}
	
\subsection{Flavour mixing}\label{Flavour}
	Using the expansion introduced in \eq{ap:FlavourExp} the matrix $M_{ij}=\psibar_i\, \Gamma\, \psi_j$, where
	$\Gamma$ is a matrix contracting the Dirac indices of $\psibar$ and $\psi$, can be written as
	\[M_{ij}=\frac{1}{2}\sum_A \lambda^{A}_{ij}\, (\psibar\, \Gamma\, \lambda^{A}\, \psi)\,,\]
	and
	\[\sum_{i,j}^N\,M_{ij}\,M_{ji} =
	    \frac{1}{2^2} \sum_{A,B} \Tr[\lambda^{A}\, \lambda^{B}]\,
	    (\psibar\, \Gamma\, \lambda^{A}\, \psi)\, (\psibar\, \Gamma\, \lambda^{B}\, \psi)\,.\]
	Since the matrices $\lambda^{A}$ are normalised to
	\[\Tr[\lambda^{A}\, \lambda^{B}]=2\delta_{AB}\]
	one ends up with
	\[\sum_{i,j}^N\,\psibar_i\, \Gamma\, \psi_j\,\psibar_j\, \Gamma\, \psi_i =
	    \frac{1}{N}(\psibar\, \Gamma\, \psi)^2 + \frac{1}{2}\sum_a\, (\psibar\, \Gamma\, \lambda^a\, \psi)^2 \,.\]

\chapter{Correlation functions}

\section{Properties of the free Wilson propagator}

\subsection{Analytically known quantities}
\label{CF:analytical}
	We work in a half Fourier transformed space, e.g. the free propagator in the interior of the SF is
  \begin{equation}
  		a\sum_{x_1}\, \e^{i k_1 (y_1-x_1)}\, [\psi(x)\psibar(y)] = \Sp(x_0,y_0,k_1)\,.
  \end{equation}

	To evaluate the zeroth and first order diagrams we need the free propagators from boundary to boundary	
  \begin{align}
      a\sum_{x_1}\, \e^{i\p (y_1-x_1)}\, [\zeta'(x_1)\zetabar(y_1)] & 
      					 = k(p_1)\,P_+ \equiv K(p_1)\,, \\
      a\sum_{x_1}\, \e^{i\p (y_1-x_1)}\, [\zeta(x_1)\zetabar'(y_1)] & = \gfive K^\dagger(p_1) \gfive\,,
  \end{align}
	and from the boundaries to the bulk (and vice versa)
  \begin{align}\label{boundary2bulk1}
      a\sum_{y_1}\, \e^{i\p (y_1-x_1)}\, [\psi(x)\zetabar(y_1)] &
      					 = \frac{2\e^{-2\omega(\pp_1)T}}{R(\pp)}\, h(x_0,p_1)\, P_+\\
      					  & \equiv H(x_0,p_1) \,, \\
      a\sum_{y_1}\, \e^{i\p (y_1-x_1)}\, [\zeta(y_1)\psibar(x)] & = \gfive H^\dagger(x_0,p_1) \gfive\,, \\
      a\sum_{y_1}\, \e^{i\p (y_1-x_1)}\, [\psi(x)\zetabar'(y_1)] &
      					 = \frac{2\e^{-2\omega(\pp_1)T}}{R(\pp)}\, h(T-x_0,p_1)\, P_-\\
      					  & \equiv H'(x_0,p_1) \,, \\
      a\sum_{y_1}\, \e^{i\p (y_1-x_1)}\, [\zeta'(y_1)\psibar(x)] & = \gfive H'^\dagger(x_0,p_1) \gfive\,.
			\label{boundary2bulk2}
  \end{align}
	From the definition of the Wilson propagator \eqref{FreePropagator1} one infers
  \begin{align}
  		k(k_1) & = -2i\oover{k}^\+_0\,\frac{A(k^\+_1)}{R(k^\+)}\,\e^{-\omega(k^\+_1)T}\,, \\
  		h(y_0,k_1) & = h_1(y_0,k_1) + i\gl h_2(y_0,k_1)\,, \\
      h_1(y_0,k_1) & = M(k^\+)\sinh(\omega(k^\+_1)(T-y_0)) - i\oover{k}^\+_0 \cosh(\omega(k^\+_1)(T-y_0))\,, \\
      h_2(y_0,k_1) & = - \oover{k}^\+_1 \sinh(\omega(k^\+_1)(T-y_0)) \,.
  \end{align}
	For $k_1=0$ and at zero bare mass we find
  \begin{align}
  		k(0)\big|_{m_0=0} & = \frac{1}{\cosh(\theta T/L)} + \rO(a^2)\,, \\
      h_1(y_0,0)\big|_{m_0=0} & = \theta/L \cosh(\theta (T-y_0)/L) + \rO(a^2)\,, \\
      h_2(y_0,0)\big|_{m_0=0} & = \theta/L \sinh(\theta (T-y_0)/L) + \rO(a^2) \,, \\
      \frac{2\e^{-2\omega(0^\+)T}}{R(0^\+)}\Bigg|_{m_0=0} & = \frac{1}{\theta/L \cosh(\theta T/L)} + \rO(a^2) \label{f41aFactor}\,.
  \end{align}
  
\subsection{The propagator at zero distance}
	Consider the free propagator
  \begin{equation}\label{Propagator}
  	S(x,y) = [\psi(x) \psibar(y)] = \ev{\psi(x) \psibar(y)}_0\,,
  \end{equation}
  which is one of the basic contractions in our computation. For coinciding arguments $y\to x$ and in the continuum theory,
  just by Euclidean invariance,	one infers that $S(x,x)$ must be diagonal in Dirac space.
	This can be made more explicit by looking at the free propagator on the lattice in a periodic box. 
	Since $S(x,x)$ it is a local quantity, it is not sensitive to the special kind of boundary conditions, once the
	continuum limit has been taken. For an infinite periodic box the free
	propagator for Wilson fermions reads
  \begin{equation}\label{PeriodicBox}
      S_{\mathrm{PB}}(x,y) = \int_{-\pi/a}^{\pi/a}\,\frac{\rd^2 p}{2\pi}\, \e^{ip(x-y)}\,\frac{-i\gmu \po_\mu + M(p)}{\po^2+M(p)^2}\\,
  \end{equation}    
	where $M(p)$ is defined as usual (cf. my notes).
	Except for $-i\gmu\po_\mu$, which is odd, and $\e^{ip(x-y)}$, which is one for $y\to x$, all terms in this expression are even.
	Thus $S_{\mathrm{PB}}(x,x)$ has no Dirac structure, i.e. is a multiple of the unit matrix in Dirac space, even at finite lattice spacing.
	This will change if we allow for general boundary conditions including a phase $\e^{i\theta}$. But it will still converge to the identity 
	in the continuum limit.
	
	For the Schr\"odinger functional free propagator the same behaviour can be shown to hold. In particular this means for the dimensionless 
	product
  \begin{equation}\label{Product}
      a\,S(x,x) = \frac{a}{L}\sum_{k_1} \Sp(x_0,x_0,k_1) =  B(x_0) + \rO(a^2/L^2\, \theta_1)\,,
  \end{equation}    
	where $B(x_0)=B_1 + B_2(x_0,\theta_1)\, a/L + \rO(a^2/L^2)$ is a c-number valued function and $\Sp(x_0,y_0,k_1)$ is the propagator
	in the half Fourier transformed space	we do the actual calculation in. Its deviation from a diagonal form in Dirac space is of order
	$a^2/L^2\, \theta_1$.
	For the diagonal part we find
  \begin{equation}\label{DiagonalPart}
      B(x_0)=0.3849001 + \li[0.5000 + \rO((T-2x_0)/L) + \rO(\theta_1)\re]\, a/L + \rO(a^2/L^2)\,,
  \end{equation}

\subsection{Bubble reduction}\label{BubbleReduction}	
\begin{figure}
		\centering
		\includegraphics[scale=0.7]{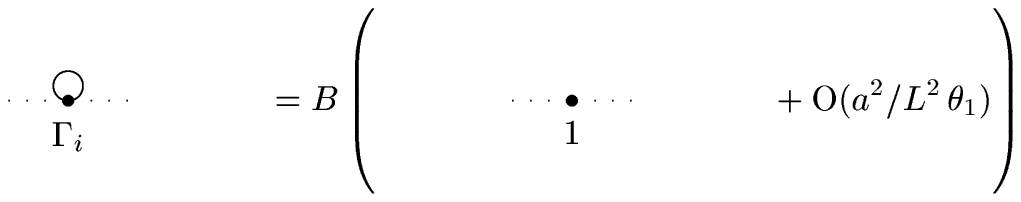}
		\caption{A four fermion interaction contracted to a bubble can be reduced to a insertion of the scalar density.}
		\label{fig:Bubble}
	\end{figure}

	In the perturbative expansion the free propagator at zero distance appears sandwiched between two Dirac structures when two spinors of one 
	and the same four fermion interaction are contracted and produce a bubble in the diagram (see also Fig. \ref{fig:Bubble})
  \begin{align}\label{Bubble}
      \Gamma_I a\,\li[\psi(x) \psibar(x)\re] \Gamma_I 
       				& = a\,S(x,x) + \Gamma_I\,\li[a\,S(x,x),\Gamma_I\re]\\
       				& = B(x_0) + F_I(x)\; a^2/L^2\, \theta_1 + \dots\,,
  \end{align}  
  where on the left hand side the brackets denote a contraction while on the right hand side they denote a commutator.
	This means that in these cases the insertion of the four fermion interactions can be reduced to the insertion of the scalar density
  \begin{multline}\label{Insertion}
      a^2\sum_x\; \ev{\dots\; \psibar(x)\Gamma_I \li[\psi(x)\; \psibar(x)\re] \Gamma_I\psi(x)\, \dots}_0 =\\
       a\sum_x\; \ev{\dots\; \psibar(x)\li(B(x_0) + F_I(x)\; a^2/L^2\, \theta_1 + \dots\re)\psi(x)\, \dots}_0 \,.
  \end{multline}    
	For $\theta_1=0$ (periodic boundary conditions in space) this reduction is exact. 
	
	The insertion of the scalar density in a correlation function in turn can be written as the derivation of this correlation function
	with respect to the mass parameter $m_0$
  \begin{equation}\label{Dm0}
      \frac{\partial}{\partial am_0} \ev{O}_0 = -a\sum_x\; \ev{O\; \psibar(x)\psi(x)}_0\,.
  \end{equation}

\section{Boundary-boundary correlation func\-tions}

\subsection{Free theory}
	In a diagrammatic expansion we find at zeroth order one diagram for the two- and the four-point function \eqref{2fermions} and
	\eqref{4fermions}. In terms of the propagators introduced in the last section they read
  \begin{align}
      f_2^{(0)} & = \TRd[K(0)] = k(0)\,, \\
      f_4^{(0)} & = \TRd[K(0) K^\dagger(0) ] = k^2(0) \label{f40Result}\,.
  \end{align}
	Note that $f_4^{(0)} = \li(f_2^{(0)}\re)^2$. At zero bare mass we find
	\begin{equation}
  		f_2^{(0)}\Big|_{m_0=0} = \frac{1}{\cosh(\theta T/L)} + \rO(a^2)\,.
  \end{equation}
	
\subsection{First order}
\label{FirstOrder}
	At first order there are two diagrams for $f_4$ and one diagram for $f_2$ (cf. Fig. \ref{fig:FirstOrderDiagramsForf4f2}).
	Only diagram a is needed in the
	computation of the ratio \eqref{RC:ratio}. That is because with the help of Lemma \ref{lemma:DiagramIdentity} 
	diagram b can be written as
	$f_2^{(1)}\cdot f_2^{(0)}$ and is hence canceled in the expansion. In terms of the above introduced propagators diagram
	a reads ($H(y_0) = H(y_0,0)$)
	\begin{equation}
  	f_{4,\text{a}}^{(1)} = \sum_I\, c_I\, f_{4,I,\text{a}}^{(1)}\,, \quad I=S,P,V\,,
  \end{equation}
	\begin{equation}\label{f41Ia}
  	f_{4,I,\text{a}}^{(1)} = \frac{a}{L}\, \sum_{x_0}\,
  		 \TRd[\Gamma_B \gfive H^\dagger(x_0) \gfive \Gamma_I H'(x_0) \Gamma'_B \gfive H'^\dagger(x_0)\gfive \Gamma_I H(x_0)]\,.
  \end{equation}
	The convention for coupling constants used in this appendix translates into the one of \eqref{LatticeCGNaction2} via
	\begin{equation}
			c_S = g^2\,,\quad  c_P = g^2- \delta_P^2 \,,\quad\text{and}\quad c_V=g_V^2\,.
	\end{equation}	
	
	Since $\{\gl,h(y_0,p_1)\}=0$ in case of diagram a, it is easier to do the computation with $\Gamma'_B = \Gamma_B = \gl$ (of
	course the result must be the same with $\Gamma'_B = \Gamma_B = \gfive$, this can be used as a check). Then
	\begin{multline}
  	f_{4,I,\text{a}}^{(1)} = \li(\frac{2\e^{-2\omega(0^\+)T}}{R(0^\+)}\re)^4\, \frac{a}{L}\, \sum_{x_0}\, \\
  		 \TRd[h(x_0)P_+ \gl P_- h(x_0) \Gamma_I h(T-x_0)P_- \gl P_+ h(T-x_0) \Gamma_I]\,,
  \end{multline}
	and
  \begin{equation}
  	\TRd[\dots] = \TRd[\gl\Gamma_I\gl\; h(x_0) P_- h(x_0)\; \Gamma_I\;  h(T-x_0) P_- h(T-x_0) ]\,.
  \end{equation}
	Using
  \begin{equation}
  	h P_- h = h_1^2 P_- - h_2^2 P_+ + i\gl h_1 h_2\,,
  \end{equation}
	this results in
  \begin{align*}
  	\Gamma_S = 1\; \to\; \TRd[\dots] = & \li(h_1(x_0)h_1(T-x_0) - h_2(x_0)h_2(T-x_0)\re)^2 \\
  	\Gamma_P = i\gfive\; \to\; \TRd[\dots] = & -\li(h_1(x_0)h_2(T-x_0) - h_2(x_0)h_1(T-x_0)\re)^2 \\
  	\Gamma_V = \go+\gl\; \to\; \TRd[\dots] = & -\Big\{\li(h_1(x_0)h_1(T-x_0) + h_2(x_0)h_2(T-x_0)\re)^2\\
  																						 & +  \li(h_1(x_0)h_2(T-x_0) + h_2(x_0)h_1(T-x_0)\re)^2\Big\} \,.
  \end{align*}
  At zero bare mass we find
  \begin{align*}
  	\Gamma_S = 1\; \to\; \TRd[\dots] = & (\theta/L)^4 \cosh^2(\theta(T-2x_0)/L) + \rO(a^2) \\
  	\Gamma_P = i\gfive\; \to\; \TRd[\dots] = & -(\theta/L)^4 \sinh^2(\theta(T-2x_0)/L) + \rO(a^2) \\
  	\Gamma_V = \go+\gl\; \to\; \TRd[\dots] = & -(\theta/L)^4 \cosh(2\theta T/L)  + \rO(a^2)\,.
  \end{align*}
  Taking the naive continuum limit of the sum in \eqref{f41Ia} for $I=S,P$ we find
  \begin{align*}
  	\frac{a}{L}\, \sum_{y_0=a}^{T-a}\, \cosh^2(\theta(T-2y_0)L) & = \frac{1}{L}\, \int_0^T\, \rd y_0 \cosh^2(\theta(T-2y_0)/L) + \rO(a)\\
  																															& = \frac{1}{4\theta}\,\sinh(2\theta T/L) + \frac{T}{2L}  + \rO(a)  \,,
  \end{align*}
  \begin{align*}
  	\frac{a}{L}\, \sum_{y_0=a}^{T-a}\, \sinh^2(\theta(T-2y_0)L) & = \frac{1}{L}\, \int_0^T\, \rd y_0 \sinh^2(\theta(T-2y_0)/L) + \rO(a)\\
  																															& = \frac{1}{4\theta}\,\sinh(2\theta T/L) - \frac{T}{2L} + \rO(a)\,.
  \end{align*}
  Finally using \eqref{f41aFactor}
	\begin{equation}\label{f41aResult}
  		f_{4,\text{a}}^{(1)}\Bigg|_{m_0=0} = 
  			\frac{T}{2L\,\li(C(\theta)\re)^2}\, \li\{ \li(c_S - c_P\re) A(\theta) + c_S + c_P - 2c_V B(\theta) \re\} + \rO(a)\,,
  \end{equation}
  \begin{equation}
  		A(\theta) = \frac{L}{2\theta T}\,\sinh(2\theta T/L) \stackrel{\theta\to 0}{\to} 1 \,,
  \end{equation}
  \begin{equation}
  		B(\theta) = \cosh(2\theta T/L) \stackrel{\theta\to 0}{\to} 1\,,
  \end{equation}
  \begin{equation}
  		C(\theta) = \cosh^2(\theta T/L) \stackrel{\theta\to 0}{\to} 1\,.
  \end{equation}

\subsection{Second order}
\label{SecondOrder}
	At second order there are nine diagrams for $f_4$ and three diagram for $f_2$ (cf. Fig. \ref{fig:SecondOrderDiagramsForf4f2}).
	But because of Lemma \ref{lemma:DiagramIdentity} we need to evaluate only the five non-reducible diagrams a-e.
	
	The second order diagrams involve one momentum loop and are therefore treated numerically. For the logarithmic divergent part
	we find	
	\begin{multline}\label{f42aResult}
  		f_{4,\text{a}}^{(2)}\Bigg|_{m_0=0} = 
  			\frac{T}{2L\,\li(C(\theta)\re)^2}\, \frac{\ln(a/L)}{2\pi}\, \Big\{ (-c_S^2 - c_P^2 + 2\,c_S\,c_P)\, B(\theta)\\
  			 - 4\, c_V^2\, B(\theta) + 4\, c_V(c_S - c_P)\, A(\theta)\Big\} + \rO(1)\,,
  \end{multline}
	\begin{multline}\label{f42bResult}
  		f_{4,\text{b}}^{(2)}\Bigg|_{m_0=0} = 
  			\frac{T}{2L\,\li(C(\theta)\re)^2}\, \frac{\ln(a/L)}{2\pi}\, \Big\{
  								2\,c_S^2\,\li(A(\theta)+1\re) - 2\,c_P^2\,\li(A(\theta)-1\re) - 4\,c_S\,c_P\\
  			  + 4\, c_V\li(c_S\,(A(\theta)+1) - c_P\,(A(\theta)-1)\re) \Big\} + \rO(1)\,,
  \end{multline}
	\begin{multline}\label{f42cResult}
  		f_{4,\text{c}}^{(2)}\Bigg|_{m_0=0} = 
  			\frac{T}{2L\,\li(C(\theta)\re)^2}\, \frac{\ln(a/L)}{2\pi}\, \Big\{ (c_S^2 + c_P^2 + 2\,c_S\,c_P)\, B(\theta)\\
  											 + 4\, c_V^2\, B(\theta) - 4\, c_V(c_S + c_P)\Big\} + \rO(1)\,,
  \end{multline}
	\begin{multline}\label{f42dResult}
  		f_{4,\text{d}}^{(2)}\Bigg|_{m_0=0} = \frac{-NT}{2L\,\li(C(\theta)\re)^2}\,
  			 \frac{\ln(a/L)}{2\pi}\, \Big\{2\,c_S^2\,\li(A(\theta)+1\re)\\
						 - 2\,c_P^2\,\li(A(\theta)-1\re)\Big\} + \rO(1)\,,
  \end{multline}
	\begin{equation}\label{f42eResult}
  		\li[4 f_{4,\text{e}}^{(2)} + am_c^{(1)} \partial_m f_{4,\text{a}}^{(1)}
  		- 2 \frac{ f_2^{(1)} + am_c^{(1)} \partial_m f_2^{(0)}}{f_2^{(0)}}\, f_{4,\text{a}}^{(1)}\re]_{m_0=0} = \rO(1)
  \end{equation}
  
\subsection{Proof of Lemma \ref{lemma:DiagramIdentity}}
\label{ProofLemma}
	A diagram $f_{2,i}^{n}$ at order $n\ge 0$ in the expansion of $f_2$ (cf. Eq. \eqref{2fermions}) takes the general form
	(the subscript $i$ is only to label the diagram)
	\begin{equation}
  		f_{2,i}^{(n)} = \li(\frac{a}{L}\re)^n\, \sum_{x_0^t\,;\,k_1^t\,;\, I_t}\,
  			\TRd[ V_n\, P_+ ]\,, \quad t=1,\dots, n\,,
  \end{equation}
  where for $n=0$
	\begin{equation}
  		V_0=K(0)\,,
  \end{equation}
  and for $n\ge 1$
  \begin{equation}
  	V_n  = F^2\, h(T-y_0)\, V(x_0^t\,;\,k_1^t\,;\, I_t)\, h(z_0)\,, \quad t=1,\dots, n\,,
  \end{equation}  
	\begin{equation}
  		y_0\,,z_0  \in \{x_0^1,\dots x_0^n\}\,, \quad F = \frac{2\e^{-2\omega(0^\+)T}}{R(0^\+)}\,,
  \end{equation}
  With $h(y_0)=h(y_0,0)$ as defined in Section \ref{CF:analytical}. For example, if $n=1$ there is only one possibility
  \begin{equation}
  	V_1  = F^2\, h(T-x_0^1)\, \Gamma_{I_1}\,\sum_{k_1}\,\Sp(x_0^1,x_0^1,k_1)\,\Gamma_{I_1}\, h(x_0^1)\,.
  \end{equation}  
	For arbitrary $n\ge 1$ the kernel $V(x_0^t\,;\,k_1^t\,;\, I_t)$ is a product of free propagators and $2n$ Dirac
	structures 	$\Gamma_{I_t}=\{1,\gfive,\go+\gl\}$ coming from the vertices.
	The free propagator is $\gfive$ hermitian $\gfive\Sp(x_0,y_0,k_1)^\dagger \gfive=\Sp(y_0,x_0,k_1)$. And since
	besides the propagators	there is an even number of $\gamma$-matices in $V$, it also is $\gfive$ hermitian. Then
	$V_n$ has the property
  \begin{equation}\label{f2property}
  	\gfive V_n^\dagger \gfive = F^2\, h(z_0)\, V(x_0^t\,;\,k_1^t\,;\, I_t)\, h(T-y_0)\,, \quad t=n,\dots, 1\,.
  \end{equation}
  Using properties of the trace and that $f_2$ is real we have the identity
  \begin{equation}\label{f2identity}
  	\TRd[ V_n\, P_+ ] = \TRd[ \gfive V_n^\dagger \gfive\, P_- ]\,.
  \end{equation}

	A reducible diagram $f_{4,i}^{(n)}$ at order $n\ge 0$ takes the general form
	\begin{equation}\label{f4nGeneral}
  		f_{4,i}^{(n)} = \li(\frac{a}{L}\re)^n \sum_{x_0^t\,;\,k_1^t\,;\, I_t}
  		 \TRd[ \Gamma_B P_- D_s P_- \Gamma'_B P_+ U_r P_+]\,,\; t=1,\dots, n\,,\; r+s=n\,.
  \end{equation}
  Because of the projectors $P_{\pm}=\tfrac{1}{2}(1\pm\go)$ only for boundary Dirac structures
  $\Gamma_B\,,\,\Gamma'_B \in \{\gl\,,\,\gfive\}$ Eq. \eqref{f4nGeneral} does not vanish and reduces to
	\begin{equation}\label{f4nGeneral2}
  		f_{4,i}^{(n)} = \li(\frac{a}{L}\re)^n\, \sum_{x_0^t\,;\,k_1^t\,;\, I_t}\,
  		 \TRd[ \Gamma_B D_s \Gamma'_B P_+ U_r P_+]\,.
  \end{equation}
  
  First we concentrate on proving the case $n=0$. For $r,s=0$ we have
	\begin{equation}
  		U_0=K(0)\,,\quad D_0=\gfive K^\dagger(0) \gfive\,.
  \end{equation}
  and there is only one diagram
	\begin{equation}\label{f4nGeneral0}
  		f_{4}^{(0)} = \TRd[ \Gamma_B \gfive K^\dagger(0) \gfive \Gamma'_B P_+ K(0) P_+]\,.
  \end{equation}
  For $\Gamma_B=\Gamma'_B$ we find 
	\begin{equation}\label{f4nGeneral02}
  		f_{4}^{(0)} = \TRd[ K^\dagger(0) P_+ K(0) P_+] = \TRd[ K^\dagger(0) P_+]\cdot \TRd[  K(0) P_+] = f_{2}^{(0)} \cdot f_{2}^{(0)}\,,
  \end{equation}
  where we used $K^\dagger(0) = K(0)$ and which proves the Lemma in this case. For mixed choices for $\Gamma_B\,,\,\Gamma'_B$ there are
  factors $\pm i$ left which can be absorbed into $\Gamma_B$ or $\Gamma'_B$.
  
  Now we turn to arbitrary $n\ge 1$. For $r,s\ge 1$ we find
  \begin{align}
  	U_r & = F^2\, h(T-y_0)\, U(x_0^v\,;\,k_1^v\,;\, I_v)\, h(z_0)\,, \quad v=1,\dots,r \\
  	y_0\,,z_0  & \in \{x_0^1,\dots x_0^r\}
  \end{align} 
  and
  \begin{align}
  	D_s & = F^2\, h(u_0)\, D(x_0^w\,;\,k_1^w\,;\, I_w)\, h(T-v_0)\,, \quad w=r+1,\dots,r+s \label{Dsdef}\\
  	u_0\,,v_0  & \in \{x_0^{r+1},\dots x_0^{r+s}\}
  \end{align}  
  Because of the projectors $P_{\pm}$ in \eqref{f4nGeneral2} and the fact that the $x_0^t$ and $k_1^t$-summations
  do not mix $U_r$ and $D_s^\dagger$ the overall trace factorises into two independent traces
	\begin{equation}\label{f4nGeneral2traces}
  		f_{4,i}^{(n)} = \li(\frac{a}{L}\re)^s\, \sum_{x_0^w\,;\,k_1^w\,;\, I_w}\, \TRd[ \Gamma_B D_s \Gamma'_B P_+]
  		  \cdot \li(\frac{a}{L}\re)^r\, \sum_{x_0^v\,;\,k_1^v\,;\, I_v}\, \TRd[ U_r P_+]\,.
  \end{equation}
  The second factor is evidently an $f_2$ diagram. Because of $\Gamma'_B\,,\Gamma_B \in \{\gl\,,\,\gfive\}$ and 
  the cyclic properties of the trace the trace of the first factor can always be written as (one may have to absorb some factor $\pm i$ into 
  $\Gamma'_B$ or $\Gamma_B$)
	\begin{equation}\label{f4identity}
  		\TRd[ \Gamma_B D_s \Gamma'_B P_+] = \TRd[D_s\, P_-]\,.
  \end{equation}
  Comparing Eqs.  \eqref{f4identity}, \eqref{Dsdef} with \eqref{f2identity}, \eqref{f2property} also the first factor in Eq.
  \eqref{f4nGeneral2traces} can be identified with an $f_2$ diagram and we arrive at
 	\begin{equation}
  		f_{4,i}^{(n)} =  f_{2,j}^{(r)} \cdot f_{2,k}^{(s)}\,.
  \end{equation}
	It should be obvious that this holds also for $s\ge 1\,, r=0$, or vice versa, which completes the proof of Lemma \ref{lemma:DiagramIdentity}.

\section{Boundary-to-interior correlation functions}
\label{boundary2interior}
\subsection{Free theory}
  
	 Here we consider tree level correlation functions of the type
   \begin{equation}\label{twofermionbulk}
       C_{\mathrm{2f,i}}(\GammaI,\GammaR;p_1,x_0) = -a^2 \sum_{y_1 z_1}\, \e^{i\p(y_1-z_1)}
                           \left\<\psibar(x)\GammaI\psi(x)\, \zetabar(y_1)\GammaR \zeta(z_1)\right\z_0\,,
   \end{equation}
   where $\GammaI$ is a matrix with Dirac and flavour indices specifying the quantum numbers of the
   current or density that is inserted at $x$.
   The disconnected part of \eq{twofermionbulk} is proportional to
   \begin{equation}
       \TRdf[S(x,x)\GammaI]\, \TRdf[ P_- S P_+ \GammaR - a P_-\gl i\po^\+_1 \GammaR]
   \end{equation}
   and the connected part
   \begin{equation}
       \TRdf[P_- S\, \GammaI\ S P_+ \GammaR ]\,.
   \end{equation}
   Both expression vanish for $\GammaR \propto 1,\go$. The disconnected part vanishes for nontrivial flavour structure
   of either $\GammaR$ or $\GammaI$. The connected part can be written in terms of propagators between boundary and interior using
   \eqsref{boundary2bulk1}{boundary2bulk2}
   \begin{equation}\label{B2I}
       \TRdf[H^\dagger(x_0,\p)\, \gfive \GammaI\, H(x_0,\p)\, \GammaR \gfive ]\,.
   \end{equation}
   The nonvanishing combinations of $\GammaR,\GammaI$ are listed in the table
   \tab{tab:CurrentDensity}. $\GammaI \propto 1,\go$ are excluded by parity.

   \begin{table*}
   \begin{center}
       \begin{tabular}{|c|c|}\hline
           $\GammaR$ & $\GammaI$ \\\hline
           $\gl$,$\gfive$ & $\gl$,$\gfive$ \\
           $\gl\lambda^a$,$\gfive\lambda^a$ & $\gl\lambda^a$,$\gfive\lambda^a$ \\\hline
       \end{tabular}
   \end{center}
       \caption{Nonvanishing Dirac and flavour structure for correlation functions with current/density insertions.}
       \label{tab:CurrentDensity}
   \end{table*}
   
	 If we introduce the shorthand
   \begin{equation}\label{BLoop1}
       L(\p)  =  a \sum_{y_1}\, \e^{i\p(y_1-z_1)}\, P_-S(a,z_1;a,y_1)P_+\,,\\       
    \end{equation}
	in the  disconnected part the the correlation function \eq{twofermionbulk} reads
   \begin{equation}
       C_{\mathrm{2f,i}}(\GammaI,\GammaR;\p,x_0) = C^{(1)}_{\mathrm{2f,i}} - C^{(2)}_{\mathrm{2f,i}}\,,
   \end{equation}
   \begin{align}
       C^{(1)}_{\mathrm{2f,i}} & = \TRdf[H^\dagger(x_0,\p)\, \gfive \GammaI\, H(x_0,\p)\, \GammaR \gfive ]\,,\\
       C^{(2)}_{\mathrm{2f,i}} & = \sum_{k_1}\, \TRdf[\Sp(x_0,x_0,k_1)\GammaI]\,\TRdf[L(p_1)\GammaR]\,.
   \end{align}
	The half Fourier transformed of the propagator $\Sp$ is defined in \eq{halfFTProp}. The minus sign infront of $C^{(2)}_{\mathrm{2f,i}}$
	is due to the additional trace of the disconnected diagram (cf. Section \ref{FeynmanRules}).
	
   Since
   \begin{equation}
       H(x_0,\p)\gfive=H(x_0,\p)i\go\gl=iH(x_0,\p)\gl\,,
   \end{equation}
   there are only two independent zero momentum correlation functions with current/density insertion and vanishing disconnected
   diagrams. Namely the insertion of the space component of the vector current $\gl$ and pseudo-scalar density $\gfive$. 
   Note that in two dimensions $\gmu\gfive = i\epsilon_{\mu\nu} \gamma_\nu$, therefore $A_\mu = i\epsilon_{\mu\nu} V_\nu$
   and $f_A=if_V$, where $f_A$ is the correlation function of the insertion of the time component of the axial current.
   Comparing \eqref{twofermionbulk} with \eqref{fXtree} we have
   \begin{equation}
       f^{(0)}_A(x_0) = \frac{1}{2N}\,C_{\mathrm{2f,i}}(\go\gfive,\gfive;\p,x_0)\,,
   \end{equation}
   and
   \begin{equation}
       f^{(0)}_P(x_0) = \frac{1}{2N}\,C_{\mathrm{2f,i}}(\gfive,\gfive;\p,x_0)\,.
   \end{equation}
   Using eqs. \eqref{FreePropagator2}, \eqref{boundary2bulk1}, \eqref{BLoop1} and performing some algebra these correlation functions
   explicitly read
   \begin{equation}
       f^{(0)}_X(x_0) = f^{(0,1)}_X + f^{(0,2)}_X\,,
   \end{equation}
   with
   \begin{align}
       f^{(0,1)}_A & = \frac{1}{R(\pp)^2}\Bigg\{ 2M_+(\pp)M_-(\pp)\e^{-2\omega(\pp_1)T} \nonumber\\
       			& - M(\pp)\li[M_-(\pp)\e^{-2\omega(\pp_1)x_0} + M_+(\pp)\e^{-2\omega(\pp_1)(2T-x_0)}\re]\Bigg\}\,,\\
       f^{(0,2)}_A & = N\,\po^+_1\,\frac{A(\pp_1)}{R(\pp)}\li(1-\e^{-2\omega(\pp_1)T} \re)\,
       					\sum_{k_1}\,\frac{-\oover{k}^+_1}{2i\oover{k}^+_0 A(k^+_1) R(k^+)}\\
       			& \li\{ M_-(k^+) + M_+(k^+)\e^{-2\omega(k^+_1)T} - M(k^+)\li(\e^{-2\omega(k^+_1)x_0}+\e^{-2\omega(k^+_1)(T-x_0)}\re) \re\}\,,
   \end{align}
   and
   \begin{align}
       f^{(0,1)}_P & = -\frac{i\po^+_0}{R(\pp)^2}\li\{ M_-(\pp)\e^{-2\omega(\pp_1)x_0} - M_+(\pp)\e^{-2\omega(\pp_1)(2T-x_0)}\re\}\,,\\
       f^{(0,2)}_P & = N\,\po^+_1\,\frac{A(\pp_1)}{R(\pp)}\li(1-\e^{-2\omega(\pp_1)T} \re)\,
       					\sum_{k_1}\, \frac{\oover{k}^+_1}{2 A(k^+_1) R(k^+)}\\
       			& \quad \li\{\e^{-2\omega(k^+_1)x_0}-\e^{-2\omega(k^+_1)(T-x_0)}\re\}\,.
   \end{align}
   Here we used the abbreviation
   \begin{equation}
       M_{\pm}(\pp) = M(\pp)\pm i\po^+_0\,.
   \end{equation}
   Note that
   \begin{equation}\label{hAtree1}
       \tfrac{1}{2}(\partial_0^* + \partial_0)\,f^{(0,1)}_A(x_0) = 2\,M(\pp)\cosh\li[a\omega(\pp_1)\re]\; f^{(0,1)}_P(x_0)\,,
   \end{equation}
   and
   \begin{multline}\label{hAtree2}
       \tfrac{1}{2}(\partial_0^* + \partial_0)\,f^{(0,2)}_A(x_0)  = N\,\po^+_1\,\frac{A(\pp_1)}{R(\pp)}\li(1-\e^{-2\omega(\pp_1)T} \re)\\
       			 \sum_{k_1}\, \frac{\oover{k}^+_1\,M(k^+)\cosh\li[a\omega(k^+_1)\re]}{A(k^+_1) R(k^+)}\,
       					 \li\{\e^{-2\omega(k^+_1)x_0}-\e^{-2\omega(k^+_1)(T-x_0)}\re\}\,.
   \end{multline}

\chapter{Perturbation theory vs. Monte Carlo simulation}

\section{Full theory with bosonic auxiliary fields}
\label{auxiliary_fields}
	Consider the chiral Gross-Neveu model in two dimensions with Wilson fer\-mi\-ons and the interaction terms $\Oss\,, \Opp\,, \Ovv$.
	In the Schr\"odinger functional set up correlation functions are calculated from the generating functional $Z$
	\begin{equation}\label{Auxiliary:CorrFunc}
	        \<O\z = \left\{ \frac{1}{Z}\, O\, Z \right\}_{\rhobar\dots\eta=0}\,,
	\end{equation}
	where
	\begin{equation}\label{Auxiliary:GenFunc}
	        Z = \int \rD[\psi]\rD[\psibar] \exp\left\{-S_0 -  S_I + (\psibar,\eta) + (\etabar,\psi) \right\}\,,
	\end{equation}
	and
	\begin{eqnarray}\label{Auxiliary:Action}
	        S_I & = & -a^2\sum_{x_0=a}^{T-a}\sum_{x_1=0}^{L-a}\,
	        	\left\{ \frac{g_S^2}{2}(\psibar\psi)^2 +  \frac{g_P^2}{2}(\psibar i\gfive\psi)^2 +  \frac{g_V^2}{2}(\psibar\gmu\psi)^2\right\}\,.
	\end{eqnarray}
	In \eq{Auxiliary:Action} the interaction terms are defined only for $0<x_0<T$. We could extent the definition to the boundary, but
	the fields there are no dynamical variables and do not contribute to the following considerations. In fact the fermionic integration in 
	\eq{Auxiliary:GenFunc} is only over fields $\psibar(x)$, $\psi(x)$ with $0<x_0<T$. The free action $S_0$ is defined in
	\eq{SF:free_action} with the Wilson-Dirac operator in presence of the boundaries \eqref{massive_wilson_dirac}. The choice of coupling
	constants translates into the one of \eqref{LatticeCGNaction2} via
	\begin{equation}
			g_S^2 = g^2\quad\text{and}\quad  g_P^2 = g^2-\delta_P^2 \,.
	\end{equation}
	
	On the right hand side of \eq{Auxiliary:CorrFunc} the operators in $O$ on the left hand side have been replaced by the corresponding
	functional derivatives. In perturbation theory one expands the 
	interaction part of the generating functional in the coupling constant $g_I^2$. In this way all sorts of vertices are added
	to the external operators in $O$.
	Here we want to introduce bosonic auxiliary fields
	absorbing the interaction terms and leaving a free fermionic theory with a modified Dirac operator.
	
	Introducing auxiliary fields defined by
	\begin{eqnarray}\label{Auxiliary:Fields}
		\sigma & \equiv & -g_{S}^2\,\psibar\psi\\
		\pi & \equiv & -g_{P}^2\,\psibar i\gfive \psi\\
		B_\mu & \equiv & -g_{V}^2\,\psibar \gmu \psi
	\end{eqnarray}
	into the generating functional via 
	\begin{equation}
		1 \propto \int \rD [X] \e^{-\tfrac{1}{2 g_X^2}\li(X+g_X^2\psibar\Gamma_X\psi\re)^2}
	\end{equation}
	the fermionic integration becomes Gaussian.

	But the generating functional now contains also the integration over the auxiliary fields and the
	free action now depends on them
	\begin{eqnarray}
	   S_0[\sigma,\pi,B_\mu] & = & \phantom{+\,} a \sum_{x_1=0}^{L-a}\, \psibar(0,x_1)P_-\{ a\gl \md_1 \psi(0,x_1) - \psi(a,x_1)\} \nonumber\\
	     &   & +\, a^2 \sum_{x_0=a}^{T-a} \sum_{x_1=0}^{L-a}\, \psibar(x)\{D+m_0+\sigma+i\gfive\pi+\gmu B_\mu\}\psi(x)\nonumber\\
	     &   & +\, a \sum_{x_1=0}^{L-a}\, \psibar(T,x_1)P_+\{ a\gl \md_1 \psi(T,x_1) - \psi(T-a,x_1)\}\,.\label{Auxiliary:FreeAction}
	\end{eqnarray}
	The the generating functional is up to constant factor
	\begin{equation}\label{Auxiliary:GenFunc2}
	        Z \propto \int \rD[\sigma]\rD[\pi]\rD[B_\mu]\, Z_0[\sigma,\pi,B_\mu]\, \e^{-S_a}\,,
	\end{equation}
	with the fermionic part of the generating functional
	\begin{equation}\label{Auxiliary:FreeGenFunc}
	        Z_0[\sigma,\pi,B_\mu] = \int \rD[\psi]\rD[\psibar]\, \exp\left\{-S_0[\sigma,\pi,B_\mu] + (\psibar,\eta) + (\etabar,\psi)\right\}\,,
	\end{equation}
	and the kinetic terms for the auxiliary fields
	\begin{equation}\label{Auxiliary:Kinetic}
	        S_a = a^2 \sum_{x_0=a}^{T-a} \sum_{x_1=0}^{L-a}\,
	        	\left\{\tfrac{1}{2g_{S}^2}\sigma^2 + \tfrac{1}{2g_{P}^2}\pi^2 + \tfrac{1}{2g_{V}^2}B_\mu^2\right\}\,.
	\end{equation}
	
	The operator $O$ contains functional derivatives only with respect to fer\-mi\-onic fields. We can therefore write the expectation
	value \eq{Auxiliary:CorrFunc}
	\begin{equation}\label{Auxiliary:CorrFunc2}
	        \<O\z = \frac{1}{Z} \int \rD[\sigma]\rD[\pi]\rD[B_\mu]\;
	        		 \<O\z_a\, Z_0[\sigma,\pi,B_\mu]_{\rhobar\dots\eta=0}\, \e^{-S_a}\,,
	\end{equation}
	where $\<\cdot\z_a$ is taken with $Z_0[\sigma,\pi,B_\mu]$ and the subscript $a$ indicates, that it is still dependent on the
	auxiliary fields
	\begin{equation}\label{Auxiliary:CorrFunc3}
	        \<O\z_a = \left\{ O\, Z_0[\sigma,\pi,B_\mu] \right\}_{\rhobar\dots\eta=0}\,.
	\end{equation}
	Introducing the Dirac operator $D_a$ depending on the auxiliary fields
	\begin{equation}\label{Auxiliary:DriacOp}
	        D_a = D+\sigma+i\gfive\pi+\gmu B_\mu\,,
	\end{equation}
	the fermionic functional integral $Z_0[\sigma,\pi,B_\mu]$ for vanishing source and boundary fields is the determinant
	\begin{equation}\label{Auxiliary:Determinant}
	        Z_0[\sigma,\pi,B_\mu]_{\rhobar\dots\eta=0} = \left(\det(D_a+m_0) \right)^{N}\,.
	\end{equation}
	
	The expectation value \eq{Auxiliary:CorrFunc3} is just a sum of tree diagrams with the propagator $S_a(x,y)$ defined through
	\begin{equation}
	   	(D_a+m_0)\,S_a(x,y) = a^{-2}\delta_{xy}\,,\quad 0<x_0<T\,.
	\end{equation}
	Using the contractions derived in Section \ref{SF:generating_functional} every combination of field operators in $O$ can be expressed in terms of the 
	propagator $S_a(x,y)$. The remaining functional integral over the c-number valued auxiliary fields is accessible by numerical methods
	\begin{equation}\label{Auxiliary:CorrFunc4}
	        \<O\z = \frac{1}{Z} \int \rD[\sigma]\rD[\pi]\rD[B_\mu]\; \<O\z_a\, \left(\det(D_a+m_0) \right)^{N}\, \e^{-S_a}\,,
	\end{equation}
	given the determinant can be well defined. Since we can choose a representation of the $\gamma$-matrices where $\gfive$ is imaginary
	and $\go\,,\gl$ real, $Q_{xy} = (D_a+m_0)(x,y)$ is a real matrix. Therefore the eigenvalues have to come in complex conjugate pairs and the
	determinant is real
	\begin{equation}
	        \det Q = (\det Q)^* = \det Q^\dagger\,.
	\end{equation}
	But because the determinant in \eq{Auxiliary:CorrFunc4} can be written as $\e^{\log(\det Q)^{N}}$, it has to be strictly positive.
	This, however, is the case for even $N$.

\section{Results}
\label{CPTMC:results}
	
We calculate at finite lattice spacing
\begin{equation}
	am(x_0) = \frac{a\tilde{\partial}_0 f_A(x_0)}{2 f_P(x_0)}
\end{equation}
in Monte Carlo simulations and in perturbation theory (PT). In the simulations we use the representation \eqref{Auxiliary:CorrFunc4} and a standard fermion
algorithm \cite{Montvay:1994cy}. Choosing
\begin{equation}
	g_P^2 = g_S^2\,, \quad a\Delta m = am_0 - am_c\,,
\end{equation}
with
\begin{equation}
	am_c =  -0.7698004(1) \; \li(N\,g_S^2-g_V^2\re) + \rO(g^4)
\end{equation}
we find in leading order perturbation theory (cf. Chapter \ref{chiral_sym_restoration})
\begin{align}
	am(x_0) & = \frac{h_0}{2 f_P^{(0)}} 
								+ \sum_I\, g_I^2\li( \frac{h_{1,I}}{2 f_P^{(0)}} - \frac{h_0\,f_{P,I}^{(1)}}{2 (f_P^{(0)})^2}\re) + \rO(g^4)\\
	                 & = \frac{h_0}{2 f_P^{(0)}}\Bigg|_{am_0=a\Delta m}\\
	                    & \phantom{xxxx}+ \sum_I\, g_I^2\Bigg( am_c^{(1)}\li( \frac{h_2}{2 f_P^{(0)}} - \frac{h_0\,\partial_{am_0}f_P^{(0)}}{2 (f_P^{(0)})^2} \re)\nonumber\\
	                 		&	  \phantom{xxxxxxxx} + \frac{h_{1,I}}{2 f_P^{(0)}} - \frac{h_0\,f_{P,I}^{(1)}}{2 (f_P^{(0)})^2}\Bigg)\Bigg|_{am_0=a\Delta m} + \rO(g^4)
\end{align}
where
\begin{equation}
	h_0=a\tilde{\partial}_0 f_A^{(0)}(x_0)\,, \quad h_1=a\tilde{\partial}_0 f_{A,I}^{(1)}(x_0)\,, \quad h_2=\partial_{am_0}h_0\,,
\end{equation}
and in the second line we expanded $am_0$ around $a\Delta m$.

For $p_1=0$ (external momentum) and $\theta_1=0$ (for these values the disconnected diagram vanishes)
the tree level amplitudes can be computed analytically, giving
\begin{equation}
	am(x_0) = aM(\pp)\cosh(a\omega(\pp_1)) + \rO(g^2)\\,.
\end{equation}.

\begin{figure}
	\centering
	\includegraphics[width=\textwidth]{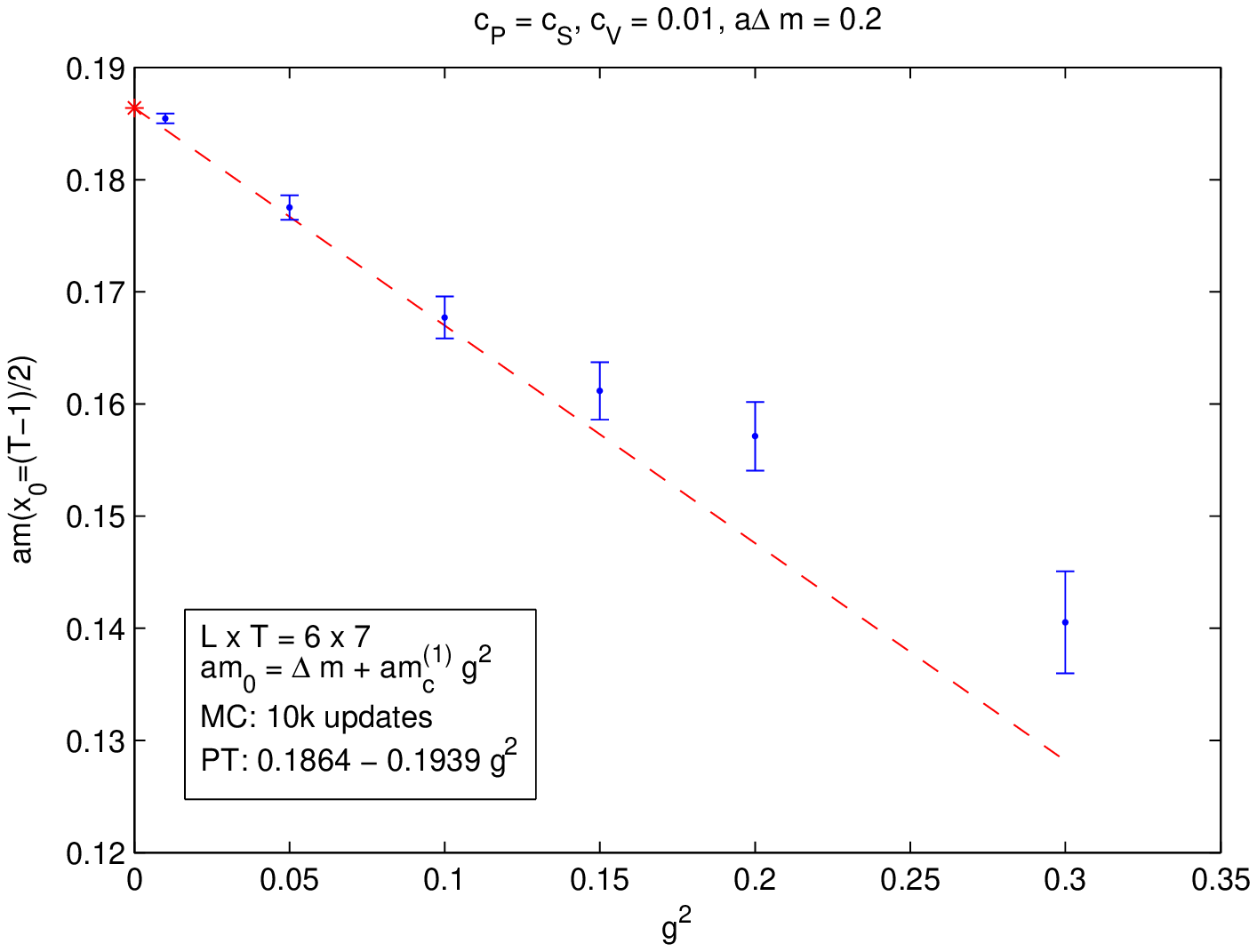}
	\caption{MC vs. PT.}
	\label{fig:MCvsPT}
\end{figure}

Monte Carlo data and PT are plotted in Fig. \ref{fig:MCvsPT}. First order PT seems to be valid at $g^2 \lessapprox 0.15$. See \cite{Korzec:2005ed}
for the same plot for a $12\times 13$ lattice.

\pagestyle{empty}

\chapter*{Acknowledgments}
I want to take the opportunity to thank the people that helped me completing this work:
\begin{itemize}
\item Many thanks to Rainer Sommer for giving me the opportunity to work with him on this project.
				 Without his advice and guidance this thesis would not have been possible. 
\item I am in dept of Ulli Wolff for being so patient at the end of writting this (it seemed like there would be no end) and
				for uncountably many fruitful discussions.
\item Thanks to all the people that were also in Zeuthen at the time, especially Damiano, Andrea, Beatrix, Isabella, Nicolas, Volker. 
\item Special thanks to the other members of the COM group in the last three years: Francesco, Tomasz, Roland, Andreas, Michele, Oliver. It was always
				fun to be there.
\item Thank you Sylvia and Tabea. I can not say, how grateful I am for having such a lovely family. They are a continuous source of energy and peace.
\end{itemize}


\chapter*{Lebenslauf}

\begin{tabular}{ll}

Name: & \dcauthorname  \dcauthorsurname \\

1985 - 91 & Otto-Engert-Oberschule Posa \\

 1991 - 97		& Friedrichgymnasium Altenburg\\

1997		&	Abitur \\

 1997 - 2003	& 	Diplomstudiengang  Physik an der\\
											& Humboldt Universität Berlin\\

 WS 00/01	 & 	Physikstudium in Großbritannien\\ 
									&  an der University of Wales Swansea\\

 Okt. 2003	&	Diplom Physik\\
 
Okt. 2002 - Dez. 2003  & 	Studentische Hilfskraft bei\\
																		& Prof. U. Wolff, HU Berlin\\
																		
Jan. 2004 - Dez. 2006	 &	Wissenschaftliche Hilfskraft,\\
																		& Theory Group, DESY Zeuthen																		
																		
\end{tabular}


\chapter*{Selbständigkeitserklärung}

\noindent Hiermit erkl\"are ich, die vorliegende Arbeit selbst\"andig ohne fremde Hilfe verfa{\ss}t und nur die angegebene Literatur und Hilfsmittel verwendet zu haben.\\[2cm]

\noindent Bj\"orn Leder\\
30. Januar 2007



\begin{thebibliography}{80}
\providecommand{\natexlab}[1]{#1}
\providecommand{\url}[1]{\texttt{#1}}
\expandafter\ifx\csname urlstyle\endcsname\relax
  \providecommand{\doi}[1]{doi: #1}\else
  \providecommand{\doi}{doi: \begingroup \urlstyle{rm}\Url}\fi

\bibitem[Yao et~al.(2006)]{Yao:2006px}
W.~M. Yao et~al.
\newblock Review of particle physics.
\newblock \emph{J. Phys.}, G33:\penalty0 1--1232, 2006.

\bibitem[Wilson(1974)]{Wilson:1974sk}
Kenneth~G. Wilson.
\newblock Confinement of quarks.
\newblock \emph{Phys. Rev.}, D10:\penalty0 2445--2459, 1974.

\bibitem[Della~Morte et~al.(2005{\natexlab{a}})]{DellaMorte:2004bc}
Michele Della~Morte et~al.
\newblock Computation of the strong coupling in qcd with two dynamical
  flavours.
\newblock \emph{Nucl. Phys.}, B713:\penalty0 378--406, 2005{\natexlab{a}}.

\bibitem[Luscher et~al.(1991)Luscher, Weisz, and Wolff]{Luscher:1991wu}
Martin Luscher, Peter Weisz, and Ulli Wolff.
\newblock A numerical method to compute the running coupling in asymptotically
  free theories.
\newblock \emph{Nucl. Phys.}, B359:\penalty0 221--243, 1991.

\bibitem[Aoki(1984)]{Aoki:1983qi}
Sinya Aoki.
\newblock New phase structure for lattice qcd with wilson fermions.
\newblock \emph{Phys. Rev.}, D30:\penalty0 2653, 1984.

\bibitem[Symanzik({\natexlab{a}})]{Symanzik:1979ph}
K.~Symanzik.
\newblock Cutoff dependence in lattice phi**4 in four-dimensions theory.
\newblock {\natexlab{a}}.
\newblock DESY 79/76.

\bibitem[Symanzik({\natexlab{b}})]{Symanzik:1981hc}
K.~Symanzik.
\newblock Some topics in quantum field theory.
\newblock {\natexlab{b}}.
\newblock Presented at 6th Int. Conf. on Mathematical Physics, Berlin, West
  Germany, Aug 11-21, 1981.

\bibitem[Symanzik(1983{\natexlab{a}})]{Symanzik:1983dc}
K.~Symanzik.
\newblock Continuum limit and improved action in lattice theories. 1.
  principles and phi**4 theory.
\newblock \emph{Nucl. Phys.}, B226:\penalty0 187, 1983{\natexlab{a}}.

\bibitem[Symanzik(1983{\natexlab{b}})]{Symanzik:1983gh}
K.~Symanzik.
\newblock Continuum limit and improved action in lattice theories. 2. o(n)
  nonlinear sigma model in perturbation theory.
\newblock \emph{Nucl. Phys.}, B226:\penalty0 205, 1983{\natexlab{b}}.

\bibitem[Knechtli et~al.(2005)Knechtli, Leder, and Wolff]{Knechtli:2005jh}
Francesco Knechtli, Bjorn Leder, and Ulli Wolff.
\newblock Cutoff effects in o(n) nonlinear sigma models.
\newblock \emph{Nucl. Phys.}, B726:\penalty0 421--440, 2005.

\bibitem[Della~Morte et~al.(2005{\natexlab{b}})Della~Morte, Hoffmann, Knechtli,
  Rolf, Sommer, Wetzorke, and Wolff]{DellaMorte:2005kg}
M.~Della~Morte, R.~Hoffmann, F.~Knechtli, J.~Rolf, R.~Sommer, I.~Wetzorke, and
  U.~Wolff.
\newblock Non-perturbative quark mass renormalization in two-flavor qcd.
\newblock \emph{Nucl. Phys.}, B729:\penalty0 117--134, 2005{\natexlab{b}}.

\bibitem[Schaefer and DeGrand(2006)]{Schaefer:2005qg}
Stefan Schaefer and Thomas~A. DeGrand.
\newblock Dynamical overlap fermions: Techniques and results. simulations and
  physics results.
\newblock \emph{PoS}, LAT2005:\penalty0 140, 2006.

\bibitem[Bernard et~al.(2006)]{Bernard:2006wx}
C.~Bernard et~al.
\newblock Update on the physics of light pseudoscalar mesons.
\newblock 2006.

\bibitem[Hasenfratz et~al.(2006)Hasenfratz, Hasenfratz, Niedermayer, Hierl, and
  Schafer]{Hasenfratz:2006xi}
Anna Hasenfratz, Peter Hasenfratz, Ferenc Niedermayer, Dieter Hierl, and
  Andreas Schafer.
\newblock First results in qcd with 2+1 light flavors using the fixed-point
  action.
\newblock \emph{PoS.}, LAT2006:\penalty0 178, 2006.

\bibitem[Del~Debbio et~al.(2006)Del~Debbio, Giusti, Luscher, Petronzio, and
  Tantalo]{DelDebbio:2006cn}
L.~Del~Debbio, L.~Giusti, M.~Luscher, R.~Petronzio, and N.~Tantalo.
\newblock Qcd with light wilson quarks on fine lattices. i: First experiences
  and physics results.
\newblock 2006.

\bibitem[Del~Debbio et~al.(2007)Del~Debbio, Giusti, Luscher, Petronzio, and
  Tantalo]{DelDebbio:2007pz}
L.~Del~Debbio, L.~Giusti, M.~Luscher, R.~Petronzio, and N.~Tantalo.
\newblock Qcd with light wilson quarks on fine lattices. ii: Dd-hmc simulations
  and data analysis.
\newblock 2007.

\bibitem[Boucaud et~al.(2007)]{Boucaud:2007uk}
Ph. Boucaud et~al.
\newblock Dynamical twisted mass fermions with light quarks.
\newblock 2007.

\bibitem[Jansen(2004)]{Jansen:2003nt}
Karl Jansen.
\newblock Actions for dynamical fermion simulations: Are we ready to go?
\newblock \emph{Nucl. Phys. Proc. Suppl.}, 129:\penalty0 3--16, 2004.

\bibitem[Gross and Neveu(1974)]{Gross:1974jv}
David~J. Gross and Andre Neveu.
\newblock Dynamical symmetry breaking in asymptotically free field theories.
\newblock \emph{Phys. Rev.}, D10:\penalty0 3235, 1974.

\bibitem[Thirring(1958)]{Thirring:1958in}
Walter~E. Thirring.
\newblock A soluble relativistic field theory.
\newblock \emph{Annals Phys.}, 3:\penalty0 91--112, 1958.

\bibitem[Destri(1988)]{Destri:1988vb}
C.~Destri.
\newblock Two loop beta function for generalized nonabelian thirring models.
\newblock \emph{Phys. Lett.}, B210:\penalty0 173, 1988.
\newblock Erratum: Phys.Lett. Erratum-ibid. B213:565,1988.

\bibitem[Bondi et~al.(1990)Bondi, Curci, Paffuti, and Rossi]{Bondi:1989nq}
Alessandro Bondi, Giuseppe Curci, Giampiero Paffuti, and Paolo Rossi.
\newblock Metric and central charge in the perturbative approach to
  two-dimensional fermionic models.
\newblock \emph{Ann. Phys.}, 199:\penalty0 268, 1990.

\bibitem[Mitter and Weisz(1973)]{Mitter:1974cy}
P.~K. Mitter and P.~H. Weisz.
\newblock Asymptotic scale invariance in a massive thirring model with u(n)
  symmetry.
\newblock \emph{Phys. Rev.}, D8:\penalty0 4410--4429, 1973.

\bibitem[Dashen et~al.(1975)Dashen, Hasslacher, and Neveu]{Dashen:1975xh}
Roger~F. Dashen, Brosl Hasslacher, and Andre Neveu.
\newblock Semiclassical bound states in an asymptotically free theory.
\newblock \emph{Phys. Rev.}, D12:\penalty0 2443, 1975.

\bibitem[Kurak and Swieca(1979)]{Kurak:1978su}
V.~Kurak and J.~A. Swieca.
\newblock Anti-particles as bound states of particles in the factorized s
  matrix framework.
\newblock \emph{Phys. Lett.}, B82:\penalty0 289, 1979.

\bibitem[Andrei and Lowenstein(1979)]{Andrei:1979sq}
N.~Andrei and J.~H. Lowenstein.
\newblock Diagonalization of the chiral invariant gross-neveu hamiltonian.
\newblock \emph{Phys. Rev. Lett.}, 43:\penalty0 1698, 1979.

\bibitem[Andrei and Lowenstein(1980)]{Andrei:1979un}
N.~Andrei and J.~H. Lowenstein.
\newblock A direct calculation of the s matrix of the chiral invariant
  gross-neveu model.
\newblock \emph{Phys. Lett.}, B91:\penalty0 401, 1980.

\bibitem[Forgacs et~al.(1992)Forgacs, Naik, and Niedermayer]{Forgacs:1991nk}
P.~Forgacs, S.~Naik, and F.~Niedermayer.
\newblock The exact mass gap of the chiral gross-neveu model.
\newblock \emph{Phys. Lett.}, B283:\penalty0 282--286, 1992.

\bibitem[Witten(1978)]{Witten:1978qu}
Edward Witten.
\newblock Chiral symmetry, the 1/n expansion, and the su(n) thirring model.
\newblock \emph{Nucl. Phys.}, B145:\penalty0 110, 1978.

\bibitem[Aoki and Higashijima(1986)]{Aoki:1985jj}
Sinya Aoki and Kiyoshi Higashijima.
\newblock The recovery of the chiral symmetry in lattice gross-neveu model.
\newblock \emph{Prog. Theor. Phys.}, 76:\penalty0 521, 1986.

\bibitem[Izubuchi et~al.(1998)Izubuchi, Noaki, and Ukawa]{Izubuchi:1998hy}
Taku Izubuchi, Junichi Noaki, and Akira Ukawa.
\newblock Two-dimensional lattice gross-neveu model with wilson fermion action
  at finite temperature and chemical potential.
\newblock \emph{Phys. Rev.}, D58:\penalty0 114507, 1998.

\bibitem[Lüscher(2006)]{Luscher:2006df}
Martin Lüscher.
\newblock The schrödinger functional in lattice qcd with exact chiral symmetry.
\newblock \emph{JHEP}, 05:\penalty0 042, 2006.

\bibitem[Ginsparg and Wilson(1982)]{Ginsparg:1981bj}
Paul~H. Ginsparg and Kenneth~G. Wilson.
\newblock A remnant of chiral symmetry on the lattice.
\newblock \emph{Phys. Rev.}, D25:\penalty0 2649, 1982.

\bibitem[Korzec et~al.(2006)Korzec, Knechtli, Wolff, and Leder]{Korzec:2005ed}
Tomasz Korzec, Francesco Knechtli, Ulli Wolff, and Bjorn Leder.
\newblock Monte-carlo simulation of the chiral gross-neveu model.
\newblock \emph{PoS}, LAT2005:\penalty0 267, 2006.

\bibitem[Weinberg(1973)]{Weinberg:1951ss}
Steven Weinberg.
\newblock New approach to the renormalization group.
\newblock \emph{Phys. Rev.}, D8:\penalty0 3497--3509, 1973.

\bibitem[Capitani(2003)]{Capitani:2002mp}
Stefano Capitani.
\newblock Lattice perturbation theory.
\newblock \emph{Phys. Rept.}, 382:\penalty0 113--302, 2003.

\bibitem[Gross and Wilczek(1973)]{Gross:1973id}
D.~J. Gross and Frank Wilczek.
\newblock Ultraviolet behavior of non-abelian gauge theories.
\newblock \emph{Phys. Rev. Lett.}, 30:\penalty0 1343--1346, 1973.

\bibitem[Politzer(1973)]{Politzer:1973fx}
H.~David Politzer.
\newblock Reliable perturbative results for strong interactions?
\newblock \emph{Phys. Rev. Lett.}, 30:\penalty0 1346--1349, 1973.

\bibitem[Jones(1974)]{Jones:1974mm}
D.~R.~T. Jones.
\newblock Two loop diagrams in yang-mills theory.
\newblock \emph{Nucl. Phys.}, B75:\penalty0 531, 1974.

\bibitem[Caswell(1974)]{Caswell:1974gg}
William~E. Caswell.
\newblock Asymptotic behavior of nonabelian gauge theories to two loop order.
\newblock \emph{Phys. Rev. Lett.}, 33:\penalty0 244, 1974.

\bibitem[Weinberg(1996)]{Weinberg:1996kr}
Steven Weinberg.
\newblock \emph{The quantum theory of fields. Vol. 2: Modern applications}.
\newblock Cambridge University Press, Cambridge, USA, 1996.

\bibitem[Montvay and Munster(1994)]{Montvay:1994cy}
I.~Montvay and G.~Munster.
\newblock \emph{Quantum fields on a lattice}.
\newblock Cambridge Univ. Pr. (Cambridge monographs on mathematical physics),
  Cambridge, UK, 1994.

\bibitem[Peskin and Schroeder(1995)]{Peskin:1995ev}
Michael~E. Peskin and D.~V. Schroeder.
\newblock \emph{An Introduction to quantum field theory}.
\newblock Addison-Wesley, Reading, USA, 1995.

\bibitem[Celmaster and Gonsalves(1979)]{Celmaster:1979km}
William Celmaster and Richard~J. Gonsalves.
\newblock The renormalization prescription dependence of the qcd coupling
  constant.
\newblock \emph{Phys. Rev.}, D20:\penalty0 1420, 1979.

\bibitem[Bode et~al.(2000)Bode, Weisz, and Wolff]{Bode:1999sm}
Achim Bode, Peter Weisz, and Ulli Wolff.
\newblock Two loop computation of the schrödinger functional in lattice qcd.
\newblock \emph{Nucl. Phys.}, B576:\penalty0 517--539, 2000.

\bibitem[Goldstone(1961)]{Goldstone:1961eq}
J.~Goldstone.
\newblock Field theories with 'superconductor' solutions.
\newblock \emph{Nuovo Cim.}, 19:\penalty0 154--164, 1961.

\bibitem[Goldstone et~al.(1962)Goldstone, Salam, and
  Weinberg]{Goldstone:1962es}
Jeffrey Goldstone, Abdus Salam, and Steven Weinberg.
\newblock Broken symmetries.
\newblock \emph{Phys. Rev.}, 127:\penalty0 965--970, 1962.

\bibitem[Nambu and Jona-Lasinio(1961)]{Nambu:1961tp}
Yoichiro Nambu and G.~Jona-Lasinio.
\newblock Dynamical model of elementary particles based on an analogy with
  superconductivity. i.
\newblock \emph{Phys. Rev.}, 122:\penalty0 345--358, 1961.

\bibitem[Lüscher(1998{\natexlab{a}})]{Luscher:1998pe}
Martin Lüscher.
\newblock Advanced lattice qcd.
\newblock 1998{\natexlab{a}}.

\bibitem[Lüscher(1998{\natexlab{b}})]{Luscher:1998pq}
Martin Lüscher.
\newblock Exact chiral symmetry on the lattice and the ginsparg- wilson
  relation.
\newblock \emph{Phys. Lett.}, B428:\penalty0 342--345, 1998{\natexlab{b}}.

\bibitem[Bochicchio et~al.(1985)Bochicchio, Maiani, Martinelli, Rossi, and
  Testa]{Bochicchio:1985xa}
Marco Bochicchio, Luciano Maiani, Guido Martinelli, Gian~Carlo Rossi, and
  Massimo Testa.
\newblock Chiral symmetry on the lattice with wilson fermions.
\newblock \emph{Nucl. Phys.}, B262:\penalty0 331, 1985.

\bibitem[Nielsen and Ninomiya(1981{\natexlab{a}})]{Nielsen:1981hk}
Holger~Bech Nielsen and M.~Ninomiya.
\newblock No go theorem for regularizing chiral fermions.
\newblock \emph{Phys. Lett.}, B105:\penalty0 219, 1981{\natexlab{a}}.

\bibitem[Nielsen and Ninomiya(1981{\natexlab{b}})]{Nielsen:1980rz}
Holger~Bech Nielsen and M.~Ninomiya.
\newblock Absence of neutrinos on a lattice. 1. proof by homotopy theory.
\newblock \emph{Nucl. Phys.}, B185:\penalty0 20, 1981{\natexlab{b}}.

\bibitem[Nielsen and Ninomiya(1981{\natexlab{c}})]{Nielsen:1981xu}
Holger~Bech Nielsen and M.~Ninomiya.
\newblock Absence of neutrinos on a lattice. 2. intuitive topological proof.
\newblock \emph{Nucl. Phys.}, B193:\penalty0 173, 1981{\natexlab{c}}.

\bibitem[Friedan(1982)]{Friedan:1982nk}
D.~Friedan.
\newblock A proof of the nielsen-ninomiya theorem.
\newblock \emph{Commun. Math. Phys.}, 85:\penalty0 481--490, 1982.

\bibitem[Wilson()]{Wilson:1975id}
Kenneth~G. Wilson.
\newblock Quarks and strings on a lattice.
\newblock New Phenomena In Subnuclear Physics. Part A. Proceedings of the First
  Half of the 1975 International School of Subnuclear Physics, Erice, Sicily,
  July 11 - August 1, 1975, ed. A.~Zichichi, Plenum Press, New York, 1977,
  p.~69, CLNS-321.

\bibitem[Hasenfratz(1998)]{Hasenfratz:1998jp}
Peter Hasenfratz.
\newblock Lattice qcd without tuning, mixing and current renormalization.
\newblock \emph{Nucl. Phys.}, B525:\penalty0 401--409, 1998.

\bibitem[Neuberger(1998)]{Neuberger:1997fp}
Herbert Neuberger.
\newblock Exactly massless quarks on the lattice.
\newblock \emph{Phys. Lett.}, B417:\penalty0 141--144, 1998.

\bibitem[Kikukawa and Noguchi(1999)]{Kikukawa:1999sy}
Yoshio Kikukawa and Tatsuya Noguchi.
\newblock Low energy effective action of domain-wall fermion and the
  ginsparg-wilson relation.
\newblock 1999.

\bibitem[Niedermayer(1999)]{Niedermayer:1998bi}
Ferenc Niedermayer.
\newblock Exact chiral symmetry, topological charge and related topics.
\newblock \emph{Nucl. Phys. Proc. Suppl.}, 73:\penalty0 105--119, 1999.

\bibitem[Sint(1994)]{Sint:1993un}
Stefan Sint.
\newblock On the schrodinger functional in qcd.
\newblock \emph{Nucl. Phys.}, B421:\penalty0 135--158, 1994.

\bibitem[Sommer(2006{\natexlab{a}})]{Sommer:2006wx}
Rainer Sommer.
\newblock Determining fundamental parameters of qcd on the lattice.
\newblock \emph{Nucl. Phys. Proc. Suppl.}, 160:\penalty0 27--31,
  2006{\natexlab{a}}.

\bibitem[Symanzik(1981)]{Symanzik:1981wd}
K.~Symanzik.
\newblock Schrödinger representation and casimir effect in renormalizable
  quantum field theory.
\newblock \emph{Nucl. Phys.}, B190:\penalty0 1, 1981.

\bibitem[Lüscher(1985)]{Luscher:1985iu}
M.~Lüscher.
\newblock Schrödinger representation in quantum field theory.
\newblock \emph{Nucl. Phys.}, B254:\penalty0 52--57, 1985.

\bibitem[Lüscher et~al.(1992)Lüscher, Narayanan, Weisz, and
  Wolff]{Luscher:1992an}
Martin Lüscher, Rajamani Narayanan, Peter Weisz, and Ulli Wolff.
\newblock The schrödinger functional: A renormalizable probe for nonabelian
  gauge theories.
\newblock \emph{Nucl. Phys.}, B384:\penalty0 168--228, 1992.

\bibitem[Narayanan and Wolff(1995)]{Narayanan:1995ex}
Rajamani Narayanan and Ulli Wolff.
\newblock Two loop computation of a running coupling lattice yang- mills
  theory.
\newblock \emph{Nucl. Phys.}, B444:\penalty0 425--446, 1995.

\bibitem[Sint(1995)]{Sint:1995rb}
Stefan Sint.
\newblock One loop renormalization of the qcd schrödinger functional.
\newblock \emph{Nucl. Phys.}, B451:\penalty0 416--444, 1995.

\bibitem[Sommer(2006{\natexlab{b}})]{Sommer:2006sj}
Rainer Sommer.
\newblock Non-perturbative qcd: Renormalization, o(a)-im\-prove\-ment and
  matching to heavy quark effective theory.
\newblock 2006{\natexlab{b}}.

\bibitem[Taniguchi(2005)]{Taniguchi:2004gf}
Yusuke Taniguchi.
\newblock Schroedinger functional formalism with ginsparg-wilson fermion.
\newblock \emph{JHEP}, 12:\penalty0 037, 2005.

\bibitem[Diehl(1997)]{Diehl:1996kd}
H.~W. Diehl.
\newblock The theory of boundary critical phenomena.
\newblock \emph{Int. J. Mod. Phys.}, B11:\penalty0 3503--3523, 1997.

\bibitem[Lüscher and Weisz(1996)]{Luscher:1996vw}
M.~Lüscher and P.~Weisz.
\newblock {O}(a) improvement of the axial current in lattice {QCD} to one-loop
  order of perturbation theory.
\newblock \emph{Nucl. Phys.}, B479:\penalty0 429--458, 1996.

\bibitem[Giusti et~al.(2003)Giusti, Hoelbling, Lüscher, and
  Wittig]{Giusti:2002sm}
L.~Giusti, C.~Hoelbling, M.~Lüscher, and H.~Wittig.
\newblock Numerical techniques for lattice qcd in the epsilon-regime.
\newblock \emph{Comput. Phys. Commun.}, 153:\penalty0 31--51, 2003.

\bibitem[Hernandez et~al.(1999)Hernandez, Jansen, and
  Luscher]{Hernandez:1998et}
Pilar Hernandez, Karl Jansen, and Martin Luscher.
\newblock Locality properties of neuberger's lattice dirac operator.
\newblock \emph{Nucl. Phys.}, B552:\penalty0 363--378, 1999.

\bibitem[Furuya et~al.(1982)Furuya, Gamboa~Saravi, and
  Schaposnik]{Furuya:1982fh}
K.~Furuya, R.~E. Gamboa~Saravi, and F.~A. Schaposnik.
\newblock Path integral formulation of chiral invariant fermion models in
  two-dimensions.
\newblock \emph{Nucl. Phys.}, B208:\penalty0 159, 1982.

\bibitem[Moreno and Schaposnik(1989)]{Moreno:1987np}
E.~Moreno and F.~A. Schaposnik.
\newblock On the issues of symmetries in the gross-neveu model.
\newblock \emph{Int. J. Mod. Phys.}, A4:\penalty0 2827--2835, 1989.

\bibitem[Korzec and Wolff(2006)]{Korzec:2006hy}
Tomasz Korzec and Ulli Wolff.
\newblock Gross-neveu model as a laboratory for fermion discretization.
\newblock 2006.

\bibitem[Gracey(1991)]{Gracey:1991vy}
J.~A. Gracey.
\newblock Computation of the three loop beta function of the o(n) gross-neveu
  model in minimal subtraction.
\newblock \emph{Nucl. Phys.}, B367:\penalty0 657--674, 1991.

\bibitem[Wetzel(1985)]{Wetzel:1984nw}
Werner Wetzel.
\newblock Two loop beta function for the gross-neveu model.
\newblock \emph{Phys. Lett.}, B153:\penalty0 297, 1985.

\bibitem[Alexandrou et~al.(2000)Alexandrou, Panagopoulos, and
  Vicari]{Alexandrou:1999wr}
C.~Alexandrou, H.~Panagopoulos, and E.~Vicari.
\newblock Lambda-parameter of lattice qcd with the overlap-dirac operator.
\newblock \emph{Nucl. Phys.}, B571:\penalty0 257--266, 2000.

\bibitem[Capitani et~al.(1999)Capitani, Gockeler, Horsley, Rakow, and
  Schierholz]{Capitani:1999uz}
S.~Capitani, M.~Gockeler, R.~Horsley, P.~E.~L. Rakow, and G.~Schierholz.
\newblock Operator improvement for ginsparg-wilson fermions.
\newblock \emph{Phys. Lett.}, B468:\penalty0 150--160, 1999.

\end{thebibliography}
\end{document}